\documentclass[preprint,floatfix
,aps,prd,superscriptaddress]{revtex4-2}

\usepackage{graphicx}
\usepackage{tikz}
\usepackage{amssymb,amsmath,amsfonts,mathtools,comment,hyperref,bbold}
\usepackage{appendix}
\usepackage[utf8]{inputenc}

\usepackage{array, makecell}
\usepackage{cellspace} %
\usepackage{extarrows}
\usepackage{relsize}

\usepackage{bm}
\usepackage{mathrsfs}
\usepackage{xcolor}

\usepackage{dcolumn}

\newcommand{\Tr}{\rm Tr}
\newcommand{\ket}[1]{\left|#1\right\rangle}
\newcommand{\bra}[1]{\left\langle #1 \right|}
\newcommand{\braket}[2]{\left\langle #1 \! \! \right. \left| #2 \right\rangle}

\newcommand{\T}{\theta}
\newcommand{\D}{\Delta}
\newcommand{\s}{\sin}
\newcommand{\C}{\cos}

\newcommand{\rs}[1]{\rho^{\rm #1}}

\newcommand{\ha}{\hat{a}}

\usepackage{enumitem}
\setlist{  
  listparindent=\parindent,
  parsep=0pt,
}

\usepackage{array}
\newcolumntype{P}[1]{>{\centering\arraybackslash}p{#1}}
\newcolumntype{M}[1]{>{\centering\arraybackslash}m{#1}}

\begin{document}

\title{
Quantum aspects of spacetime:
\\
A quantum optics view of acceleration radiation and black holes}

\author{C. R. Ord\'o\~nez}
\affiliation{Department of Physics, University of Houston, Houston, Texas 77024-5005, USA}

\author{A. Chakraborty}
\affiliation{Institute for Quantum Computing, Department of Physics and Astronomy, Waterloo, 
University of Waterloo, ON N2L 3G1, Canada\looseness=-1}

\author{H. E. Camblong}
\affiliation{Department of Physics and Astronomy, University of San Francisco, San Francisco, California 94117-1080, USA\looseness=-1}

\author{Marlan O. Scully}
\affiliation{Institute for Quantum Science and Engineering, Department of Physics and Astronomy, Texas A\&M University, College Station, Texas 77843, USA\looseness=-1}

\author{William G. Unruh}
\affiliation{Institute for Quantum Science and Engineering, Department of Physics and Astronomy, Texas A\&M University, College Station, Texas 77843, USA\looseness=-1}
\affiliation{Department of Physics and Astronomy, University of British Columbia, Vancouver, British Columbia V6T 1Z1, Canada\looseness=-1}

\begin{abstract}
For the {\it centennial of quantum mechanics\/}, we offer an overview of the central
role played by quantum information and thermalization in problems involving fundamental properties 
of spacetime and gravitational physics.
This is an open area of research still a century after the initial development of formal quantum mechanics, 
highlighting the effectiveness of quantum physics in the description of all natural phenomena.
These remarkable connections can be highlighted with the tools of modern quantum optics,
which effectively addresses the three-fold interplay of interacting atoms, fields, and spacetime backgrounds
describing gravitational fields and noninertial systems.
In this review article, we select aspects of these 
phenomena centered on quantum features of the acceleration radiation of particles in the presence of black holes.
The ensuing horizon-brightened radiation (HBAR) provides a case study of the role played by quantum physics
in nontrivial spacetime behavior, and also shows a fundamental correspondence with black hole thermodynamics.

\end{abstract}

\maketitle
\newpage
\tableofcontents

\section{Quantum physics: Introduction and historical overview}
\label{sec:introduction}

A century of developments in {\it quantum mechanics\/},
following the successful establishment of the matrix and wave mechanics frameworks 
in 1925 and 1926~\cite{QM-conceptual,QM-historical, QM-sources, Schrodinger-wave-mech-papers},
has provided the foundations for most areas of science and technology, leading to a vast range of theoretical and practical applications, and a complete redesign of our view of nature~\cite{Weinberg_QM}. 
This radical transformation offers a unified view of a universal {\it quantum dynamics governing all particles and 
fields}~\cite{Weinberg_QM,Weinberg_QFT}.

\subsection{Quantum framework}

One fundamental approach to quantum dynamics is the
 matrix mechanics framework of Heisenberg, Born, and Jordan, developed  
in 1925~\cite{Heisenberg_matrix-mech, Born-Jordan_matrix-mech, Born-Heisenberg-Jordan_matrix-mech},
where physical observables are described by matrices subject to a set of time evolution equations.
An alternative framework was discovered in late 1925~\cite{Schrodinger_SEq-1926,Schrodinger_SEq-1926_PR}:
the Schr\"{o}dinger picture, in which the evolution of a quantum state $\ket{\Psi}$ with respect to the given time 
$t$ is governed by the Schr\"{o}dinger equation 
\begin{equation}
 \mathrm {i} \hbar {\frac {d}{dt}}|\Psi (t)\rangle ={\hat {H}}|\Psi (t)\rangle 
\; ,
\label{eq:Schrodinger-eq}
\end{equation}
driven by a Hamiltonian operator ${\hat {H}}$ (using the modern notation developed
by Dirac~\cite{Dirac_QM-book,Dirac_QM-notation}), and with a physical scale in terms of
the reduced Planck or Dirac constant $\hbar= h/2\pi$ (where $h$ is the ordinary Planck constant).
Correspondingly, the physical observables are represented by operators or matrices $\hat{A}$ 
satisfying commutation relations, with the primary canonical commutators,
\begin{equation}
[\hat{q},\hat{p}] \equiv \hat{q}\hat{p} -\hat{p}\hat{q} =i \hbar
\; ,
\label{eq:primary-canonical-commutators}
 \end{equation}
between pairs of conjugate position-momentum variables $\hat{q}$ and $\hat{p}$.
In this quantum context,
the commutator between two operators or matrices $\hat{A}$ and $\hat{B}$ is generally defined by 
$[\hat{A},\hat{B}] = \hat{A}\hat{B} -\hat{B}\hat{A} $, and the hats denote the operator nature of these quantities.
The primary relation~(\ref{eq:primary-canonical-commutators}) 
is introduced to accommodate the Heisenberg uncertainty principle~\cite{Heisenberg_uncertainty},
describing the incompatibility of simultaneous measurements of position and momentum, 
as well as of any pair of canonically conjugate variables.

In terms of a distinct but equivalent approach, 
the observables of matrix mechanics are subject to a set of time evolution equations that
 give the same physical outcome as Eq.~(\ref{eq:Schrodinger-eq}), as first shown in
Ref.~\cite{Schrodinger_SEq-1926_equivalence}.
  The complete equivalence of these two dynamical approaches was further established 
  in a generalized transformation theory by Dirac~\cite{Dirac_QM-transf-theory}, 
  in what became the modern framework of quantum mechanics~\cite{Dirac_QM-book}. 
  In this setting, the original matrix mechanics can be displayed in the Heisenberg picture
  by starting with the evolution of states given by the dynamical Eq.~(\ref{eq:Schrodinger-eq})
  of the Schr\"{o}dinger picture and recasting the operator dynamics (representable as matrices)
  in terms of the Heisenberg equation
\begin{equation}
 {\frac {d}{dt}}\hat{A}_{_{\text{H}}} (t)={\frac {i}{\hbar }}
 [H_{_{\text{H}}}(t), \hat{A}_{_{\text{H}}}(t)]+\left({\frac {\partial \hat{A}_{_{\text{S}}}}{\partial t}}\right)_{\! \! \!_{\text{H}}}
 \; ,
\label{eq:Heisenberg-eq}
   \end{equation}  
  which highlights that the Hamiltonian governs the time evolution via quantum-mechanical commutators.
The Schr\"{o}dinger and Heisenberg picture quantities are labeled with subscripts $\text{S}$ and $\text{H}$. The 
relation between the quantities $\hat{A}_{_{\text{H}}}$ and $\hat{A}_{_{\text{S}}}$ in the two pictures
is defined by a similarity transformation with 
respect to the time evolution of Eq.~(\ref{eq:Schrodinger-eq})---this
 guarantees identical outcomes for the predicted values of any observable quantities. 
 In what follows, we will omit the hats for the notation of operators. 
 
 The underlying formal mathematical structure of quantum mechanics 
was fully spelled out in two groundbreaking treatises, by Dirac~\cite{Dirac_QM-book} and 
by von Neumann~\cite{vonNeumann_QM-book}, both based on their earlier series of 
seminal papers from 
1927~\cite{Dirac_QM-transf-theory,vonNeumann_QM-foundation, vonNeumann_density-matrix, vonNeumann_entropy}.
These also
included an alternative and more general description of {\em quantum states as statistical mixtures\/}
 in terms of a density matrix or density operator $\rho$~\cite{vonNeumann_density-matrix,Landau_density-matrix},
subject to the von Neumann equation~\cite{vonNeumann_density-matrix}
\begin{equation}
   i \hbar \frac{d\rho}{dt} = 
   [H,\rho]
    \; ,
    \label{eq:vonNeumann-eq-00}
    \end{equation}
     which generalizes Eq.~(\ref{eq:Schrodinger-eq}). In addition, 
 the quantum-information measure of states can be expressed 
 in terms of the von Neumann entropy~\cite{vonNeumann_entropy}
\begin{equation}
S = - k_{B} \mathrm{Tr} \left[  {\rho} \ln {\rho} \right] 
    \; ,
    \label{eq:vonNeumann-entropy}
\end{equation}
which, as the quantum-mechanical counterpart of the Gibbs entropy, physically scales 
in terms of Boltzmann's constant $k_{B}$; 
in Eq.~(\ref{eq:vonNeumann-entropy}), 
$\mathrm{Tr}$ stands for the operator trace.
The ensuing comprehensive, information-based framework, centered on Eqs.~(\ref{eq:vonNeumann-eq-00})
and (\ref{eq:vonNeumann-entropy}),
has become central to the ongoing developments in 
{\em quantum information and quantum computing\/}~\cite{nielsen00, Qinfo-review_LewisSwan}.
These revolutionary beginnings, including the establishment of a consistent universal framework and its initial success in atomic physics and chemistry~\cite{QM-conceptual,QM-historical, QM-sources},  
were followed by an impressive sequence of transcendental discoveries and applications of the 
quantum principles to an ever expanding set of systems over the next one hundred 
years~\cite{Weinberg_QM}.

\subsection{Quantum field theory}

One of the most consequential applications was the 
development of {\em quantum field theory\/}, which also followed naturally by 1927, in a formulation of the quantization of 
 the ubiquitous electromagnetic fields~\cite{Dirac_QFT}.
 This approach, the canonical quantization of fields, became a paradigm for all field theories,
  generalizing the dynamics of Eqs.~(\ref{eq:Schrodinger-eq}) and (\ref{eq:Heisenberg-eq})
for field operators $\Phi$ using a Hamiltonian framework, 
with conjugate field momenta and commutation relations~\cite{Weinberg_QFT}.
 The simplest realization of this framework is provided by the 
 quantization of a real scalar (spin-zero) field, to be used in this review for the sake of simplicity.
 This can be expressed in terms of a complete set of orthonormal modes:
$\left\{ \phi_{\boldsymbol{s}} (t,\mathbf{r}), \phi^{*}_{\boldsymbol{s}} (t,\mathbf{r})\right\}$ 
(including the 
complex-conjugate modes $\phi^{*}_{\boldsymbol{s}}$),
 via the expansion~\cite{Weinberg_QFT,birrell-davies}
\begin{equation}
    \Phi(t, \mathbf{r}) 
    = \sum_{\boldsymbol{s}} \bigl[ a_{\boldsymbol{s}} 
     \phi_{\boldsymbol{s}} (t,\mathbf{r})
      +  a_{\boldsymbol{s}}^{\dagger}  \phi^{*}_{\boldsymbol{s}} (t,\mathbf{r}) 
      \bigr]
    \; ,    \label{eq:field_expansion}
        \end{equation}
where $(t, \mathbf{r})$
 stand for the spacetime coordinates,
 and the field modes $\phi_{\boldsymbol{s}} $ are identified by  
 the mode frequency $\omega$ and a set of quantum numbers
 (collectively labeled by the symbol ${\boldsymbol{s}}$).
  For example, in flat spacetime, there is a natural choice of Fourier modes
 $\phi_{\boldsymbol{s}} (t,\mathbf{r}) \propto e^{-i\omega_{\boldsymbol{s}} t } \, \phi_{\boldsymbol{s}} (\mathbf{r})$ 
 associated with the time $t$ of a given inertial frame; this is the common procedure used
 in ordinary quantum field theory, dating back to the seminal paper~\cite{Dirac_QFT} on the 
 quantization of the electromagnetic field.
 (An elaboration of this procedure is shown in Sec.~\ref{sec:scalar-field_curved-ST} and in 
 Appendix~\ref{app:QOptics}).
In addition, the operator coefficients $  a_{\boldsymbol{s}} $ and $  a_{\boldsymbol{s}}^{\dagger}$
 in Eq.~(\ref{eq:field_expansion}), known as mode operators,
  are associated with particle annihilation and creation
 (where $ a_{\boldsymbol{s}}^{\dagger}$ is the adjoint of $a_{\boldsymbol{s}}$),
 and satisfy the canonical commutator relations of the field-operator algebra
 (generalizing $[q,p] = i \hbar$):
 \begin{equation}
[a_{\boldsymbol{s}}^{} ,  a_{\boldsymbol{s'}}^{\dagger}]
= \delta_{ {\boldsymbol{s}}, {\boldsymbol{s'}} }
\; , \; \; \;
[a_{\boldsymbol{s}}^{} ,  a_{\boldsymbol{s'}}^{}] = 0
\; , \; \; \;
[a_{\boldsymbol{s}}^{\dagger} ,  a_{\boldsymbol{s'}}^{\dagger}] = 0
\; .
\label{eq:field-operator-algebra}
\end{equation}
Similar expressions involving anticommutators are applicable to fermion fields. 
The quantum-field mode expansion is ordinarily written as in Eq.~(\ref{eq:field_expansion})
for a field $  \Phi(t, \mathbf{r}) $ in the Heisenberg picture, 
with time dependence such that the dynamics satisfies the Heisenberg equation~(\ref{eq:Heisenberg-eq}).
This choice has several advantages, including an intuitive correspondence with the classical regime, where the
 dynamics is similarly described in terms of Poisson brackets---this played an important
role in the early development of quantum field theory~\cite{Weinberg_QFT, Cao_QFT-development}.
Moreover, this general technique can be appropriately generalized
 to all quantum fields~\cite{Weinberg_QFT}, though
a complete understanding of the subtleties of quantum field theory and associated
technical tools took decades of development~\cite{Cao_QFT-development},
 culminating with the successful Standard Model of particle physics by the 1970s, 
 and leading the way to the current frontiers of particle physics~\cite{Weinberg_SM}.  
 In addition, the canonical approach, based on Eqs.~(\ref{eq:field_expansion})--(\ref{eq:field-operator-algebra})
 has been shown to be equally valid in curved spacetime~\cite{birrell-davies}, as emphasized 
 in the next paragraph and throughout this review article.
 Specifically, a similar procedure applies with modes
 $\phi_{\boldsymbol{s}} (t,\mathbf{r}) \propto e^{-i\omega_{\boldsymbol{s}} t } \, \phi_{\boldsymbol{s}} (\mathbf{r})$, 
 provided that the given spacetime is endowed with time-translation symmetry, 
 i.e., for stationary spacetimes; in such cases, Eq.~(\ref{eq:field_expansion}) effectively 
separates the expansion into positive-frequency and negative-frequency modes
[$ \phi_{\boldsymbol{s}} (t,\mathbf{r})$ and $ \phi^{*}_{\boldsymbol{s}} (t,\mathbf{r})$ respectively]. 
One can still use Eq.~(\ref{eq:field_expansion}) with more 
general modes in other instances, though additional subtleties need to be considered~\cite{birrell-davies}. 
 The technical details of quantum field theory, including how to define an inner product for orthonormality, are 
 introduced in Sec.~\ref{sec:QOptics-interactions}.

 Parenthetically, in all the applications of quantum physics,
 there is an alternative and powerful path-integral or functional formulation due to 
 Dirac~\cite{Dirac-PI} and Feynman~\cite{Feynman-PI},
based on the classical Lagrangian, and equivalent to the canonical quantization described above.
Even though we do not further elaborate on the path-integral formalism in this review article,
 it is noteworthy that it provides both deep insight and computational efficiency 
 for a variety of problems~\cite{Weinberg_QFT, Kleinert-PI, Grosche-PI}.

\subsection{Quantum information and quantum field theory in curved spacetime}

\begin{figure}[h]
    \centering
    \includegraphics[width=0.5\linewidth]{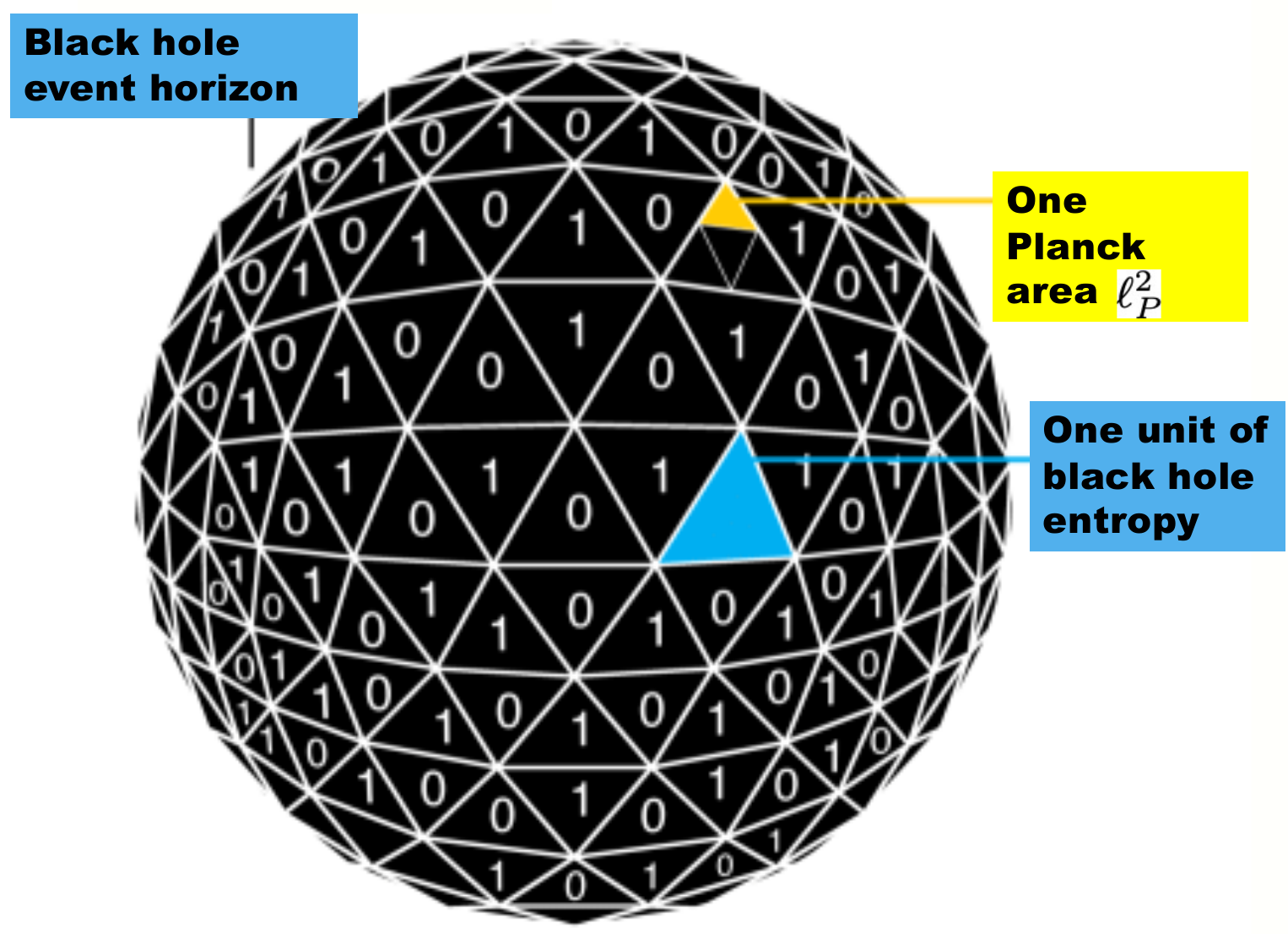}
    \caption{         
    A quantum information representation of the event horizon as a quantum computer, in units
    proportional to the Planck area $\ell_{P}^2 =   \hbar G/(k_{B}c^3)$.}
           \label{fig:BH-as-Qcomputer}
            \vspace*{-0.15in}
       \hspace*{0in}
\end{figure}

Even in recent decades, 
profoundly surprising implications in its foundations and technological applications have been discovered,
most notably in terms of its relation to {\it quantum information theory
and quantum computing\/}~\cite{nielsen00, Qinfo-review_LewisSwan},
and their interplay with spacetime behavior~\cite{peres2004quantum}.
In its more general context, now known as {\it relativistic quantum 
information\/}~\cite{peres2004quantum, Mann_RQI-review},
the concepts of vacuum, particles, and particle detectors have been reexamined, 
and a wealth of surprising phenomena related to thermodynamic properties have been uncovered. 
The basic framework to tackle spacetime quantum behavior has been 
{\it quantum field theory in curved spacetime\/}~\cite{birrell-davies}, where the canonical
quantization is enforced as in Eqs.~(\ref{eq:field_expansion})--(\ref{eq:field-operator-algebra})
with an appropriate definition of orthonormality and choice of a complete set of modes adapted to the given spacetime geometry. 
In this framework, the role of horizons in governing quantum effects via information properties 
is instrumental in altering the nature of physical states and generating thermal behavior.
An outstanding manifestation of these concepts is
  the thermodynamics of black holes~\cite{hawking76,BH-thermo_reviews},
where a network of relations linking quantum theory with gravitation and thermodynamics include:
 (i)
 the Hawking effect~\cite{hawking74,hawking75}, 
with thermal radiation at the Hawking temperature
 \begin{equation}
T_H =
\frac{1}{2 \pi} \frac{\hbar }{ k_{B} c } \kappa
\;  ,
\label{eq:Hawking-temperature}
\end{equation}
  proportional to the surface gravity $\kappa$~\cite{GR_Carroll-2003};
  and
 the Bekenstein-Hawking entropy
$S_{\mathrm{BH}}$ of the black hole~\cite{bekenstein-S_1972, bekenstein-S_1973},
   \begin{equation}
  S_{\mathrm{BH}}
   = \frac{1}{4} 
   \frac{ k_{B}c^3}{ \hbar G }
\, A
    \; , 
    \label{eq:BH-entropy}
  \end{equation} 
proportional to its event horizon area $A$,
along with a generalized second law of thermodynamics (GSL)~\cite{bekenstein-S_1972, bekenstein-S_1973}
 and the four laws of black hole mechanics~\cite{bardeen-carter-hawking1973}.
 This thermodynamic behavior suggests a quantum-information interpretation of the entropy 
 as encoded on the black hole event horizon, in units of the Planck area $\ell_{P}^2$,
 with the Planck length $\ell_{P} =    \sqrt{\hbar G/(k_{B}c^3)}$,
 as depicted in Fig.~\ref{fig:BH-as-Qcomputer}.
 %
%
Similar thermodynamic behavior is displayed by accelerated particles, leading to 
Unruh effect~\cite{Fulling-1973_CFQ,Davies-1975_Unruh,unruh-notes,ufd},
with acceleration radiation, 
and an associated Unruh temperature~\cite{Fulling-1973_CFQ,Davies-1975_Unruh,unruh-notes,ufd},
  \begin{equation}
T_U 
= \frac{1}{2 \pi} \frac{\hbar }{ k_{B} c } a
\;  ,
\label{eq:Unruh-temperature}
\end{equation}
proportional to the acceleration $a$; in this case, the apparent horizon emergent in the noninertial frame
of the accelerated observer also encodes nontrivial quantum-information properties.
   These nontrivial spacetime effects are quantum-relativistic in nature, a fact that which is highlighted
   in Eqs.~(\ref{eq:BH-entropy})--(\ref{eq:Unruh-temperature})
   via the Planck constant $\hbar$ and the speed of light $c$; and they are also 
   thermal, as revealed by the Boltzmann constant $k_{B}$. In addition,
   they provide further realizations of the quantum nature of all physical systems in a manner 
   that points to a transition towards a {\it theory of quantum gravity\/}~\cite{QGravity-review}.
   
   \subsection{Quantum optics}
   
In parallel with the development of quantum information and
curved spacetime field theory, 
the discipline of {\it quantum optics}~\cite{scullybook,meystrebook}, 
with its foundations on {\it quantum electrodynamics (QED)\/}~\cite{PikeSarkar_QTRad},
has revealed a remarkable effectiveness in explaining the same phenomena. 
Arguably, the great experimental success of quantum optics is intertwined with the development of laser physics and
other applications in atomic physics~\cite{scullybook,meystrebook}.
However, its domain of applicability has been greatly expanded as a theoretical tool,
adding insights from the {\it quantum theory of matter-field interactions\/}~\cite{Compagno-etal_atom-field-int}, which has been the trademark of quantum physics since its early beginnings~\cite{Weinberg_QM}.
Specifically, a direct use of quantum optics techniques in the context of relativistic spacetime,
with applications to various configurations dealing with accelerated systems, can
be traced back to Refs.~\cite{scully2003,belyanin2006}, 
where the relevance of the conversion of virtual to real atomic transitions is highlighted.
In these spacetime quantum developments, the ``physical reality'' of acceleration radiation was theoretically 
confirmed~\cite{Scully-etal_Unruh-reality} from first principles, extending earlier results 
on the excitation of accelerated particle detectors of Ref.~\cite{Audretsch-Muller_stimulated-acceleration}.
Recently, these ideas were further extended to study the acceleration radiation in the presence of black holes, 
which has been called ``horizon-brightened acceleration radiation'' (HBAR); see below~\cite{Scully_2018_HBAR}.
These techniques have verified and further developed the extensive earlier literature on moving mirror 
models~\cite{Fulling-Davies_mirror-1, Davies-Fulling_mirror-2, Anderson-etal_mirror-3, Good-etal_mirror-4,dynamical-Casimir-curved}.
In particular, important conceptual problems---including the interpretation of various configurations of detectors 
and mirrors as well as the equivalence principle---have been easily tackled with related
quantum-optics treatments~\cite{Fulling-Wilson_2018_EP,Ben-Benjamin-etal_2019_Unruh-rev}, 
which are applicable to a variety of spacetime configurations~\cite{Scully_2020_laser-review} 
and further discussed in this review article.
Other closely related developments involve the Casimir~\cite{Casimir_1948,Bordag-etal_Casimir-effect}
and dynamical Casimir~\cite{Moore_dynamical-Casimir-effect, Dodonov_dynamical-Casimir-effect}
effects---with {\it quantum virtual processes\/}
 and particle creation in a vacuum as a common denominator.
Several aspects of most of these phenomena are comprehensively summarized in recent
reviews~\cite{Dodonov_dynamical-Casimir-effect,Crispino-et-al_2008_Unruh-effect,Pasante_2018_dispersion-vacuum, Masood-etal_2024_atom-field-rev}
that have a significant overlap with this article.

For our current purposes, in this review article, we select a subset of concepts 
of relativistic quantum information and highlight the effectiveness of quantum optics techniques to
shed light on the nature of spacetime
and surprising {\it spacetime and gravitational quantum effects\/}.
 Specifically, our focus will be on the {\it quantum effects associated with the 
 motion of particles in black hole backgrounds\/}, summarizing and extending the work of Ref.~\cite{Scully_2018_HBAR}.
 
 \subsection{Organization of this article}
 
This article is organized as follows.
An extensive discussion of the quantum optics description of interactions and particle detectors 
is given in Sec.~\ref{sec:QOptics-interactions}, including 
general overviews of {\it quantum field theory\/} in curved spacetime (for scalar fields) and 
the two-level atom in the form of the {\it quantum Rabi model\/} (QRM) and generalizations.
In Sec.~\ref{sec:QOptics-density-matrix},
we offer an introduction to the {\it quantum-optics density matrix\/} approach.
In addition, in Sec.~\ref{sec:spacetime_CQM}
we discuss the setup of quantum aspects of spacetime geometry, as well as 
the quantum-mechanical framework known as 
{\it conformal quantum mechanics\/} (CQM)~\cite{dAFF,CQM-renormalization-PLA}, 
which is governed by an inverse 
square potential~\cite{renormalization_ISP}, and generically offers a versatile description of
scale-invariant physics 
applications~\cite{Qanomaly-molecular,Qanomaly-molecular-EFT,Qanomaly-molecular-to-BH}, including its central role 
 in all near-horizon black hole geometries~\cite{nhcamblong,nhcamblong-sc,nhcamblong-heat-kernel,nhcamblong-conformal-tightness,acceler-rad-Schwarzschild,acceler-rad-Kerr,acceler-rad-Qopt-1,acceler-rad-Qopt-2}.
With this comprehensive background, 
in Sec.~\ref{sec:HBAR}
we derive the most important results of acceleration radiation (HBAR)~\cite{Scully_2018_HBAR,acceler-rad-Schwarzschild,acceler-rad-Kerr,acceler-rad-Qopt-1,acceler-rad-Qopt-2},
leading to an HBAR-black hole thermodynamics correspondence in Sec.~\ref{sec:HBAR-thermo}.
After some concluding remarks on the implications of these nontrivial quantum effects
in Sec.~\ref{sec:conclusions}, we supplement the article with
two appendices dealing with (i) additional technical and historical background on quantum optics, and 
(ii) other aspects of spacetime physics and black holes.

Some remarks on conventions and units are in order. 
First, in this introductory section, standard general units have been 
useful to highlight the interplay of scales across physics.
By contrast, for the remainder of this article (starting with Secs.~\ref{sec:QOptics-interactions} and 
\ref{sec:QOptics-density-matrix}), we will switch to units with $c=1$, so that spatial and temporal
measurements have the same dimensions, and the relativity formulas are easier to read.
Moreover, throughout the paper, the notation $ x^{\mu} = (t,\mathbf{r})$ is being used 
to denote an arbitrary set of spacetime coordinates adapted to the geometry, with a splitting 
into temporal and spatial components.
 Subsequently, in the technical presentations of the geometry of spacetime 
 and conformal quantum mechanics in Sec~\ref{sec:spacetime_CQM},
and for quantum thermodynamics in Secs.~\ref{sec:HBAR} and \ref{sec:HBAR-thermo},
we will mostly switch to a full-fledged set of natural units, with 
all of the universal constants $c$, $\hbar$, $k_{B}$, and $G$ equal to one (except where
 stated otherwise).
Finally, for the spacetime geometry, we will use a ``mostly plus metric'' with signature $(-,+,+, \ldots )$,  
having only one negative sign for time, along with the other metric spacetime conventions of 
Refs.~\cite{GR_Carroll-2003, MTW-gravitation}.

\section{Quantum field theory and quantum optics of atom-field interactions and particle detectors}
\label{sec:QOptics-interactions}
  \begin{figure}[h]
    \centering
    \includegraphics[width=0.85\linewidth]{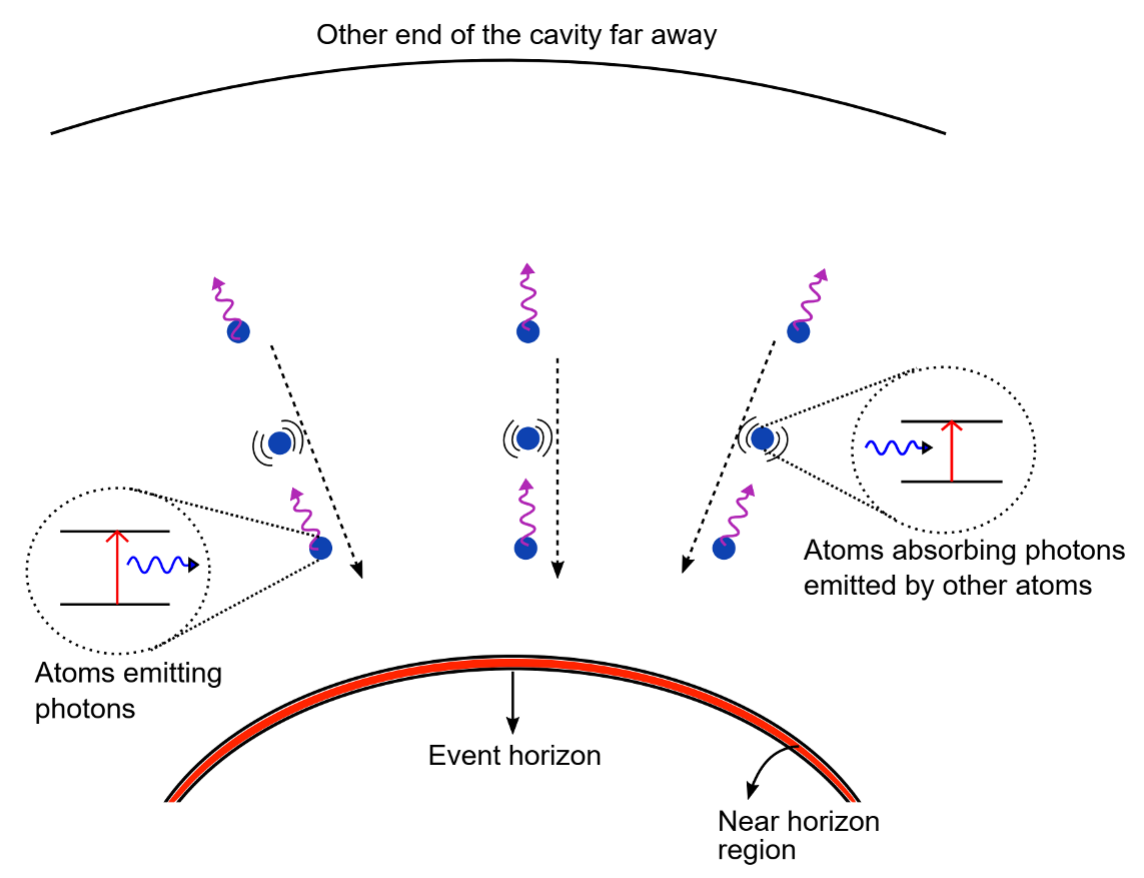}
    \caption{The thought experiment for the HBAR model, where atoms freely fall
    into a black hole in a Boulware vacuum, simulating an analog quantum-optics system with boundary mirrors. This ``optical cavity model" is only a conceptual device to represent the vacuum setup. The dashed lines show the direction of the free-fall motion of the atoms, which radiate in all directions (but only the outgoing radiation, to be measured far away, is shown for clarity). As the radiation goes up the gravity well, gravitational redshift makes its wavelength increase.}
       \label{fig:HBAR-setup}
       \hspace*{0in}
\end{figure}

In this section, we review the quantum behavior of matter and fields, and their interactions.
 It is for these interactions that the techniques of quantum optics become most insightful.
While the natural setting of quantum optics is usually for systems in the lab in flat 
 spacetime~\cite{scullybook,meystrebook}, the same concepts apply more generally to
 arbitrary spacetime configurations.
 Thus, for our current purposes, we will consider the interaction between atoms 
 (representing an atomic cloud) and a quantum field, in a generic 
 gravitational spacetime background.
 Specific spacetime geometries are described
in Sec.~\ref{sec:spacetime_CQM} and Appendix~\ref{app:spacetime}
for generalized Schwarzschild and Kerr metrics~\cite{GR_Carroll-2003}, respectively.

In a particular experimental setup, an initial state should be specified. 
For the HBAR model~\cite{Scully_2018_HBAR} of acceleration radiation in black hole backgrounds,
a thought experiment consists of atoms that are randomly injected, following free-fall
 paths in the given gravitational background around a black hole, 
 as shown in Fig.~\ref{fig:HBAR-setup}.
  And the quantum field is set up in a Boulware-like vacuum, which is defined with
 modes adapted to stationary coordinates~\cite{birrell-davies,Boulware_1975}.
 The experimental arrangement of Fig.~\ref{fig:HBAR-setup} corresponds to the way in which quantum optics experiments
 with fixed mirrors are used in the laboratory. In that sense, it is an ``optical cavity model'' that simulates the Boulware-like vacuum most naturally. Unlike ordinary lab experiments, the presence of a gravitational background has nontrivial
effects; for example, the top and bottom of the cavity are accelerating in a general relativistic sense, and the
 radiation emitted near the black hole's horizon has a wavelength that increases as it travels in the outgoing direction due to the
gravitational redshift. 

\begin{figure}[h]
    \centering
    \includegraphics[width=0.375\linewidth]{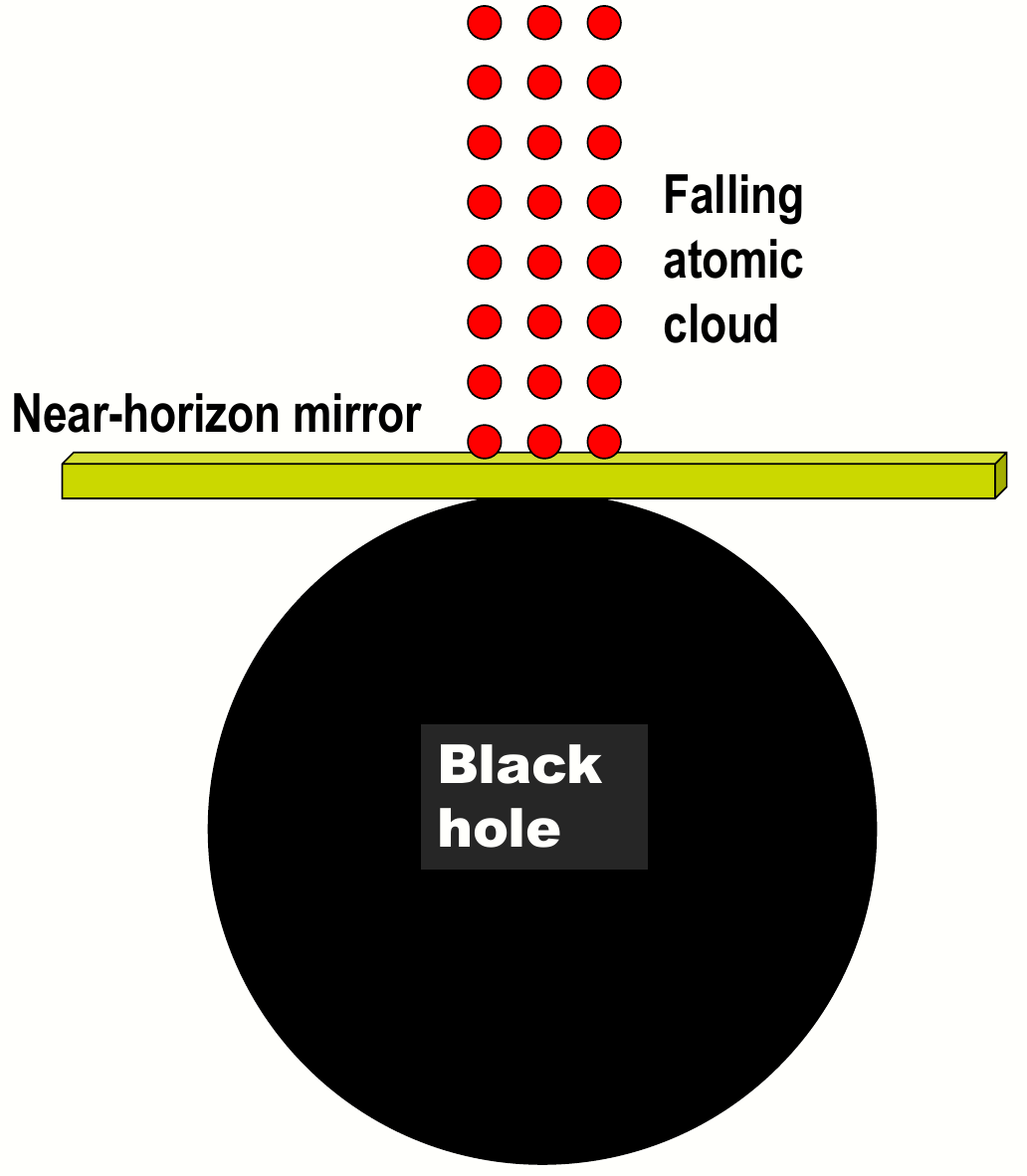}
    \caption{In the thought experiment for the HBAR model, a mirror is an operational
    quantum-optics type device that enforces a boundary condition to guarantee 
    the presence of a Boulware vacuum.}
       \label{fig:HBAR-setup-with-mirror}
       \hspace*{0in}
\end{figure}

However, cautionary remarks on the cavity model are in order. First, 
the setup involves an appropriate boundary condition at the event horizon, which can be enforced with a mirror, as displayed in Fig.~\ref{fig:HBAR-setup-with-mirror}.
Now, 
a physical model of a mirror an infinitesimal distance above the horizon would be probably be unstable under reasonable assumptions of causal sound wave propagation. Second, as a result, the cavity should only be viewed as an auxiliary construction within a thought experiment that simulates a cavityless black hole with a Boulware state. Third, even the Boulware vacuum would not be the physically reasonable assumption if one considered an ordinary black hole formed from the usual processes of gravitational collapse, as this system would normally end up with an Unruh vacuum instead~\cite{QBH-primer_Brout-etal_1995, BH-evap_Fabbri-NavarroSalas_2005}.
Despite these limitations, if caution is exercised in its interpretation, the proposed HBAR thought experiment remains a powerful  device to probe the interplay between fundamental quantum effects and strong gravitational fields, and provides an alternative probe of black hole thermodynamics.

\subsection{Quantum field theory: Scalar field in curved spacetime}
\label{sec:scalar-field_curved-ST}

 For the sake of simplicity, a scalar (spin-zero) 
 field $\Phi$ provides the essential ingredients of the relevant physics for a variety of
 fundamental questions. Moreover, the results can be easily generalized to fields with nonzero spin. 
 In fact, quantum optics was originally developed for and is most commonly applied to spin-one electromagnetic fields~\cite{scullybook,meystrebook,PikeSarkar_QTRad}
(see Appendix~\ref{app:QOptics}), 
but its methodology generically works in a similar manner for scalar 
 and other fields. In other words,
 ordinary vector, spin-one photons can be replaced in this model by scalar, spin-zero ``photons.''
 Likewise, we can consider a simplified treatment with a two-level atom capturing the essential
 features of atomic electron transitions.
When the field is described by a Boulware vacuum,
which corresponds to stationary coordinates~\cite{birrell-davies,Boulware_1975},
  there exits a relative acceleration between the atoms and the field, 
 which is the physical source of the ensuing acceleration 
 radiation~\cite{Scully_2018_HBAR,acceler-rad-Schwarzschild,acceler-rad-Kerr,acceler-rad-Qopt-1,acceler-rad-Qopt-2}.
 
    \subsubsection{Field action and quantization}

 A real scalar field  $\Phi$, with mass $\mu_\Phi$, in the geometric background of a spacetime metric 
$g_{\mu \nu}$, is defined by an action~\cite{birrell-davies,Boulware_1975}
\begin{equation}
S [\Phi]
=
-
\frac{1}{2}
\int
d^{D} x
\,
\sqrt{-g}
\,
\left[
g^{\mu \nu}
\,
\nabla_{\mu} \Phi
\, 
\nabla_{\nu} \Phi
+ 
\mu_{\Phi}^{2} \Phi^{2}
+  \xi R \Phi^{2}
\right]
\; 
\label{eq:scalar_action}
\end{equation}
(with spacetime dimensions $D \geq 4$), where $g$ is the determinant of the metric and
the coupling of $\Phi$ to the metric $g_{\mu \nu}$ is via
 its covariant derivatives $\nabla_{\mu} \Phi$ and with the curvature scalar $R$ via the nonminimal coupling 
 constant $\xi$.
At the classical level, the action principle $\delta S[\Phi]/\delta \Phi = 0$ gives the Euler-Lagrange equations for the action~(\ref{eq:scalar_action}), 
which govern the dynamics in a spacetime background and
 take the form of the Klein-Gordon equation 
\begin{equation}
\left[ \Box - \left( \mu_\Phi^{2} + \xi R
 \right)  \right]\Phi 
\equiv 
\frac{1}{ \sqrt{-g} }\partial_{\mu} \left(\sqrt{-g} \,g^{\mu \nu}\,\partial_{\nu} \Phi\right)- (\mu_\Phi^{2} + \xi R) \Phi= 0
\; .
\label{eq:Klein_Gordon_basic}
\end{equation}
Here, we will use the spacetime conventions of Refs.~\cite{GR_Carroll-2003, MTW-gravitation}, including units with 
$c=1$ [as stated at the end of Sec.~(\ref{sec:introduction})];
and the field theory conventions of Refs.~\cite{Crispino-et-al_2008_Unruh-effect, Takagi:1986}.
Additional details, related to the metric and near-horizon behavior, are described in Sec.~\ref{sec:spacetime_CQM}
and Appendix~\ref{app:spacetime}. 

The quantization of the theory is established by the usual canonical Hamiltonian rules.
The canonical quantization procedure involves promoting the classical field and its conjugate momentum 
to quantum operators satisfying the canonical commutation relations similar to the primary commutators of 
Eq.~(\ref{eq:primary-canonical-commutators}).
Given the Lagrangian density $\mathcal{L}$ as the integrand of the action integral~(\ref{eq:scalar_action}),
\begin{equation}
    \mathcal{L} = -\sqrt{-g}
    \left[
g^{\mu \nu}
\,
\nabla_{\mu} \Phi
\, 
\nabla_{\nu} \Phi
+ 
\mu_{\Phi}^{2} \Phi^{2}
+  \xi R \Phi^{2}
\right]
\; ,
\label{eq:scalar_Lagrangian}
\end{equation}
the conjugate momentum is 
\begin{equation}
    \Pi  = \frac{\partial \mathcal{L}}{
    \partial \left( \nabla_{0}{\Phi} \right) }
    = - \sqrt{-g} \,  \nabla^{0}{\Phi} 
    \; .
\end{equation}
 Upon quantization, the field $\Phi (t,\mathbf{r})$ and its canonical momentum 
 $\Pi (t,\mathbf{r})$ are operators that satisfy the equal-time canonical commutation relations similar to
 Eq.~(\ref{eq:primary-canonical-commutators}),
\begin{equation}
\begin{aligned}
&    \left[ {\Phi} (t,\mathbf{r}), {\Phi}(t,\mathbf{r}') \right] = 0
   \; \; , \; \; \;
     \left[ {\Pi} (t,\mathbf{r}), {\Pi}(t,\mathbf{r}') \right] = 0  
  \\
&    \left[ {\Phi} (t,\mathbf{r}), {\Pi}(t,\mathbf{r}') \right] = i \hbar \delta^{(D-1)} (\mathbf{r} -\mathbf{r}') 
   \; ,
    \end{aligned}
    \label{eq:field-canonical-commutators}
\end{equation}
with a delta-function distribution defined via
$\displaystyle \int_{\Sigma} d^{D-1} x'  \, 
w(\mathbf{r}' ) \, \delta^{(D-1)} (\mathbf{r} -\mathbf{r}') = w(\mathbf{r})$,
for all Schwartz test functions $ w(\mathbf{r})$ 
(density of weight one in the second argument) on the $(D-1)$-dimensional 
spacelike hypersurface $\Sigma$, which is essentially a slice in spacetime usually described as ``space.'' 

 \subsubsection{Field modes}

Formally, the quantization is reduced to the problem of finding a complete set of solutions
$\bigl\{  \phi_{\boldsymbol{s}} (t,\mathbf{r}), \phi^{*}_{\boldsymbol{s}} (t,\mathbf{r}) \bigr\}$
of the classical equation~(\ref{eq:Klein_Gordon_basic}) and
expanding the quantum field theory operator $\Phi$ as in Eq.~(\ref{eq:field_expansion}),
which reads
\begin{equation}
    \Phi(t, \mathbf{r}) 
    = \sum_{\boldsymbol{s}} \bigl[ a_{\boldsymbol{s}} 
     \phi_{\boldsymbol{s}} (t,\mathbf{r})
     + \mathrm{H.c.} \bigr]
    \; ,    \label{eq:field_expansion_2}
        \end{equation}
where $\mathrm{H.c.}$ is the Hermitian conjugate.  
The modes are labeled with the subscript ${\boldsymbol{s}}$, specifying a complete set 
of quantum numbers and the mode frequency $\omega$.
In addition, the modes are assumed to satisfy the orthonormality conditions
 \begin{equation}
 (\phi_{\boldsymbol{s}},  \phi_{\boldsymbol{s'}} ) 
 = - (\phi^{*}_{\boldsymbol{s}},  \phi^{*}_{\boldsymbol{s'}} ) 
 =   \delta_{ {\boldsymbol{s}}, {\boldsymbol{s'}} }
   \; \; , \; \; \; 
    (\phi^{*}_{\boldsymbol{s}},  \phi_{\boldsymbol{s'}} ) 
 = (\phi_{\boldsymbol{s}},  \phi^{*}_{\boldsymbol{s'}} ) 
   = 0
\; , 
   \label{eq:KG-orthonormality}
   \end{equation}
where  the standard inner product in the given geometry~\cite{Crispino-et-al_2008_Unruh-effect, Takagi:1986},
\begin{equation}
(\Phi_1,\Phi_2) 
= 
 i \int_\Sigma 
\left(  \Phi_1^{*} \partial_\mu \Phi_2  -  \Phi_1 \partial_\mu \Phi_2^{*}  \right)
 d\Sigma^\mu
 \; ,
 \label{eq:KG-inner-product}
\end{equation}
is consistent with the Klein-Gordon equation.
In Eq.~(\ref{eq:KG-inner-product}), the integral is performed on a $(D-1)$-dimensional 
spacelike hypersurface $\Sigma$, with  
``volume'' element $d \Sigma^{\mu} = n^{\mu} \sqrt{\gamma}\,  d^{D-1} x$ is along 
the normal, future-directed ``time direction'' $n^{\mu}$ (with a corresponding induced metric $\gamma_{ij}$). 
This product is independent of the chosen hypersurface $\Sigma$, thus it is applicable to any spatial slice,
as follows from Gauss's divergence theorem.
For example, in flat (Minkowski) spacetime, the inner product is simply:
$(\Phi_1,\Phi_2)  = i \int \left(  \Phi_1^{*} \partial_{t} \Phi_2  -  \Phi_1 \partial_{t} \Phi_2^{*}  \right) d^{D-1} x$
integrated over ordinary space.

In the quantization of the theory, for the interpretation of particle excitations of the field,
the functions
$\bigl\{  \phi_{\boldsymbol{s}} (t,\mathbf{r}), \phi^{*}_{\boldsymbol{s}} (t,\mathbf{r}) \bigr\}$ 
are identified as positive/negative frequency modes. This classification is naturally suggested by the form
it takes in the simplest geometry: flat (Minkowski) spacetime,
where the condition $\partial_{t}  \phi_{\boldsymbol{s}}= -i \omega \phi_{\boldsymbol{s}}$ (with $\omega> 0$)
 gives the familiar time dependence $e^{-i \omega t}$,
with a frequency $\omega$ such that $\hbar \omega$ is the positive 
energy of the corresponding quantum particle excitation. 
This relation can be generalized to spacetimes that have time-translation symmetry, so that 
 a timelike vector field ${\boldsymbol{\xi}} $ akin to  $\partial_{t} $
 is available, where the positive frequency can be similarly identified with the energy.
 Such a generalization involves a Killing vector field~\cite{GR_Carroll-2003, MTW-gravitation, GR_Wald-1984}, 
  which is an infinitesimal generator of a symmetry of the metric, so that
 the distances between points on the manifold are invariant along the Killing-field direction; 
 see Sec.~\ref{sec:spacetime_CQM}.
 In such spacetimes, the positive frequency modes $ \phi_{\boldsymbol{s}}$ 
 can be identified in a coordinate-independent manner by the equation
 \begin{equation}
 \xi^{\mu}  \partial_{\mu}  \phi_{\boldsymbol{s}}= -i \omega \phi_{\boldsymbol{s}}
 \end{equation}
  ($\omega> 0$),
 where the left-hand side is an example of a Lie derivative~\cite{GR_Carroll-2003, MTW-gravitation, GR_Wald-1984} 
 along the flow of the Killing vector field ${\boldsymbol{\xi}} $. 
 Moreover, the condition $ \xi^{\mu}  \partial_{\mu} t = 1$ formally defines the Killing time
 as the preferred time associated with the symmetry. 
 The corresponding conjugate modes $ \phi^{*}_{\boldsymbol{s}}$ 
 satisfy the negative-frequency geometric condition 
  \begin{equation}
 \xi^{\mu}  \partial_{\mu}  \phi^{*}_{\boldsymbol{s}}= i \omega \phi^{*}_{\boldsymbol{s}}
 \; ,
 \end{equation}
 confirming the frequency-sign classification.
 Operationally, the particle interpretation can be probed 
 considering a detector following orbits of the Killing vector field; then, the Killing time and the proper time $\tau$
 are proportional, so that the positive and negative frequencies correspond to what the detector actually registers,
 with
 $ \partial_{\tau}  \phi_{\boldsymbol{s}}= -i \omega \phi_{\boldsymbol{s}}$, and
the modes can be used to compute the number of particles.

 \subsubsection{Fock space and Hamiltonian}
 
The field canonical commutators~(\ref{eq:field-canonical-commutators})
 are equivalent to a set of the commutation relations for the annihilation/creation operators.
 These are the canonical commutator relations of the field-operator algebra~(\ref{eq:field-operator-algebra}), i.e.,
 $
  [a_{\boldsymbol{s}}^{} ,  a_{\boldsymbol{s'}}^{\dagger}]
= \delta_{ {\boldsymbol{s}}, {\boldsymbol{s'}} }$
,
$[a_{\boldsymbol{s}}^{} ,  a_{\boldsymbol{s'}}^{}] = 0$, and
$[a_{\boldsymbol{s}}^{\dagger} ,  a_{\boldsymbol{s'}}^{\dagger}] = 0$.
The proof of this statement involves a straightforward substitution  
of the field expansion~(\ref{eq:field_expansion_2}) in Eq.~(\ref{eq:field-canonical-commutators}), along with the use
of the Klein-Gordon inner product~(\ref{eq:KG-inner-product}).
These operator-algebra commutators have the same form as for a simple one-dimensional problem 
in quantum mechanics, with the interpretation that Eq.~(\ref{eq:field_expansion_2}) represents an expansion 
in a set of quantum harmonic oscillators corresponding to all the modes with configuration labels $\boldsymbol{s}$.
 Once this is established, with the annihilation/creation canonical commutators, 
 one can build the states of the quantum theory as follows.
 The lowest energy state, known as the vacuum $\ket{0}$, is defined by the set of annihilation conditions
 \begin{equation}
 a_{\boldsymbol{s}} \ket{0} = 0
 \; \; , \; \; \; 
 \text{for all} \;  {\boldsymbol{s}}
 \label{eq:vacuum}
 \; .
 \end{equation}
The vacuum state can then be used to generate all other states 
 by repeated action with the creation operators. Thus, starting with a single mode labeled by ${\boldsymbol{s}}$,
 \begin{equation}
\ket{ n_{\boldsymbol{s}} }
= \frac{1}{ \sqrt{ n_{\boldsymbol{s}}!   } }
 \left(  a_{\boldsymbol{s}}^{\dagger} \right)^{n_{\boldsymbol{s}}}
 \ket{0}
 \; 
  \end{equation}
 gives a state with $n_{\boldsymbol{s}} $ excitations or ``photons.''
 From the commutator relations, this excitation number or ``occupation number'' is the eigenvalue of the number operator 
 $N_{\boldsymbol{s}} = a^{\dagger}_{\boldsymbol{s}}  a_{\boldsymbol{s}} $. 
  Repeating the procedure for all modes, 
  a basis for the states of the quantum-field system can be established with the 
  tensor products of single-particle Hilbert spaces,
 displaying all excitation numbers in the form
 \begin{equation}
\ket{ 
n_{\boldsymbol{s}_{1}} ,  n_{\boldsymbol{s}_{2}} \ldots n_{\boldsymbol{s}_{j}}  \dots}
=
\ket{ n_{\boldsymbol{s}_{1}} } \otimes \ket{ n_{\boldsymbol{s}_{2}} } \otimes  \ldots
\otimes \ket{ n_{\boldsymbol{s}_{j}} } \ldots
 \; .
 \label{eq:Fock-occupation-rep}
  \end{equation}
   This is often called the occupation number representation and the general framework is the so-called 
   ``second quantization.''
An alternative shorthand notation for the occupation number representation, 
which we use and extend in Sec.~\ref{subsec:density-matrix_HBAR} is
$ \boldsymbol{ \left\{  \right. } n  \boldsymbol{\left. \right\}}    \equiv
\boldsymbol{ \left\{  \right.  } n_{1}, n_{2}, \ldots , n_{j } , \ldots \boldsymbol{\left. \right\}  }$,
where $n$ refers to the collection of excitation numbers.
  Finally, with this basis, and considering all possible states with different
  numbers of excitations for all modes, this procedure amounts to the construction of  
  the Fock space of states as the direct sum of 
  tensor products of single-particle Hilbert spaces~\cite{Weinberg_QFT}.
A simple example of this construction is given in Appendix~\ref{app:QOptics} 
for a spin-one electromagnetic field in Minkowski spacetime.   
In conclusion, 
Fock space and the occupation number representation 
provide the framework to describe quantum states with a variable number of particles, thus forming the foundation 
for quantum field theory and many-body physics~\cite{MBP_Mahan_2000}.
These powerful tools were also developed in Dirac's 1927 seminal paper~\cite{Dirac_QFT}, 
and were subsequently extended by Jordan and Wigner~\cite{Jordan-Wigner_1928}, 
and Fock~\cite{Fock_1932}.

In order to describe the dynamics, 
the Hamiltonian density 
\begin{equation}
\mathcal{H} = \Pi \,  \nabla_{0}{\Phi} - \mathcal{L}
\end{equation}
is needed, leading to a field Hamiltonian $H= \int d^{D} x \, \mathcal{H} $.
 For the free scalar field Lagrangian~(\ref{eq:scalar_Lagrangian}), 
 this gives  $\displaystyle \mathcal{H} = \frac{1}{2} \left(
 -\nabla^{0} \Phi  \nabla_{0} \Phi +  \nabla^{j} \Phi  \nabla_{j} \Phi + \mu^{2}_{\Phi}  \Phi^2 + \xi R \Phi^2
 \right)$, where the index $j$ (with implicit summation of derivatives) refers to the spatial coordinates; 
 and this can be restated as a generic Hamiltonian quadratic in the canonical variables. 
 Any such Hamiltonian has the following mode decomposition, which can be derived
 using the field expansion~(\ref{eq:field_expansion_2}), the orthonormality 
 relations~(\ref{eq:KG-orthonormality}), and the classical equation~(\ref{eq:Klein_Gordon_basic}) 
for the modes:
\begin{subequations}
\label{eq:H_field_global}
\begin{align}
H_{\rm field} 
&
= \sum_{{\boldsymbol{s}}}^{}
\hbar \omega_{\boldsymbol{s}}
\left(
a_{\boldsymbol{s}}^{\dagger} a_{\boldsymbol{s}} 
+
a_{\boldsymbol{s}} a_{\boldsymbol{s}}^{\dagger} 
\right)
= \sum_{{\boldsymbol{s}}}^{}
\hbar \omega_{\boldsymbol{s}}
\bigl(
a_{\boldsymbol{s}}^{\dagger} a_{\boldsymbol{s}} 
+
\tfrac{1}{2}
 \bigr)
 \; ,
 \label{eq:H_field_wZPE}
\\
H_{\rm field} 
& \approx
\hbar \omega_{\boldsymbol{s}}
\,
a_{\boldsymbol{s}}^{\dagger} a_{\boldsymbol{s}} 
\; ,
\label{eq:H_field}
\end{align}
\end{subequations}
where the notation $H_{\rm field} $ will be used to distinguish this physical system from the other parts
of the Hamiltonian of an interacting system of fields and atoms.
In its final form, the field Hamiltonian~(\ref{eq:H_field_global}) 
is identical to the sum of mode-specific quantum harmonic oscillator Hamiltonians.
Here, the symbol $\omega_{\boldsymbol{s}}$ is used redundantly for the sake of clarity 
(as, in the notation used here, 
$\omega$ is part of the mode label ${\boldsymbol{s}}$).
For computational convenience, the expression~(\ref{eq:H_field})
is written with the $ \approx$ symbol to implement the subtraction of the zero-point energy, i.e.,
the energy of quantum fluctuations in the vacuum,
which has no direct impact on the relevant physics, excluding questions of gravitational interactions via a 
cosmological constant~\cite{Weinberg_QM}, or for the Casimir effect in systems 
with finite boundaries~\cite{Casimir_1948, Bordag-etal_Casimir-effect, Masood-etal_2024_atom-field-rev}. 
This is the expression to be used in practical calculations for the great majority of
experimental realizations.
Additional details on the physics of quantum fields,
when the system is modeled by the usual spin-one electromagnetic fields,
are discussed in Appendix~\ref{app:QOptics}. 
 
  \subsubsection{Subtleties of quantum field theory in curved spacetime}
 
Finally, it should be highlighted that the equations outlined in this section are applicable to any state
 of the quantum field. However, the interpretation of mode functions and the associated 
 definition of ``positive frequency'' modes encounter significant challenges in curved spacetime, 
 due to the absence of global time-translation invariance~\cite{birrell-davies}.
 Unlike flat spacetime, where the Poincaré group provides a unique vacuum state, 
 general curved spacetimes lack such a preferred definition. 
 Thus, as the definition of vacuum and the number of particle excitations relies on a
 chosen set of mode functions with a specific positive-frequency characterization, this
 means that the different different choices of time slicing can lead to different,
 observer-dependent results.
Nonetheless, there is a well-defined procedure for comparison of particle
measurements among observers: 
Bogoliubov transformations~\cite{birrell-davies, GR_Carroll-2003, Crispino-et-al_2008_Unruh-effect}.
In this framework, two different sets of modes, $\left\{  f_{\boldsymbol{s}}(x) ,  f^{*}_{\boldsymbol{s}} (x)\right\}$
and $\left\{  g_{\boldsymbol{s}} (x),  g^{*}_{\boldsymbol{s}} (x)\right\}$, 
with their corresponding operator algebras 
$\left\{  a_{\boldsymbol{s}},  a^{\dagger}_{\boldsymbol{s}} \right\}$
and
$\left\{  b_{\boldsymbol{s}},  b^{\dagger}_{\boldsymbol{s}} \right\}$ respectively,
are linearly related by
\begin{equation}
 g_{\boldsymbol{s}}
 = \sum_{\boldsymbol{s'}} \left(
 \alpha_{{\boldsymbol{s}} {\boldsymbol{s}'} }  f_{\boldsymbol{s}'}
 + \beta_{{\boldsymbol{s}} {\boldsymbol{s}'} }  f^{*}_{\boldsymbol{s}'} \right)
 \; ,
 \label{eq:Bogoliubov}
 \end{equation}
 as follows by their linear completeness.
 Then, Eq.~(\ref{eq:Bogoliubov}) can be used to fully predict all well-posed questions on particle excitations.
  This can be done by establishing the whole network of linear relations:
(i) the inverse of the transformation~(\ref{eq:Bogoliubov});
 (ii) the Bogoliubov coefficients in terms of inner-product projections:
$ \alpha_{{\boldsymbol{s}} {\boldsymbol{s}'} }
 = \left( f_{\boldsymbol{s}'},  g_{\boldsymbol{s}} \right)$
and $ \beta_{{\boldsymbol{s}} {\boldsymbol{s}'} }
 = - \left( f^{*}_{\boldsymbol{s}'},  g_{\boldsymbol{s}} \right)$;
 and (iii) f the corresponding operator-algebra relations:
$ b_{\boldsymbol{s}}
 = \sum_{\boldsymbol{s'}} \left(
 \alpha^{*}_{{\boldsymbol{s}} {\boldsymbol{s}'} }  a_{\boldsymbol{s}'}
 - \beta^{*}_{{\boldsymbol{s}} {\boldsymbol{s}'} }  a^{\dagger}_{\boldsymbol{s}'} \right)$
 and their inverses.
 For example, the excitation number of $g$-mode particles is the expectation value of
 the number operator  $N_{(g), \boldsymbol{s}} = b^{\dagger}_{\boldsymbol{s}}  b_{\boldsymbol{s}} $;
 now, if this is computed for the vacuum state 
 $ \ket{0_{f}}$
 of the other $f$-mode set, then straightforward algebra yields
  \begin{equation}
  \bra{0_{f} }
  N_{(g), \boldsymbol{s}} \ket{0_{f}}
  = \sum_{\boldsymbol{s'}} 
   \left| \beta_{{\boldsymbol{s}} {\boldsymbol{s}'} }   \right|^{2}
   \; .
  \end{equation}
  Thus, an empty vacuum in one set of modes appears populated with $\bra{0_{f} } N_{(g), \boldsymbol{s}} \ket{0_{f}} \neq 0$
  particle excitations in another set when at least some of the 
$  \beta_{{\boldsymbol{s}} {\boldsymbol{s}'} } $ coefficients are nonzero,
i.e, $\beta_{{\boldsymbol{s}} {\boldsymbol{s}'} } $ measures the non-overlap between the two sets of positive-frequency 
modes, yielding an observable nontrivial mismatch.
 These properties highlight the observer-dependent nature of the vacuum and particle concepts in curved spacetime.
Extraordinary predictions from this framework include 
the Unruh and Hawking effects, and other phenomena~\cite{birrell-davies}.
  
 For our purposes, in the HBAR setup, we will use modes 
 adapted to stationary coordinates in a black hole geometry, thus defining a
 Boulware vacuum~\cite{birrell-davies,Boulware_1975}.

\subsection{Particle-field interactions: Atom Hamiltonian, field coupling, and allowed transitions}
\label{sec:QOptics-interactions_Hamiltonian}

In the previous section, we discussed the formulation of quantum field theory in generic spacetime backgrounds, 
centered on the details of the quantization of a scalar field. Aside from some subtle technical issues, 
such framework is straightforward. 
By contrast, the physics of atoms and their interactions with the field can be considerably more complex. 
Thus, we will consider a simplified framework for the atom-field interactions.

 \subsubsection{Generalized quantum Rabi model}
 
In our analysis of the HBAR problem,  
 we will treat the atom-field interactions by using a simplified 
model of the atom as a two-level system. This model 
does capture the essence of the relevant physics and is of widespread use in quantum optics and quantum information
science. 

As a two-state system, the relevant model is a generalization of the quantum Rabi 
model (QRM), whose semiclassical version was established by Rabi in the seminal Refs.~\cite{Rabi-1936, Rabi-1937}, 
and further extended with a coupling of the atom to a single quantized electromagnetic or bosonic field mode in the 
1960s~\cite{JCM_Jaynes-Cummings-1963}. 
A particular case of the quantum Rabi model is the celebrated  
Jaynes-Cummings model (JCM)~\cite{JCM_Jaynes-Cummings-1963},
which can be solved analytically in a closed form.
(For details on the quantum optics related to these models,
see Refs.~\cite{scullybook,meystrebook}; 
also, Refs.~\cite{JCM_Shore-etal--1993, JCM_Greentree-etal-2013, JCM_Larson-etal-2024}
for extensive reviews and discussions of the JCM and applications; 
and Ref.~\cite{QRM-rev_Braak-et-al-2016} 
 for a recent review of the more general QRM.)
The standard QRM, which is of widespread use 
in quantum optics, condensed matter physics, quantum information science, and other fields, consists of: 
(i) a two-level atom described by the Hamiltonian $H_{\rm at}$ as a two-state system; 
(ii) a field mode of the form $\hbar \omega_{0} a^{\dagger} a$ [cf.\ $H_{\rm int}$ in Eq.~(\ref{eq:H_field})];
(iii) an interaction $ H_{\rm int}$ as a linear monopole or dipole coupling
of the two-level atom with the bosonic field.
 Despite its apparent simplicity, 
it continues to be an active area of theoretical and applied research.
In fact, a completely general solution of the QRM has eluded researchers for decades, but it is now
regarded as a quasi-exactly-solvable model~\cite{QRM-rev_Braak-et-al-2016}, following
a series of partial solutions found in the past 
decade~\cite{QRM-sol_Braak-2011, QRM-sol_Chen-etal-2012, QRM-sol_Zhong-etal-2013, QRM-sol_Maciejewski-etal-2014, QRM-sol_Xie-etal-2017}.
However, this solution is complex, and most of the applications rely on numerical 
approximations~\cite{QRM_Bishop-etal-1996, QRM_Bishop-et-al1999, QRM_Bishop-2001, QRM_Fessatidis-etal-2002, QRM_Emary-2003}
and approximate analytical solutions~\cite{QRM_Pereverzev-2006, QRM_Debergh-2007, QRM_Irish-2007}.
 In addition to the well-known quantum optics applications~\cite{scullybook},
 this model also appears in similar formats in a variety of problems: 
 the Holstein model for the electron-phonon interaction in crystal lattice~\cite{polaron_Holstein-1959},
 in superconducting 
 qubits~\cite{SC-QRM_Chiorescu-etal-2004, SC-QRM_Wallraff-etal-2004, SC-QRM_Johansson-etal-2006, SC-QRM_Schuster-etal-2007},
in quantum dots~\cite{SC-QRM_Hennessy-etal-2007},
and in coupled nanomechanical oscillators oscillators~\cite{Qosc_Irish-Schwab-2007},
and other more recent applications~\cite{QRM-rev_Braak-et-al-2016}, including growing interest 
within quantum information science and circuit QED~\cite{circuit-QED_Blais-etal-2021}.

For the problems under the discussion in this article, and for similar studies of atom-field interactions in nontrivial spacetime 
configurations, a multimode form of the QRM is used, adapted to a specific spacetime background
with field modes that solve the classical equation~(\ref{eq:Klein_Gordon_basic}),
as in Sec.~\ref{sec:scalar-field_curved-ST}.
The two-level atom of the QRM and its generalizations has an energy spectrum 
with a ground state and an excited state: $\ket{b}\equiv \ket{E_{-}}$ and $\ket{a}\equiv \ket{E_{+}}$ respectively.
These are orthonormal energy eigenstates $\ket{E_{\mp}}$, with
\begin{equation}
E_{a} \equiv E_{+} > E_{b} \equiv E_{-}
\; \; , \; \; \; 
E_{+} - E_{-} = \hbar \nu
\; \; , \; \; \; 
E_{+}+ E_{-} = 2 \overline{E}
\; ,
\end{equation}
 such that
$\braket{E_{\pm}}{E_{\pm}} = 1$ and $\braket{E_{\pm}}{E_{\mp}} = 0$.
Then, the atom Hamiltonian is given by	
\begin{equation}
H_{\rm at} 
=  E_b \ket{b} \bra{b} + E_a \ket{a} \bra{a} 
=  
 \overline{E} \, \mathbb{1}_{2}
+
\frac{1}{2} \hbar \nu  \sigma_{z}
\; ,
\label{eq:H_atom}
\end{equation}
where $\sigma_{z} = \text{diag} (1,-1)$ is the diagonal Pauli matrix operator of a two-state system; and the first term 
proportional to the identity matrix $ \mathbb{1}_{2}$ (with the average energy
$\overline{E} $) can be dropped by choosing the energy reference level.
Then, the total Hamiltonian of the field-atom system is
\begin{equation}
H = H_{\rm at} + H_{\rm field} + H_{\rm int}
\;,	\label{eq:total-hamiltonian}
\end{equation}
where the field Hamiltonian of the full-fledged quantum field version 
 takes the form of Eq.~(\ref{eq:H_field}).
Correspondingly, for the atom-field interaction ${H}_{\rm int} \equiv V_{\rm int} $,
as shown in Appendix~\ref{app:QOptics}, 
this is the monopole analog of a dipole coupling for a spin-one photon field.
The scale of this simplified model can be adjusted by considering
a coupling $g =\mu E/\hbar$, where $\mu$ is the atomic dipole moment and E is the electric field.
 With a given coupling strength $g$, which we will assume to be weak,
this interaction yields the Hamiltonian ${H}_{\rm int} \equiv V_{I} $
given by
\begin{equation}
{H}_{\rm int} 
=
g \, \Phi ( {\boldsymbol{r }} (\tau), t(\tau) ) \, \sigma (\tau)
 \; ,
\label{eq:QO_interaction_potential}
\end{equation}
where $\sigma$ is the atomic-state transition operator defined below,
and both the field operator $\Phi$ and $\sigma$
are evaluated at the proper time $\tau$ of the atom. 
The atomic state transitions are described by the evolving linear-combination operators
\begin{equation}
\sigma (\tau) = \sigma_- e^{-i\nu \tau} + \sigma_+ e^{i\nu \tau}
=
\sigma_- e^{-i\nu \tau} + \mathrm{H.c.} 
\; , 
\label{eq:QO_atomic-transitions}
\end{equation}
where $\mathrm{H.c.}$ is the Hermitian conjugate and
$\sigma_{\mp}$ are the atomic lowering and raising operators
\vspace{-0.5em}
	\begin{equation}
	{\sigma}_{-} \equiv {\sigma}_{ba} = \ket{b}\bra{a} 
	\;, \qquad\qquad\qquad
	{\sigma}_{+} \equiv {\sigma}_{ab} = \ket{a}\bra{b} 
	\label{eq:sigma-operators}
	\; 
 	\end{equation}
(as lowering/raising Pauli matrices);
notice that these can be defined via $\sigma (0) \ket{ E_{\pm}} = \ket{ E_{\mp}}$.

 \subsubsection{Perturbation theory and the interaction picture}

The expressions above for the Hamiltonian~(\ref{eq:QO_interaction_potential}) 
are written in the interaction picture, where the operators acquire an extra time dependence from the 
 free Hamiltonians of the field and atom. 
 More generally, in the interaction picture, of widespread use for perturbative calculations in
 quantum field theory, the action and the Hamiltonian are separated into a free part (``unperturbed'') associated
 with the free fields and a part associated with the interactions (both
 self-interactions or interactions among the different fields).
 As a result, this picture involves: 
 (i) operators evolving in time according to the Heisenberg equation~(\ref{eq:Heisenberg-eq})
  associated with only the free part of the Hamiltonian; 
 and (ii) states evolving in time according to Eq.~(\ref{eq:Schrodinger-eq})
 associated with the interaction terms.
  In this hybrid interaction-picture form, the expressions in Eq.~(\ref{eq:sigma-operators})
    are ideally suited for standard quantum field theory calculations.
 Instead, in atomic physics, quantum optics, condensed matter physics, quantum information science and other fields,
 the QRM interaction Hamiltonian is written in the Schr\"{o}dinger picture,
 without the time dependence and for a single mode, i.e., as 
 $H_{\rm int} = g \left(a + a^{\dagger} \right)
\left( \sigma_{-} + \sigma_{+} \right)$, with the Hamiltonian being
$\displaystyle H=\frac{1}{2} \hbar \nu \sigma_{z}  
+ \hbar \omega_{0} a^{\dagger} a + 2 g \left(a + a^{\dagger} \right) \sigma_{x} $.
For our applications, we will use
 the Hamiltonian expressions (\ref{eq:H_field}) and (\ref{eq:H_atom})--(\ref{eq:sigma-operators})
  as the anticipated multimode generalization of the QRM.
Moreover, the product $\hat{m} = g \sigma$ acts as a monopole operator implementing 
the coupling with the field $\Phi$, and can also be used as the basis for a model detector.

With the expansions~(\ref{eq:field_expansion}) and (\ref{eq:QO_atomic-transitions})
 of the field and atomic operators, 
the interaction Hamiltonian~(\ref{eq:QO_interaction_potential}) 
in the interaction picture takes the explicit form 
\begin{equation}
    V_{I} \equiv H_{\rm int, I}   = 
    \sum_{ \boldsymbol{s} } g_{ \boldsymbol{s} }
     \left[
     a_{ \boldsymbol{s} }  \sigma_{+} e^{-i (\omega_{\boldsymbol{s}}t  -\nu \tau)}
     +
 a_{ \boldsymbol{s} }^{\dagger}  \ \sigma_{-} e^{i (\omega_{\boldsymbol{s}}t  -\nu \tau)}
     +
          a_{ \boldsymbol{s} }^{\dagger}  \sigma_{+} e^{i (\omega_{\boldsymbol{s}} t +\nu \tau )}
          +
               a_{ \boldsymbol{s} }  \sigma_{-} e^{-i (\omega_{\boldsymbol{s}} t +\nu \tau)}
                     \right]
    \; , 
     \label{eq:multimode-interaction-QO}
\end{equation}
in the notation we will use in subsequent sections. (Here, the symbol $I$ stands for interaction picture.)
Incidentally, the products of field and atomic transition operators 
in Eqs.~(\ref{eq:QO_interaction_potential}) and (\ref{eq:multimode-interaction-QO}) denote tensor products
of factors in separate spaces
(e.g., the term $a_{\boldsymbol{s}} \sigma_+$ stands for
$a_{\boldsymbol{s}} \otimes \sigma_+$ affecting two distinct systems).

 \subsubsection{Interpretation of the interaction terms}
 
\begin{figure}
    \centering
  \includegraphics[width=1\linewidth]{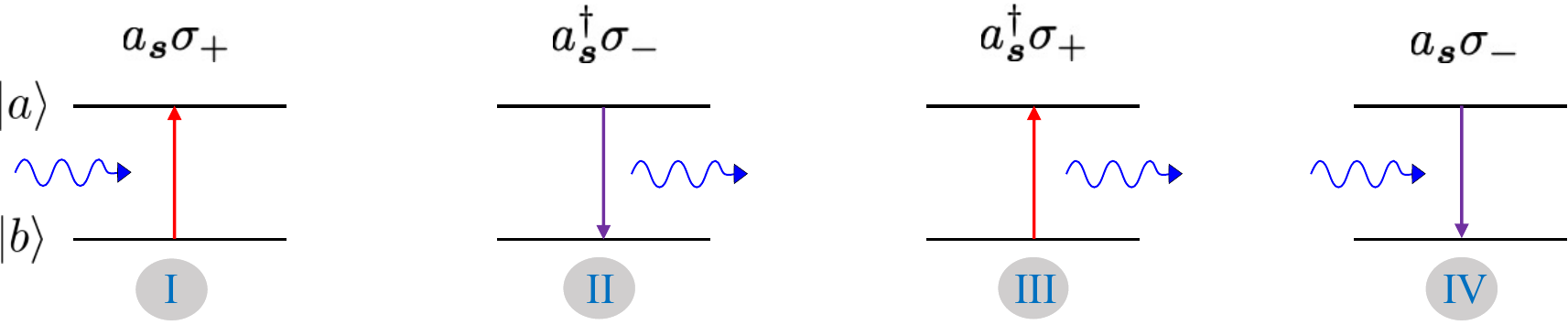}
  \caption{Schematics of the emission and absorption 
  processes corresponding to the four terms in the interaction Hamiltonian of Eq.~(\ref{eq:multimode-interaction-QO}). Here, $\sigma_{\pm} $ are the atomic raising and lowering operators defined in Eq.~(\ref{eq:sigma-operators}). 
    The specific couplings of I and II give the rotating terms, while those of III and IV give the counter-rotating terms.
    The latter can be neglected as the rotating-wave approximation (RWA) under a broad range of ordinary lab conditions (near resonance and in the weak-coupling regime), but they are 
critically important in relativistic setups with accelerated particles and/or horizons.}
\hspace*{0in}
 \label{fig:QO-processes}   
    \end{figure}

 For each frequency, Eq.~(\ref{eq:multimode-interaction-QO})
involves four terms, with each one
representing a particular photon creation/annihilation process with atomic excitation/de-excitation, as shown 
in Fig.~\ref{fig:QO-processes}. 
The interaction terms in Eq.~(\ref{eq:multimode-interaction-QO}) 
are usually classified in two pairs: (i) rotating terms, labeled I and II in Fig.~\ref{fig:QO-processes}; 
and (ii) counter-rotating terms, labeled III and IV.
The rotating terms are
$ a_{ \boldsymbol{s} }  \sigma_{+} $,
where the field loses a photon with atom excitation,
and
$ a^{\dagger}_{ \boldsymbol{s} }  \sigma_{-} $,
where the field gains a photon with atom de-excitation.
   By contrast, the counter-rotating terms are
$ a^{\dagger}_{ \boldsymbol{s} }  \sigma_{+} $,
where the field gains a photon with atom excitation,
and
$ a_{ \boldsymbol{s} }  \sigma_{-} $,
   where the field loses a photon with atom de-excitation.
  The rotating interaction terms are so called 
 because of their oscillatory phases with respect to time, whereas
 the faster time oscillations of the counter-rotating terms tend to have them suppressed near resonance.
 Thus, the rotating terms are considered the dominant building blocks for the resonant energy exchange
 with weak coupling between the atom and the field; as such, they are the ones 
 that appear in the Jaynes-Cummings model---this selective use of terms is called the
  rotating-wave approximation (RWA)~\cite{scullybook,meystrebook}.
 In that context, the counter-rotating terms are usually considered virtual processes, but this only means that 
    their effects less are significant in specific regimes under ordinary lab conditions.
    However, in the strong coupling regime
 and far from resonance, the counter-rotating terms do become important and the full-fledged QRM is mandatory.
     The remarks made about the interaction terms so far apply to the usual discussions
     of the physics of  inertial observers in flat spacetime.
However, in the presence of acceleration or nontrivial spacetime backgrounds, there is no physical rationale
 to consider regimes where processes III and IV would be neglected, 
 and they do become instrumental in nontrivial effects associated with radiation acceleration.
 This is often described intuitively as virtual processes turning to real ones under special conditions, e.g.,
 with accelerated particles. 
We will see how this is physically realized for acceleration radiation,
 starting in Sec.~\ref{sec:QOptics-interactions_HBAR}.

\subsection{Particle detectors in quantum field theory}
\label{sec:QOptics-interactions_UDW}

The use of simple models to describe photon detectors 
has a long tradition in quantum optics~\cite{Q-coh_Glauber-1963}.
More generally, a particle detector is a 
 controllable quantum system locally coupled to quantum fields.
In this generalized sense,
detector models can be used as 
probes for a large variety of effects in
the presence of a nontrivial spacetime structure, including various relativistic quantum information properties. 
Thus, they have become standard tools in modern quantum field theory.
The first type of such models, by Unruh~\cite{unruh-notes},
involves a small-box detector with a 
particle transitioning from the ground state to an excited state.
 DeWitt~\cite{DeWitt_QG-1979} introduced the
 point-source two-level detector, commonly known as the Unruh-DeWitt (UDW) 
detector in the literature~\cite{Crispino-et-al_2008_Unruh-effect}.
The UDW detector is based on the point-like two-level atom discussed in the previous section,
with monopole coupling~(\ref{eq:QO_interaction_potential});
essentially, it couples a qubit to a quantum field via a monopole interaction.
 Specifically, in this form, the UDW detector was originally conceived as a basic probe of the thermal nature 
  of the vacuum in accelerated frames, where
  it can be used to identify a mixed, thermal state~\cite{Unruh-Wald_Rindler-particle-1984} 
  at the Unruh temperature~(\ref{eq:Unruh-temperature}).
Additional varieties of this concept have been extensively studied in the literature;
 for example, particle detectors with finite spatial extent~\cite{detector_Grove-Ottewill-1983},
 and more recently, harmonic-oscillator 
 detectors~\cite{HO-detector_Lin-Hu-2007, HO-detector_Brown-etal-2013, HO-detector_Bruschi-et-al-2013},
which are ideally suited for nonperturbative calculations.
The study and applications of particle detectors is an ongoing area of research, where quantum correlations can be studied 
in detail~\cite{Hu-etal_RQI-UDWdetector-2012,Foo-etal_RQI-UDWdetector-2020}.

For an UDW monopole detector, the detector-field interaction is described by the Hamiltonian of
Eqs.~(\ref{eq:QO_interaction_potential}) and (\ref{eq:multimode-interaction-QO}).
As discussed in Sec.~\ref{sec:QOptics-interactions_Hamiltonian},
the terms in Eq.~(\ref{eq:multimode-interaction-QO}), which are depicted in Fig.~\ref{fig:QO-processes}, correspond to the four combinations
of photon absorption/emission and atomic excitation/deexcitation allowed by the physics.
Ordinarily, only the terms I and II are consistent with conservation of energy,
but the virtual processes III and IV can be realized in nontrivial spacetime configurations
with acceleration and/or gravitational fields. For the current purposes,
the initial state of the detector-field system is a tensor product state $\ket{0_M,b}\equiv \ket{0_M}\otimes \ket{b}$, where $\ket{b}$ is the ground state of the detector. 
 Thus, when the detector clicks (``detects a particle''), this signifies its transition to the excited state $\ket{a}$
via the interaction~(\ref{eq:QO_interaction_potential})
 with the field; and the field goes to an excited state $\ket{\psi}$.

\subsection{Particle-field interactions: Transition probabilities of HBAR radiation}
\label{sec:QOptics-interactions_HBAR}

The horizon-brightened acceleration radiation (HBAR) discussed in this paper is one
of the most interesting applications of the conversion of virtual into real processes via the counter-rotating terms in Eq.~(\ref{eq:multimode-interaction-QO}) and Fig.~\ref{fig:QO-processes}.
In HBAR 
radiation~\cite{Scully_2018_HBAR,acceler-rad-Schwarzschild,acceler-rad-Kerr,acceler-rad-Qopt-1,acceler-rad-Qopt-2},
a black hole provides a nontrivial spacetime background, where one can consider a field $\Phi$ prepared 
with a configuration corresponding to stationary coordinates, in what is known as a Boulware-like state.
One simple operational approach to set up this configuration is the introduction of mirrors at specific boundaries,
with one boundary right outside the event horizon---see Sec.~\ref{app:spacetime}.
This setup amounts to a thought experiment involving an atom or atoms interacting with a field, as in 
Secs.~\ref{sec:QOptics-interactions_Hamiltonian} and \ref{sec:QOptics-interactions_UDW}.
In essence, this is {\it a scaled-up version of an optical cavity in a lab, but with the
ability to probe near-horizon black hole behavior\/}. 
See the qualifying remarks on the ``optical cavity model'' 
in the introductory part of Sec.~\ref{sec:QOptics-interactions}.

{\em The atoms are initially in their ground state\/} $\ket{b}$: each one acts as an UDW detector, but collectively they 
produce a radiation field. The calculations outlined below refer to this radiation field.
The basic description of the field $\Phi$ follows the theory reviewed in Sec.~\ref{sec:scalar-field_curved-ST},
with the field states labeled by their occupation numbers $n_{\boldsymbol{s}}$.
In particular, the state with no scalar photons is the vacuum or ground state $\ket{0}$, 
such that $a_{\boldsymbol{s}} \ket{0} = 0$ for all 
modes ${\boldsymbol{s}}$; and
 the state $\ket{1_{\boldsymbol{s}}}$ has only one photon in mode ${\boldsymbol{s}}$.
Then, the coupling~(\ref{eq:QO_interaction_potential}) allows for:
(i) the emission of a scalar photon 
with the simultaneous transition of the atom to its excited state $\ket{a}$;
 and (ii) the absorption of a scalar photon 
 with the atom also transitioning to its excited state $\ket{a}$. 
 These correspond to the counter-rotating term III and  
 the rotating term I, respectively, as illustrated in Fig.~\ref{fig:QO-processes}. 
 For the term III, the field is in the vacuum configuration, and 
 the first-order perturbation probability amplitude is
$-(i/\hbar) \, 
I_{{\mathrm e}, {\boldsymbol{s}} }$, where
$I_{{\mathrm e}, {\boldsymbol{s}} } =
\int d\tau \;\bra{1_{\boldsymbol{s}},a}V_I(\tau)\ket{0,b}$.
Similarly, 
for the term I, the field is in the one-photon mode configuration $\ket{1_{\boldsymbol{s}}}$, with 
the absorption probability amplitude being
$-(i/\hbar) \, 
 I_{{\mathrm a}, {\boldsymbol{s}} }$, where
$I_{{\mathrm a}, {\boldsymbol{s}} } =
\int d\tau \;\bra{0,a}V_I(\tau)\ket{1_{\boldsymbol{s}},b}$.

Therefore, up to first order in perturbation theory, 
and using the operators~(\ref{eq:sigma-operators}),
the emission probability $P_{{\mathrm e}, {\boldsymbol{s}} }$ for the field mode ${\boldsymbol{s}}$,
 is given by
\begin{equation}
P_{{\mathrm e}, {\boldsymbol{s}} } 
= \left| \int d\tau \;\bra{1_{\boldsymbol{s}},a}V_I(\tau)\ket{0,b}\right|^2 
  = g^2 \left|\int\; d\tau\; \phi_{\boldsymbol{s}}^*(\mathbf{r} (\tau),t(\tau)) \, e^{i\nu\tau}\right|^2
        \; ;
        \label{eq:P_ex_explicit}
\end{equation}
and  the absorption probability is 
\begin{equation}
    P_{{\mathrm a}, {\boldsymbol{s}} }
     = \left|\int d\tau \;\bra{0,a}V_I(\tau)\ket{1_{\boldsymbol{s}},b}\right|^2
     = g^2 \left|\int\; d\tau\; \phi_{\boldsymbol{s}}(\mathbf{r} (\tau),t(\tau)) \, e^{i\nu\tau}\right|^2
        \; .
         \label{eq:P_ab_explicit} 
\end{equation}

    The expressions in Eqs.~(\ref{eq:P_ex_explicit}) and (\ref{eq:P_ab_explicit}) 
     are critical in finding the field configuration generated by the falling atomic cloud in the 
     HBAR thought experiment. 
     With the reasonable assumption that the coupling is weak, they are dominant 
    first-order approximations; as such, this involves no serious physical restrictions.
    Otherwise, the spacetime background can be completely general. But, as we will see below,
   a near-horizon black hole background leads to a universal outcome for HBAR radiation.
   In order to assess this effect, we have to specify the
   spacetime configuration and evaluate the probabilities~(\ref{eq:P_ex_explicit}) and (\ref{eq:P_ab_explicit})    
   for the corresponding field modes and spacetime geodesic worldlines (with proper time $\tau$).
   As we will see in the next section,
   the field density matrix will characterize the thermal properties of the radiation leading to HBAR thermodynamics.

   \section{Quantum states and density matrix: 
   From open quantum systems and quantum optics to HBAR radiation}
\label{sec:QOptics-density-matrix}

   The standard description of the time evolution of physical states in quantum dynamics, in the form 
   of the Schr\"{o}dinger equation~(\ref{eq:Schrodinger-eq}) 
   only applies to isolated systems described by a pure quantum state $\ket{\Psi}$.
In that case, the unitary time evolution operator
$\hat{U}(t,t_{0}) = 
T \exp \left[ -(i/\hbar)\int_{t_{0}}^{t}  dt' \hat{H}(t') \right] $
(where $T$ is the time-ordering operator giving the Dyson series~\cite{Weinberg_QM,Sakurai-QM}),
governs the fundamental dynamics of the state 
$\ket{\Psi (t)}$ at a any given time $t$ through the Hamiltonian,
in such a way that
$\ket{\Psi (t)} = \hat{U}(t,t_{0}) \ket{\Psi (t_{0})}$.

 However, a more general treatment is needed for any system that is not closed, but
 interacting with an environment, or part of a larger system to which it may entangled.
  In their most general form, this defines the larger class 
  of {\it open quantum systems\/}~\cite{Breuer-Petruccione_OQS},
  whose description requires a density matrix (density operator).
    
 We briefly outline next the definition and main concepts associated with the density matrix, and summarize
 the particular form of this operator that has been developed for the theory of lasers in quantum optics.  We
 further consider analog systems, including HBAR radiation and its reduced density matrix.

   \subsection{Density matrix in quantum physics: Basic definitions and properties}
\label{subsec:density-matrix}

The necessity to extend the standard pure-state approach based on the Schr\"{o}dinger equation~(\ref{eq:Schrodinger-eq})
became apparent as soon as formal quantum mechanics was born.
In 1927,
the density matrix---also called density operator---was 
introduced as a more general characterization of a physical state in quantum physics
and its associated dynamic evolution. 
This is a remarkable extension,
due to von Neumann~\cite{vonNeumann_density-matrix} and Landau~\cite{Landau_density-matrix},
 that brings to completion the quantum dynamics program 
of Eqs.~(\ref{eq:Schrodinger-eq}) and (\ref{eq:Heisenberg-eq}).
Over the following decades,
this approach has been gradually applied to a variety of problems
in fundamental quantum physics and statistical mechanics. 

One of the earliest findings was the realization that
the density matrix is needed even in simple cases where one 
looks at subsystems of larger system~\cite{Landau_density-matrix}.
This concept has developed into a central tool in the description of quantum
entanglement and quantum information more generally~\cite{nielsen00}.
 In addition, attempts to understand the role played by the environment in quantum systems have
 stressed the need to use states consisting of statistical mixtures. This has led to a
  systematic development of the theory
of open quantum systems as a more recent quantum development that requires 
a density matrix as a conceptually distinct description of their states~\cite{Breuer-Petruccione_OQS}.
    
   \subsubsection{Probabilistic definition of the density matrix}
\label{subsubsec:def-density-matrix}

The quantum dynamics program 
of the Schr\"{o}dinger and Heisenberg pictures, implemented via Eqs.~(\ref{eq:Schrodinger-eq}) and (\ref{eq:Heisenberg-eq}), involves {\em pure states\/}.
These are states described by normalized vectors in Hilbert space, which can be prepared when one
has maximal amount of information about the given physical system.
A natural extension is to consider states about which only partial information is available. 
These {\em mixed states\/} can be described as
statistical mixtures of pure states properly weighted with a probability distribution 
 associated with an incomplete knowledge of the system.
Such extensions encompass
the types of states used in statistical-mechanical and thermodynamic treatments of systems, and in physical situations where subsystems need to be analyzed independently from the environment or in the presence of quantum entanglement: see the examples discussed below. 

With these ideas in mind, the density matrix can be defined 
as the mathematical description of a general quantum state that exists as {\em a statistical
mixture of an ensemble of normalized pure states\/}
 $\bigl\{ \ket{\psi_i} \bigr\}$.
 If this general state has a probability $p_i$ of being state $ \ket{\psi_i} $, 
 the general definition of the density matrix is~\cite{Weinberg_QM, Breuer-Petruccione_OQS},
\begin{equation}
    \rho = \sum_i p_i \ket{\psi_i}\bra{\psi_i}
    \; 
     \label{eq:density-def-1}
\end{equation}
where the real nonnegative probabilities $p_i$  (i.e., $0 \leq p_i \leq 1$) are
 subject to the normalization condition
 \begin{equation}
 \sum_{i} p_i = 1
 \; .
     \label{eq:density-def_prob}
\end{equation} 
Several qualifying remarks are in order regarding this definition.
\begin{itemize}
\item
First, the 
states  $\bigl\{ \ket{\psi_i} \bigr\}$ are normalized but otherwise arbitrary: they do not need to be orthogonal.
\item
Second, 
a pure state 
$ \ket{\psi}$ can be represented by only one term in the sum~(\ref{eq:density-def-1}), 
as a projection operator  $\ket{\psi}\bra{\psi}$.
\item
Third,
the definition of the density matrix~(\ref{eq:density-def-1}),
with Eq.~(\ref{eq:density-def_prob}),
 is a convex weighted sum of the projection operators 
of the constituent pure states
$\bigl\{ \ket{\psi_i} \bigr\}$.
\item
Fourth, the decomposition of a mixed state into a mixture of pure states of the form~(\ref{eq:density-def-1}) 
 is not unique. 
\end{itemize}

The general definition of a quantum state~(\ref{eq:density-def-1}) gives a rigorous representation of the two
types of probabilities that can enter in quantum mechanical systems: 

\noindent
(i) the 
intrinsic probabilistic nature of the quantum state vectors (quantum uncertainty);
\\
\noindent
(ii) 
 the incomplete nature of the information
about the initial state of the system, as described by the weighted sum~(\ref{eq:density-def-1}).

 \subsubsection{Properties and operator definition of the density matrix}
\label{subsubsec:properties-density-matrix}

The explicit probabilistic definition of a general quantum state via  Eq.~(\ref{eq:density-def-1})
can be replaced by an equivalent operator statement that defines the
density matrix $\rho$ as 
a self-adjoint positive semidefinite operator of trace one~\cite{Weinberg_QM, Breuer-Petruccione_OQS}.
Specifically, the density operator $\rho$ is characterized by the following three properties.
\begin{enumerate}
\item
Self-adjointness:
$
    \rho = \rho^{\dagger}
   $.
   
\item
Positive semi-definiteness:
all of the eigenvalues of $\rho$ are non-negative (\(\ge 0\)).

\noindent
[This is required by the probabilities $p_i$ in Eq.~(\ref{eq:density-def-1})
being nonnegative.]

\item
$\Tr \left( \rho \right)= 1$.

\noindent
[This trace normalization is equivalent to the probability normalization condition~(\ref{eq:density-def_prob}).]

\end{enumerate}
Moreover, the eigenvalues and corresponding operator trace of $\rho$
 can be used to characterize the nature of the quantum state, as follows.
\begin{itemize}
\item
{\em Pure State\/}.
A quantum pure is a state with no classical uncertainty. This amounts to
  all the eigenvalues being zero, except for a single eigenvalue equal to unity.
  It is characterized by the idempotent condition $\rho = \rho^2$.
 As an operator trace relation, this is equivalent to a purity
$\text{Tr}(\rho ^{2})=1$.

\item
{\em Mixed State\/}.
A proper quantum mixed state is a state that is not pure. As such, it is represented by a nontrivial
 statistical ensemble, as in Eq.~(\ref{eq:density-def-1}) 
 with more than one nonzero probability. 
 This implies it has multiple nonzero eigenvalues. 
  As an operator trace relation, this is equivalent to a purity
 \(0< \text{Tr}(\rho ^{2})<1\). 
\end{itemize} 
 
As a result,
 the operator trace gives an invariant definition of probabilistic statements, including:
 \begin{itemize}
 \item 
  the primary definition of
 expectation value of an operator $A$ as 
 $\left\langle A \right\rangle_{\rho}  = \Tr \left( \rho A \right)$;
\item 
the density matrix of subsystems of larger system in terms of
partial traces: see Eq.~(\ref{eq:reduction-of-density-matrix}) in Sec.~\ref{subsec:density-matrix_OQS}.
\end{itemize}
In addition, the time evolution is given by the von Neumann equation,
as shown in Sec.~\ref{subsec:density-matrix_OQS}.

   \subsubsection{Density matrix: Examples}
\label{subsubsec:density-matrix-examples}

\begin{itemize}

\item
{\em  System in thermodynamic equilibrium with a thermal reservoir.}

   A familiar example of a mixed state is that of a system in thermal equilibrium. This is 
   critical 
  for our discussion of the remainder of this review article.
  
  In what is called the canonical ensemble, a system exchanges energy with an environment or heat bath
  at temperature  $T$; thus, it is not in a pure state.
 We can then write the mixed state in the form of Eq.~(\ref{eq:density-def-1}),  
in terms of the eigenstates $\ket{n}$ of the system 
Hamiltonian ${H^{S}}$, 
i.e., $    \rho = \sum_n p_n \ket{n}\bra{n}$, with normalized
probabilities 
\begin{equation}
p_{n} = Z^{-1} e^{-\beta E_n} \; , \; \; \text{where} \; \;
\beta= 1/(k_{B}T)
\end{equation}
 is the inverse temperature parameter,
$ Z \equiv Z(\beta) = \sum_n e^{-\beta E_n}= \Tr \left( e^{-\beta {H^{S}}} \right)$,
and 
 $E_n$ is the energy of the state $\ket{n}$.
  Then, thermal-equilibrium density matrix given by
    \begin{equation}
    \rho = Z^{-1} e^{-\beta {H^{S}}}
    \; .
    \end{equation}
It should be noted that the operator ${H^{S}}$ is associated with the given system $S$
(and not with the combined system Hamiltonian that includes the environment). 

\item
{\em Spin 1/2 examples.}

Spin 1/2 systems provide simple examples of density matrices, with beams of particles having 
different polarizations. In the examples below, with a slight abuse of notation, 
we use both the Dirac notation and matrix representations; for the latter, the usual standard Pauli-matrix 
representation is used~\cite{Weinberg_QM}.

(i) A completely polarized beam with a given orientation of spin.
For example, the
$z$ component of spin: $S_{z} =-\hbar/2$ gives 
\begin{displaymath}
\rho = \ket{+}\bra{+} 
=
\begin{pmatrix} 0\;  \; & \; \; 0  \\ 0 \; \; & \; \; 1 \end{pmatrix}
\; .
\end{displaymath}
And the $x$ component of spin: $S_{x} =\pm\hbar/2$ gives 
\begin{displaymath}
\rho = \ket{\pm}_{x}\bra{\pm}_{x}  = \frac{1}{2}
\begin{pmatrix} 1\;  \; & \; \; \pm 1   \\ \pm 1 \; \; & \; \; 1 \end{pmatrix}
\; ,
\end{displaymath}
as $ \ket{\pm}_{x} = (1/\sqrt{2}) \left( \ket{+}\pm \ket{-} \right)$
For both examples, as these are pure states, 
the density matrices satisfy $\rho = \rho^2$ and ${\rm Tr} ( \rho^2) =  1$.

(ii) An unpolarized beam, which is maximally mixed, i.e., an incoherent mixture or 
 statistical ensemble with the same proportions of spin up and spin down; thus,
  $p_{+}=p_{-}=1/2$ in the definition of Eq.~(\ref{eq:density-def-1})]:
\begin{displaymath}
\rho = 
\frac{1}{2} \left( \ket{+}  \bra{+} \right) + \frac{1}{2} \left( \ket{-}  \bra{-} \right)
  = \frac{1}{2}
\begin{pmatrix} 1\;  \; & \; \; 0  \\ 0 \; \; & \; \; 1 \end{pmatrix} = \frac{1}{2} \,  \mathbb{1}_{2}
\; .
\end{displaymath}
As this is a mixed state, $\rho \neq \rho^2=  (1/4) \mathbb{1}_{2}$ and ${\rm Tr } (\rho^2) = 1/2 < 1$.
This is the purity characterization of a maximally mixed state.
And as the state is completely unpolarized, all spin components have zero expectation values:
$\left\langle S_{j} \right\rangle = {\rm Tr} (\rho S_{j}) = 0 $ (with $j=x,y,z$), as $\rho \propto   \mathbb{1}_{2}$ 
and the spin-matrix traces are zero.
Moreover, $\rho$ can be trivially 
rewritten as the incoherent sum (with equal weights)
in terms of spin plus and minus for arbitrary orientations of the quantization axis.

(iii) Finally, partially polarized beams representing arbitrary mixed states can represented 
as $\rho = p_{1} \rho_{1} + p_{2} \rho_{2}$, as in Eq.~(\ref{eq:density-def-1}), with $p_{1}+p_{2}=1$ and $\rho_{j}$ 
(with $j=1,2$) being a generic density matrix of a pure state with arbitrary choice of 
spin orientation. 

\end{itemize}

   \subsection{Density matrix in quantum physics: Open quantum systems}
\label{subsec:density-matrix_OQS}

In this section, as in the introductory Sec.~\ref{sec:introduction}, we are using $H$ to denote a generic global
 Hamiltonian of a possibly combined system.
 
 \subsubsection{Quantum dynamics: von Neumann equation}
 
The quantum dynamics in the density-matrix generalized framework is governed 
by the von Neumann equation~(\ref{eq:vonNeumann-eq-00}), i.e.,
\begin{equation}
   \frac{d\rho}{dt} = 
   \frac{1}{i \hbar} [H,\rho]
    \; .
    \label{eq:vonNeumann-eq}
\end{equation}
Equation~(\ref{eq:vonNeumann-eq})
 is also referred to as the Liouville-von Neumann 
or quantum Liouville equation, as it can be viewed as generalizing the
classical Liouville equation for the phase-space distribution function~\cite{Breuer-Petruccione_OQS}:
according to the formal classical-quantum correspondence principle~\cite{Dirac_correspondence}, 
the classical Poisson brackets are promoted to quantum-mechanical commutators~\cite{Weinberg_QM}.
Constructively, Eq.~(\ref{eq:vonNeumann-eq}) can be established 
   from the Schr\"odinger equation~(\ref{eq:Schrodinger-eq}) and the definition~(\ref{eq:density-def-1}); but, in its final form,
   it captures the more general evolution of quantum-mechanical systems,
   with the qualifications discussed in the next two paragraphs.
 
It should be noted that the von Neumann equation~(\ref{eq:vonNeumann-eq})
 describes the evolution for the whole system, including the environment, 
and can be used as the starting point for approximation schemes 
leading to master equations that extend the Schr\"odinger equation~(\ref{eq:Schrodinger-eq})
as a predictive tool for the evolution of all states.
In this sense, Eq.~(\ref{eq:vonNeumann-eq}) serves as the generator of
a variety of derived master equations tailored to specific physical applications, 
as discussed below.
An example of a master equation originally used in quantum optics
 is given in the next section.

 \subsubsection{Open quantum systems}
 
 The dynamics described by the von Neumann equation is especially insightful in 
open quantum systems, when applied along with the separation of a system $S$
within a larger, combined system. 
In this case, the global Hamiltonian is
\begin{equation}
 H=H^{{S}}+H^{{E}}+H^{{SE}}
 \end{equation}
where $ {\displaystyle H^{{S}}}$ is the given system Hamiltonian, $ {\displaystyle H^{{E}}}$
 is the Hamiltonian of the complementary system or ``environment''
 (whose specific state is typically unknown),
 and $ {\displaystyle H^{{SE}}}$ describes their interaction.
 
In this formalism, the system $S$ is in a state 
given by a reduced form of the density matrix, via the partial trace
\begin{equation}
 \rho^{{S}}(t)
 = \Tr_{{E}}
 \left[ \rho ^{{SE}}(t) \right]
\; .
\label{eq:reduction-of-density-matrix}
 \end{equation}
The partial trace effectively sums over the degrees of freedom of the environment, 
leaving only the relevant degrees of freedom of the system $S$ by itself. 
This is a reduction process, usually described as ``averaging out'' the states of the complementary system; then,
the reduced dynamics involves an effective treatment of the system subject to a quantum master equation.
%
 \begin{figure}[h]
    \centering
    \includegraphics[width=0.95\linewidth]{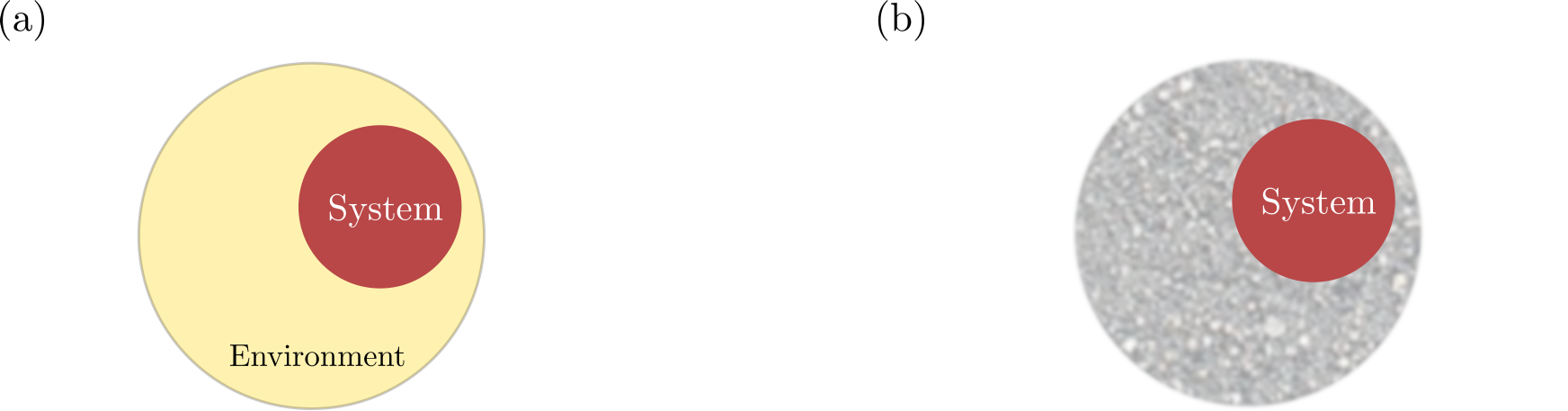}
    \caption{Schematic representation of the exchange of physical information 
    of the system ($S$) with the environment ($E$).
    The corresponding mathematical procedure is that of the partial trace, Eq.~(\ref{eq:reduction-of-density-matrix}).}
       \label{fig:system-environment}
      \hspace*{0in}
\end{figure}
%
The corresponding time evolution, unlike that for
the original von Neumann equation~(\ref{eq:vonNeumann-eq}), 
becomes nonunitary. The loss of unitarity is associated with the physical exchange of information of the given system $S$ with the environment $E$; see Fig.~\ref{fig:system-environment}.
 It is this physical exchange that is formally implemented
by the mathematical prescription~(\ref{eq:reduction-of-density-matrix}).

As a result of the response to the
external conditions to which the given system $S$ is subject to,
there exist a variety of master equations~\cite{Breuer-Petruccione_OQS}.
 In this sense, {\it a quantum master equation
is the most general equation for time evolution of an open quantum system\/}, extending the applicability
of quantum dynamics beyond the original Schr\"{o}dinger equation~(\ref{eq:Schrodinger-eq}).
Our primary example of this reduction process and use of a practical master equation,
which we will address in the next section,
is a quantum-optics master equation that has a broad range of applications, including for 
HBAR radiation.

 \subsubsection{Perturbation theory of density matrix}

As a final step, the density matrix can be evaluated using perturbation theory
 provided that the coupling is sufficiently weak. This is an important approximation needed
 to get practical results as the combined system presents a formidable problem.
 The successive orders are given by repeated commutator iterations with the Hamiltonian, 
 leading to a sequence of terms that can be evaluated 
 explicitly, and for which additional approximations and averages are available.
 
 Using this approach to second-order perturbation theory within the interaction picture
 (i.e., with $H_{\rm int} = H_{{\rm int},I}$),
 the perturbative expansion is
\begin{equation}
    \rho(t) 
    = \rho(0) 
    - (i/\hbar) 
    \int_0^{t} dt'\; [H_{\rm int},\rho(0)] 
    +
    (-i/\hbar)^2  \int_0^{t} dt' \int_0^{t'} dt''\; 
    \left[H_{\rm int},[H_{\rm int},\rho(0)]\right] + 
    \ldots
    \; ,
    \label{eq:QLE-perturbation-order2}
\end{equation}
 and this pattern continues to all orders.
 This is the density-matrix generalization for mixed states 
 of the Dyson series of pure states~\cite{Sakurai-QM} mentioned above,
 and the generalization of the well-known perturbation theory approach within canonical quantum field 
 theory~\cite{Weinberg_QFT}.

      In essence, the derivation of useful master equations follows a two-step reduction procedure:
\begin{displaymath}
\begin{pmatrix}
\text{Perturbation theory}
\\
\text{Eq.~(\ref{eq:QLE-perturbation-order2})}
\end{pmatrix}
\Longrightarrow
\begin{pmatrix}
\text{Reduction to system $S$}
\\
\text{Partial trace, Eq.~(\ref{eq:reduction-of-density-matrix})}
\end{pmatrix}
\end{displaymath}
  Next, we will show how these general results, including the 
  two-step reduction process,
  Eqs.~(\ref{eq:reduction-of-density-matrix}) and (\ref{eq:QLE-perturbation-order2}),
  are used for a particular quantum-optics form of the reduced density matrix.

      \subsection{Reduced density matrix in quantum optics: From lasers to curved spacetime}
\label{subsec:density-matrix_lasers}

   The quantum optics approach to the density matrix was developed in the 1960s for 
   an improved understanding of the operation of lasers and masers~\cite{scullybook,meystrebook}.
   This practical field evolved from Einstein's original work on the stimulated emission of 
   electromagnetic radiation~\cite{Einstein_AB-radiation}, allowing for 
   the possibility of population inversion and subsequent invention of lasers and 
   masers~\cite{Maiman-laser_1960,Bertolotti_masers-lasers}.
   
   \subsubsection{Scully-Lamb master equation}
   
   Initially, for the laser-related applications, a semiclassical approach was used: a quantum treatment of the atoms combined with a classical picture for the laser field, along the lines of the simple model described in 
  Sec.~\ref{sec:QOptics-interactions}.
   But in the early 1960s, Glauber~\cite{Glauber_laser_1964}
    pointed out that only a fully quantum-mechanical description would be reliable, requiring the
    development of a density matrix for the laser electromagnetic field.
    In a sequence of seminal papers~\cite{Scully-Lamb-rho_1964,Scully-Lamb-rho_1965}, 
    this problem was solved for the reduced density matrix of the single-mode laser field
    under a specific set of assumptions (see below), leading to the
    first laser master equation, whose diagonal elements are given by 
\begin{equation}
    \dot{\rho}_{n,n} =
   \underbrace{ 
   -  \big\{ [\alpha - \beta (n+1) ](n+1) \, {\rho}_{n,n} 
   -
   (\alpha - \beta n )n\, {\rho}_{n-1,n-1} 
  \big\}  
   }_{ \text{pumping} }
   -  
    \underbrace{
  \gamma \big[n \, {\rho}_{n,n}-(n+1) \, {\rho}_{n+1,n+1} \big] 
   }_{ \text{damping} }
    \; ,
  \label{eq:master_equation_Scully-Lamb_diagonal}
   \end{equation}
in terms of the optical parameters $\alpha$, $\beta$, and 
 $\gamma$, known as the linear gain, saturation, and loss respectively. 
In the Scully-Lamb master equation,
 while the most interesting elements are the diagonal ones, ${\rho}_{n,n} $, 
its more general form also includes off-diagonal elements~\cite{Scully-Lamb-rho_1964,Scully-Lamb-rho_1965}.
This equation relies on several assumptions that make it useful and versatile
even beyond lasers, but not universal.
 More generally, the master equations for laser systems comprise 
 a vast topic in quantum optics~\cite{scullybook,meystrebook}. 
 Among the assumptions leading to the master 
 equation~(\ref{eq:master_equation_Scully-Lamb_diagonal})
 are:
 \begin{itemize}
 \item
  the two-level atom model and electromagnetic coupling of Section~\ref{sec:QOptics-interactions}, 
  with interaction Hamiltonian~(\ref{eq:multimode-interaction-QO}) and associated probabilities;
\item
the injection of atoms at a specific rate;
\item
the particular use of the optical parameters $\alpha$, $\beta$, and $\gamma$;
\item
and a Markovian approximation (memoryless property):
the future evolution is solely determined by its current state.
\end{itemize}
 This master equation is further described in the next section, where it is generalized to be used 
 in specific spacetime backgrounds in the presence of black holes.

\begin{figure}[h]
    \centering
    \includegraphics[width=0.95\linewidth]{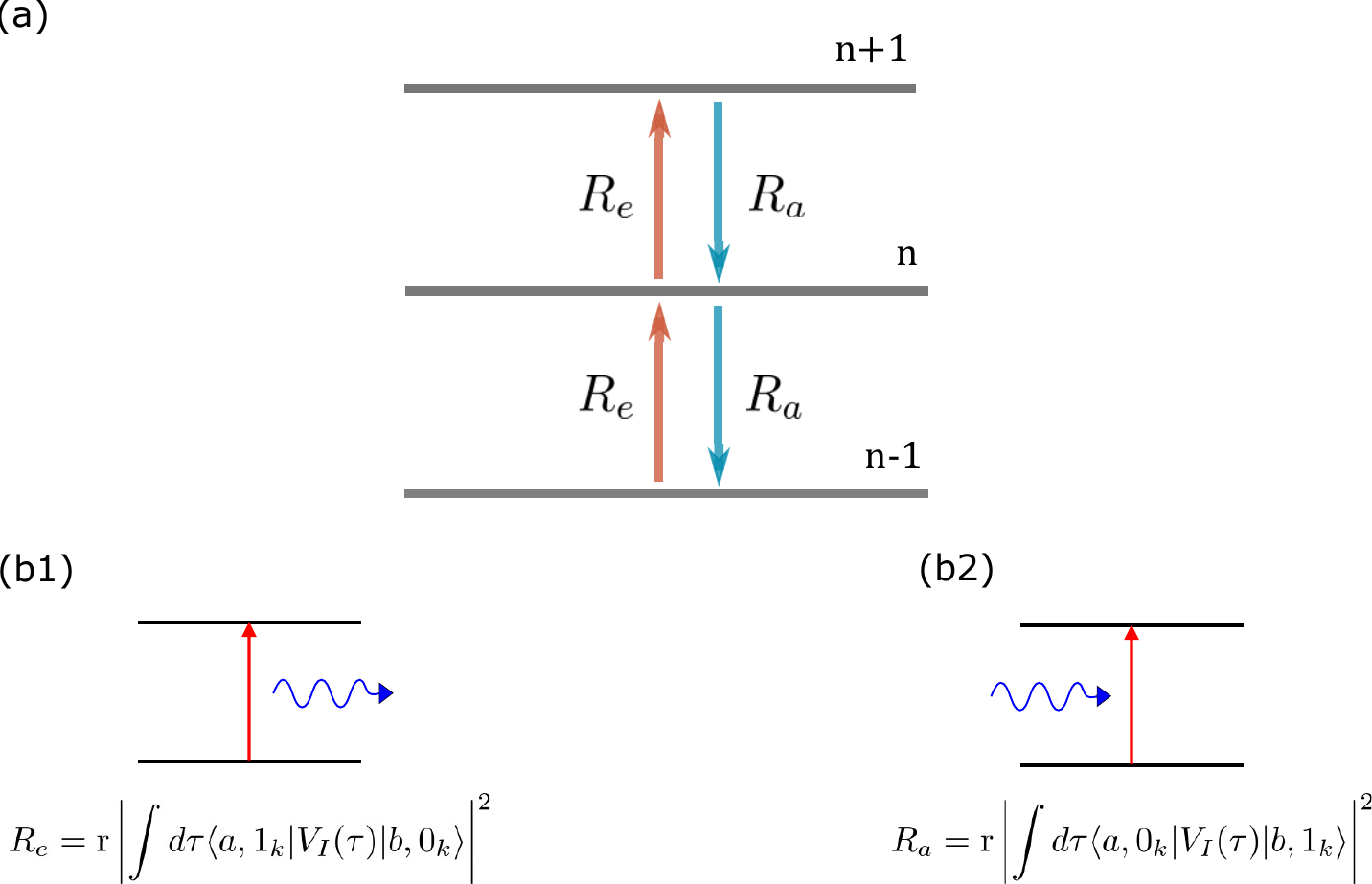}
    \caption{Probability flows governed by the generalized Scully-Lamb master 
    equation~(\ref{eq:master_equation_final_multimode}). Part (a) describes the flows between  
    the three relevant levels of the field mode,
    with specific transition rates $R$. The modes are labeled with $k \equiv \mathbf{s}$.    
    Parts (b1) and (b2) display the two relevant transition processes, for the given initial condition
    starting from the atomic ground state.}
       \label{fig:detailed-balance}
       \hspace*{0in}
\end{figure}

The Scully-Lamb laser model, with 
Eq.~(\ref{eq:master_equation_Scully-Lamb_diagonal}),
in addition to leading to a deeper understanding of this technology, has generated a vast literature of research in:
    quantum optics and open quantum systems (for 
    example~\cite {Carmichael_1974, Gea-Banacloche_1997, Arkhipov-etal_PT-symmetry, Minganti-etal_L-collapse}, 
    and references therein);
    Bose-Einstein condensates (BEC) as ``atom laser'' 
    analogs~\cite{BEC-laser-analog_DeGiorgio-Scully-1970, BEC-laser-analog_Scully-1999, BEC-laser-analog_Scully-etal-2000, BEC-laser-analog_Scully-etal-2022};
    and black hole physics (as further discussed in this review article)~\cite{Scully_2018_HBAR}.
The terms in Eq.~(\ref{eq:master_equation_Scully-Lamb_diagonal}) 
     describe probability flows between the photon occupation number $n$ and the adjacent numbers 
   $(n-1)$ and  $(n+1)$, 
   having a structural form
   \begin{equation}
    \dot{\rho}_{n,n} =
   \underbrace{ 
   -  \big[ R_{{\mathrm e}, n } \, (n+1) \, {\rho}_{n,n} - R_{{\mathrm e}, n-1 } \, n \,  {\rho}_{n-1,n-1}\big]  
   }_{ \text{emission} }
   -  
    \underbrace{
\big[    R_{{\mathrm a}, n} \,
    n \, {\rho}_{n,n}
    -
    R_{{\mathrm a}, n+1} \,
    (n+1) \, {\rho}_{n+1,n+1} \big] 
   }_{ \text{absorption} }
    \; ,
  \label{eq:master_equation_Scully-Lamb-gen_diagonal}
   \end{equation}
   in which $R_{{\mathrm e}, n } $ and $R_{{\mathrm a}, n } $
   are emission and absorption probabilities associated with 
   pumping and damping in the laser model---in effect,
the pumping terms are primarily governed by stimulated emission 
via the excitation rate of atoms in the cavity, mode frequency, and dipole matrix element, while the 
     damping is due to cavity losses described by a quality factor~\cite{scullybook}.   
  However, written in the form of Eq.~(\ref{eq:master_equation_Scully-Lamb-gen_diagonal}), 
   the master equation leads to a broader class of applications for analog systems, including the ones 
   mentioned above. 
   This general form 
    is shown in part (a) of Fig.~\ref{fig:detailed-balance}.
   (The complete figure will be discussed further below; the parts labeled (b) 
   and the notation are specific to the 
   HBAR radiation analog system of the next section.)

 \subsubsection{Analog systems}

For the BEC analog system, one typically has a dilute system 
forming a gas of $N$ ideal bosons, which are realized experimentally with atoms; and the
$N$ particles are 
confined in three-dimensional harmonic trap~\cite{BEC-laser-analog_DeGiorgio-Scully-1970}.
In addition,
these are in equilibrium at temperature $T \ll T_{c}$ (with $T_{c}$ being the
BEC transition temperature).
Under these conditions, the master equation for the diagonal elements,
with $n \equiv n_{0}$ being the number of bosons in the ground state,
 has the same form as 
Eq.~(\ref{eq:master_equation_Scully-Lamb_diagonal}), with the following analog 
replacements~\cite{BEC-laser-analog_Scully-1999}:
$\alpha \rightarrow \kappa (N+1)$,
$\beta \rightarrow \kappa$,
and $\alpha \rightarrow \kappa N (T/T_{c})^3$
(and $\kappa$ a rate constant); and $R_{e}$ and $R_{a}$ represent cooling and heating rates.
This master equation has been generalized for arbitrary temperatures $T$ in Ref.~\cite{BEC-laser-analog_Scully-etal-2000} (see the review and additional results in Ref.~\cite{BEC-laser-analog_Scully-etal-2022}).

    The laser in cavity, and the analog systems described by Eq.~(\ref{eq:master_equation_Scully-Lamb-gen_diagonal})
    require the use of a density matrix because part of the system 
    has an element of randomness that is treated as a reservoir.
    In the laser, the model involves excited atoms emit photons in a cavity at a given mode; 
a density matrix is needed for the electromagnetic laser field
as the atoms are injected into the cavity, and their degrees of freedom are averaged out.
This averaging process leading to a reduced density matrix is described in the next section for the analog system
of atoms falling into the black hole: HBAR radiation~\cite{Scully_2018_HBAR}, 
where the resulting master equation is described by a particular important case of  
Eq.~(\ref{eq:master_equation_Scully-Lamb-gen_diagonal}).

      \subsection{Reduced density matrix for HBAR field}
\label{subsec:density-matrix_HBAR}

We are now ready to review the results on the analog density matrix 
of the HBAR radiation emitted by a cloud of atoms falling into
a black hole~\cite{Scully_2018_HBAR}. 
A complete derivation of these 
results can be seen in Ref~\cite{acceler-rad-Qopt-1}.
In short, the properties of the density matrix derived in this way are essential to the analysis of the state and thermal properties of the HBAR radiation field.

 \subsubsection{Composite system and reduction procedure}
 
In what follows, we will refer to the field as the photon system (labeled with ${\mathcal P}$),
where the choice of a scalar field yields a simple model of emission of scalar photons. 
This is the result of its interaction 
 with an atom (labeled with ${\mathcal A}$).
Then, the quantum master equation has indeed the form outlined in Eq.~(\ref{eq:master_equation_Scully-Lamb-gen_diagonal})
and applies to the reduced density matrix 
    ($\rs{\mathcal P}$) of the field, due to the random injection of atoms.
This density matrix can be obtained by the general reduction 
procedure: via partial tracing (over the atomic degrees of freedom) from the density matrix of the composite 
system,
 \begin{equation}
 \rs{\mathcal P}  =  \mathrm{Tr_{\mathcal A}} \, \left(  \rs{{\mathcal P}{\mathcal A}} \right)
 \; .
 \end{equation}
The time evolution of the combined atom-field system is governed 
by the von Neumann equation for the density matrix $\rs{{\mathcal P}{\mathcal A}}$.
As in Eq.~(\ref{eq:QLE-perturbation-order2}),
 in second-order perturbation theory within the interaction picture, 
\begin{equation}
\!  \! \!  \! \!  \! 
\rs{{\mathcal P}{\mathcal A}}(\tau) 
=  \rs{{\mathcal P}{\mathcal A}}(\tau_0) 
- \frac{ i }{ \hbar } 
 \! \int_{\tau_0}^{\tau} \! \! d\tau' [V_I(\tau'),\rs{{\mathcal P}{\mathcal A}}(\tau_0)] 
+   \left( - \frac{ i }{ \hbar } \right)^2  \! 
 \! \int_{\tau_0}^{\tau} \! \! \!  d\tau'  \! \!  
 \int_{\tau_0}^{\tau'} \! \! \! \!  d\tau'' [V_I(\tau'),[V_I(\tau''),\rs{{\mathcal P}{\mathcal A}}(\tau_0)]] 
\label{eq:density-matrix_evolution}
\; ,
\end{equation}
where $V_{I} = H_{{\rm int},I}$ is the interaction potential or Hamiltonian in the interaction picture.
For the initial state of the combined system, one can consider 
the tensor product $\rs{{\mathcal P}{\mathcal A}}(\tau_0) = \rs{\mathcal P}(\tau_0) \otimes \rs{\mathcal A} (\tau_0)$, 
where the atoms and field are initially uncorrelated. 
The required time parameter $\tau$ in Eq.~(\ref{eq:density-matrix_evolution}) 
is the proper time of free-fall trajectories. 

 \subsubsection{Experimental procedure and injection averaging}
 
The standard experimental setup in quantum optics involves an optical cavity.
A cloud of atoms acts as a reservoir within the cavity. 
The cavity can be defined with appropriate mirrors in a given spacetime geometry.
This setup, with the averaging procedure, as depicted in Figs.~\ref{fig:HBAR-setup} and \ref{fig:HBAR-setup-with-mirror},
is the black-hole analog of a standard quantum engineering approach that
is useful for experimental studies of quantum 
information~\cite{Qreservoir_Pielawa-et-al_2007, Qreservoir_Pielawa-et-al_2010}.
In such studies of ``reservoir computing''~\cite{Qreservoir_Zhu-et-al_2025},
 atomic quantum reservoirs are physical systems used to process information in a 
similar way to how neural networks work---they are often composed of atoms in a cavity, 
in designs that have also been modeled with the QRM of Sec.~\ref{sec:QOptics-interactions_Hamiltonian}.
For the HBAR problem, the atoms are injected in their ground state, 
the initial atomic density matrix (at time $\tau_0$) is $ \rs{\mathcal A} = \ket{b}\bra{b}$. 
The evolution of the radiation field is averaged over a distribution of injection times. 
In this model, which generalizes the original Scully-Lamb model of a laboratory laser cavity,
 a Markovian property is assumed.
As an operational procedure,
 the effective evolution of the field is obtained by:
(i)
tracing over the atomic degrees of freedom (with $\mathrm{Tr_{\mathcal A}}$),
leading to the reduced field density matrix $\rs{\mathcal P}  =  \mathrm{Tr_{\mathcal A}} \, \left(  \rs{{\mathcal P}{\mathcal A}} \right)$; and (ii) 
implementing an averaging procedure with 
 a time scale larger than the reservoir's memory time
(time scale for a representative distribution of injection times). 
As a result, a master equation of the 
form~(\ref{eq:master_equation_Scully-Lamb-gen_diagonal})
is obtained: this is the equation for an approximate coarse-grained reduced density matrix $\rs{\mathcal P}$.  

The cavity analog for a spacetime geometry corresponds to a spatial region bounded by constant values 
of coordinates adapted to stationary configurations. (As mentioned in the introductory part of 
Sec.~\ref{sec:QOptics-interactions}, the ``optical cavity model'' is a useful device to simulate the Boulware vacuum---see
Fig.~\ref{fig:HBAR-setup-with-mirror}---but caution should be exercised in its interpretation.)
For example, for the Schwarzschild geometry, the spatial coordinates are adapted to a set
of static observers, with the coordinate time $t$ being a convenient parameter to label the ``cavity time''
(experienced at given, fixed locations of an optical cavity)---the actual cavity time, which is a static proper time,  
experiences gravitational time dilation with a factor $\sqrt{- g_{tt} (r)} = \sqrt{f(r)}$.
If an atom is injected at an initial coordinate time $t_{0} =t_{i,a}$,
the subsequent geodesic motion is given by the equations in Sec.~\ref{sec:spacetime_CQM}. 
As $\tau = \tau(t)$,
the time parameter $\tau$ can be replaced by $t$, and  
the corresponding  ``microscopic'' change in the field density matrix
is
$\delta  \rs{\mathcal P}_{a} \equiv \delta \rs{\mathcal P} (t; t_{i,a}) 
  =  \rs{\mathcal P}(t) - \rs{\mathcal P}(t_{i,a})$. 
  Thus,
  the corresponding course-grained  or ``macroscopic'' change is 
\begin{equation}
\dot{\rho}^{\mathcal P} \equiv \frac{\Delta \rs{\mathcal P}}{\Delta t} = \mathfrak{r}  \, \frac{1}{\Delta N} 
\, \sum_{a} \delta \rs{\mathcal P} (t; t_{i,a}) 
=  \mathfrak{r} \,  \overline{\delta \rs{\mathcal P} }
\; ,
\label{eq:dot-rho_coarse-grained}
\end{equation}
where 
the overdot notation gives the rate of change~\cite{scullybook} with respect to the cavity 
time $t$,
$\mathfrak{r}= \Delta N/\Delta t $ is the injection rate,
 and $\overline{\delta \rs{\mathcal P}}$ is the average microscopic change 
 with respect to particle injection. 
The statistical average is defined starting
with a number of atoms 
$\Delta N $ during a time interval $T$, in the form
$(1/\Delta N ) \sum_{a} X^{\mathcal P} =  \int d t_{i,a} f (t_{i,a}) X^{\mathcal P}$
(when applied to a field quantity $X^{\mathcal P}$), where $f (\xi)$ is the probability distribution of the random variable. Then,
\begin{equation}
\overline{\delta \rs{\mathcal P}}
=
 \int d t_{i} \, f ( t_{i}) 
\,
  \delta \rs{\mathcal P} (t; t_{i}) 
\; .
\label{eq:injection-average}
\end{equation}
For a completely random distribution of injection times, a uniform distribution $ f ( t_{i}) = 1/T$ can be chosen.
 The resulting coarse-grained field density matrix
 satisfies the multimode form of the generalized Scully-Lamb master equation~\cite{acceler-rad-Qopt-1} 
\begin{equation}
\begin{aligned}
 \dot{\rho}_{\rm diag}(  \boldsymbol{ \left\{  \right. } n  \boldsymbol{\left. \right\}  } )  
 =
    - 
     \sum_{j}
     &
     \left\{
     R_{{\rm e},\, j}  \big[(n_j+1) \,
   {\rho}_{\rm diag} (  \boldsymbol{ \left\{  \right. } n  \boldsymbol{\left. \right\}  } )
      - n_j \,
   {\rho}_{\rm diag} (  \boldsymbol{ \left\{  \right. } n  \boldsymbol{\left. \right\}  }_{n_j -1} )
            \big] \right.
          \\
           &
            \left.
            +  
           R_{{\rm a},\, j} \big[ n_j \,
    {\rho}_{\rm diag} (  \boldsymbol{ \left\{  \right. } n  \boldsymbol{\left. \right\}  } )
      - (n_j + 1)  \,
     {\rho}_{\rm diag} (  \boldsymbol{ \left\{  \right. } n  \boldsymbol{\left. \right\}  }_{n_j +1} )
                           \big] \right\}
    \label{eq:master_equation_final_multimode}
    \; ,
\end{aligned}
\end{equation}
which is valid under the assumption 
 that only the diagonal elements are relevant;
this is the case for {\em random injection times\/}.
In Eq.~(\ref{eq:master_equation_final_multimode}),
the emission and absorption rate coefficients are
 $R_{e, j } = \mathfrak{r} \, P_{e, j }$
and
$ R_{a, j}= \mathfrak{r} \, P_{a, j} $,
with $\mathfrak{r}$ being the atom injection rate; and
the index $j$ is shorthand for a given mode ${\boldsymbol{s}}_{j}  $,
with the single-mode quantum numbers $ {\boldsymbol{s}} $ chosen in an ordered sequence.
In addition,
the diagonal elements of the density matrix are denoted by
$  
  {\rho}_{\rm diag} (  \boldsymbol{ \left\{  \right. } n  \boldsymbol{\left. \right\}  } )
   \equiv
  {\rho}_{  n_1,n_2, \ldots  ;   n_1,n_2, \ldots  }
$,
where we further developed the notation of Sec.~\ref{sec:scalar-field_curved-ST}:
$ \boldsymbol{ \left\{  \right. } n  \boldsymbol{\left. \right\}}    \equiv
\boldsymbol{ \left\{  \right.  } n_{1}, n_{2}, \ldots , n_{j } , \ldots \boldsymbol{\left. \right\}  }$ 
 for the occupation number representation, along with
$
  \boldsymbol{ \left\{  \right. } n  \boldsymbol{\left. \right\}}_{n_j + q}  
  \equiv
 \left\{  \right.  
 n_{1}, n_{2}, \ldots , n_{j }+ q , \ldots \boldsymbol{\left. \right\}  }
 $ (with $q$ an integer-number shift).
   In Sec.~\ref{sec:conformal_steady_state},
    we will use Eq.~(\ref{eq:master_equation_final_multimode}) to establish the thermal nature of 
    the HBAR radiation field.

\section{Quantum aspects of spacetime: Black holes, horizons, and conformal quantum mechanics (CQM)}
\label{sec:spacetime_CQM}

This is a mainly self-contained section where 
 we address the essential geometric features of the spacetime background.
 It serves as an introduction to spacetime properties related to black holes, as well as to the near-horizon approximation and conformal quantum mechanics.
 In the main text of this article, we will consider a generalization of Schwarzschild spacetime geometries;
further generalization to include black hole rotation is discussed in Appendix~\ref{app:spacetime}.
 
However, one should also keep in mind that this background is of direct relevance for the 
computation of the probabilities~(\ref{eq:P_ex_explicit}) and (\ref{eq:P_ab_explicit}) for HBAR radiation, 
which requires geometry-specific field modes and spacetime worldlines.
These probabilities will be computed in the next section, where 
we will derive the emission of radiation fields due to the fall of particles through the event
horizon, as well the associated black hole thermodynamics.

For the remainder of this article, 
with the exception of the first paragraph of the next Sec.~\ref{sec:BH-geom_horizons} or
unless stated otherwise, we will use
Planck natural units ($\hbar =1$, $c=1$, $k_{B} =1$, $G=1$).

\subsection{Black hole geometry and horizons}
\label{sec:BH-geom_horizons}

We first introduce a general class of static spacetime geometries with black holes, defined via the metric
\begin{equation}
ds^{2}=- f (r) \,  dt^{2}+\left[ f(r) \right]^{-1} \, dr^{2}+ r^{2} \, d \Omega^{2}_{(D-2)}
\; 
\label{eq:RN_metric}
\end{equation}
in $D$ spacetime dimensions.
Inspection of Eq.~(\ref{eq:RN_metric}) 
reveals a spherically symmetric metric written in coordinates $(t,r, \Omega)$, where the 
time and radial metric elements are $-g_{tt} = g^{rr} = f(r)$, and $\Omega$ describes 
the usual angular spherical coordinates of the unit $(D-2)$-sphere
 with metric $d \Omega^{2}_{(D-2)}$.
 The prime example is the ordinary four-dimensional ($D=4$) Schwarzschild 
 metric~\cite{GR_Carroll-2003, MTW-gravitation, GR_Wald-1984}, which is due to a black hole
 of mass $M$,
for which $f(r) = 1 -2GM/r$, where Newton's gravitational constant $G$ is restored. 
 By extension, a metric of the form~(\ref{eq:RN_metric}) describing a spacetime 
 gravitational background with an event horizon, may be 
called a ``generalized Schwarzschild geometry.''
One such generalization is the Reissner-Nordstr\"{o}m (RN)
black hole~\cite{GR_Carroll-2003, MTW-gravitation, GR_Wald-1984}, with 
 mass $M$ and electric charge $Q$, for which
 $f(r) = 1 -2GM/r + K_{e} GQ^2/r^2$ (withn $K_{e}$ being Coulomb's constant).
More general spacetime realizations
include extensions to any number of dimensions $D \geq 4$~\cite{mye:86}.
Specifically,
for the particular case of an RN black hole
of mass $M$ and electric charge $Q$, 
the factor $f(r)$ in Eq.~(\ref{eq:RN_metric})
becomes
$f(r) =
1
-
\left(  R_{M}/{r} \right)^{D-3}
 +
\left(R_{Q}/{r} \right)^{2(D-3)} 
$
with the characteristic length scales $R_{M}$ and $R_{Q}$ 
given similarly in terms of $M$ and $Q^2$.
These extensions also allow for 
combinations of Schwarzschild, 
Reissner-Nordstr{\"o}m, 
and de Sitter geometries
with a cosmological constant $\Lambda$,
and black hole solutions with additional quantum charges~\cite{ortin}; for example, 
the inclusion of a cosmological constant 
is achieved with an extra term
$- 2 \Lambda r^{2}/[(D-1)(D-2)]$ in $f(r)$.
 Further extensions with angular momentum, for rotating black holes, can be treated similarly,
as addressed in Appendix~\ref{app:spacetime}.

The generalized Schwarzschild geometries of Eq.~(\ref{eq:RN_metric}) 
are static and spherically symmetric, i.e., the metric
has invariance under time translations and spatial rotations. 
Metric invariance is an example of an isometry:
a transformation that preserves distances between points in a metric space.
A Killing vector field is an infinitesimal generator of an isometry;  
distances between points on the manifold remain unchanged when following the flow of the Killing vector.
For the generalized Schwarzschild metric, time and rotational invariance are described by the 
corresponding Killing vectors~\cite{GR_Carroll-2003, MTW-gravitation, GR_Wald-1984}
\begin{equation}
\boldsymbol{\xi}_{(t)} = \partial_t 
\; \; , \; \; \; \; \;
\boldsymbol{\xi}_{(\phi)} = \partial_\phi
\; ,
\label{eq:Killing-vectors}
\end{equation}
where $\phi$ is a rotational angle around a generic plane. 
For a particular coordinate choice in Eq.~(\ref{eq:RN_metric}), 
spherical symmetry refers to the symmetries of the sphere with metric $d \Omega^{2}_{(D-2)}$;
these involve a complete set of angular Killing vectors,
e.g., with respect to the standard azimuthal angle $\phi$ and 2 additional orientations in 
4-dimensional spacetime~\cite{GR_Carroll-2003, MTW-gravitation, GR_Wald-1984}).
Generally, one can use the norms and inner products of Killing vectors to set up
geometrical interpretations in coordinate-free forms. In particular, for questions
related to time evolution, the product
 $g_{{t}{t}}  = \boldsymbol{\xi}_{({t})} \cdot   \boldsymbol{\xi}_{({t})}$ plays an important role
 because the flow of the time-translation Killing vector $  \boldsymbol{\xi}_{({t})} $
generates time evolution. (For example, 
in a Schwarzschild background, $  \boldsymbol{\xi}_{({t})} $ is proportional to the spacetime velocity
of a static observer, i.e., one that remains at a fixed position with constant spatial coordinates.)

For our discussion of acceleration radiation, 
the most relevant features of the geometries defined by Eq.~(\ref{eq:RN_metric})
are related to the existence of an event horizon ${\mathcal H}$.
This is a hypersurface that is the interior boundary 
of the region in spacetime from which a light ray can travel to infinity.
For a static spacetime [as described by Eq.~(\ref{eq:RN_metric})], 
it can be identified via the polynomial roots of the metric component
 \begin{equation}
 g_{{t}{t}}  = \boldsymbol{\xi}_{({t})} \cdot   \boldsymbol{\xi}_{({t})}
 = f(r) = 0
\; .
\label{eq:g_tt-null}
 \end{equation}
 Thus, at every point on the hypersurface ${\mathcal H}$,
  the norm of the time-translation Killing vector
  $  \boldsymbol{\xi}_{({t})} $ becomes null, defining a null tangent direction,
    while the Killing vectors associated with the other (angular) directions remain spacelike.
  By definition, this makes ${\mathcal H}$ a null hypersurface. 
Now, the light cones built at every point on a null hypersuface are tangent to it and 
 confined to one side: thus, the geodesics cross the surface in only one direction (inward).
 Therefore, the null condition~(\ref{eq:g_tt-null}) signals the presence of an event horizon as
a boundary where all timelike or null geodesics are ingoing and cannot go back to infinity.
The event horizon itself is generated by light rays that cannot escape the black hole.
Moreover, this a general result for stationary spacetimes that
can be further confirmed by a detailed analysis of the geodesics and spacetime causal 
structure~\cite{GR_Carroll-2003, MTW-gravitation, GR_Wald-1984}.
Stationary geometries that are not static: they are typically
associated with a black hole with angular momentum or rotation and exhibit
additional subtleties; see Appendix~\ref{app:spacetime}.

Given the existence of an event horizon for the generalized Schwarzschild geometries of Eq.~(\ref{eq:RN_metric}), 
a near-horizon analysis can be carried out,
 centered on the functional dependence of the external
nongravitational fields in the neighborhood of the outer event horizon ($r=r_{+}$).
As we show in the next two sections, the near-horizon behavior gives crucial insights into the emergence 
of conformal symmetries and black-hole thermodynamics.

Two important quantities related to the event horizon are closely linked with the 
black hole's thermodynamic and quantum features: the surface gravity and the black hole horizon area.
First, the surface gravity is defined from the timelike Killing vector 
$\boldsymbol{\xi} \equiv \boldsymbol{\xi}_{(t)} $ in Eq.~(\ref{eq:Killing-vectors})
as the horizon value $\kappa$ from
\begin{equation}
\kappa^2 = 
\left.
-\frac{1}{2} \left( \nabla_{\mu} {\xi}_{\nu} \right) 
\left(  \nabla^{\mu} {\xi}^{\nu} \right)
\right|_{r=r_{+}}
\; ,
\label{eq:surface-gravity}
\end{equation}
for a normalized vector $\boldsymbol{\xi} $ 
that conforms to an operationally defined notion of gravitational 
acceleration~\cite{GR_Carroll-2003, GR_Wald-1984, frolov}.
A normalization is required due to the multiplicative ambiguity in the definition of the Killing vector.
In asymptotically flat and static spacetimes, 
  including the ones described the generalized Schwarzschild metrics~(\ref{eq:RN_metric}), 
this can be enforced with a unit normalization via the inner product,
\begin{equation}
\bigl. \boldsymbol{\xi} \cdot \boldsymbol{\xi} \,
 \bigr|_{r \rightarrow \infty} = -1
 \; .
\label{eq:Killing_normalization}
\end{equation}
In such spacetimes, the acceleration $\kappa$ defined via Eqs.~(\ref{eq:surface-gravity}) 
and (\ref{eq:Killing_normalization}) can be shown to be constant over the entire event horizon, and
agrees with an operational value needed to keep a test particle in the near-horizon region
as measured from infinity.
Appendix~\ref{app:spacetime} shows how to generalize this definition centered on Eq.~(\ref{eq:surface-gravity}).
For the generalized Schwarzschild black holes from Eq.~(\ref{eq:RN_metric})
(with $r =r_{+}$), it takes the form 
\begin{equation}
\kappa= \frac{ f'_{+} }{ 2  }
\; ,
\label{eq:surface-gravity_gen-Schwarzschild}
\end{equation}
where $f'_{+} = f'(r_{+}) \neq 0$ for nonextremal geometries.
This geometrical quantity~(\ref{eq:surface-gravity})--(\ref{eq:surface-gravity_gen-Schwarzschild}) 
 is proportional to the Hawking temperature, as given in Eq.~(\ref{eq:Hawking-temperature}).
Second, the horizon area, generally defined via integration with the metric of the angular coordinates
(on the unit sphere), takes the $D$-dimensional form~\cite{mye:86,nhcamblong}
\begin{equation}
A
= 
\Omega_{(D-2)} 
\,
r_{+}^{D-2}
\; ,
\label{eq:EH-area}
\end{equation}
in terms of  the solid angle $\Omega_{(D-2)}$
in $D$ dimensions; 
in 4D, this reduces to the familiar area $A = 4 \pi r_{+}^2$.
This geometrical area~(\ref{eq:EH-area}) 
is, in all cases, proportional to the Bekenstein-Hawking entropy~(\ref{eq:BH-entropy}).
In particular,
this implies that the area changes are governed by
\begin{equation}
  \delta A  = 8 \pi   \, \frac{ \delta {M} }{ \kappa}
  \; 
    \label{eq:BH-area-changes-geom}
  \end{equation}
in the pure Schwarzschild case (mass $M$ only), and similar expressions where $\delta M$ is replaced by a subtracted energy associated with electric fields or rotation 
in the RN and Kerr geometries respectively,
suggesting a thermodynamic analogy (see Appendix~\ref{app:spacetime}).
 In Sec.~\ref{sec:HBAR-thermo},
  we show that these are not just analogies but represent a genuine quantum thermodynamic framework 
  both for the intrinsic properties of black holes and the related properties of horizon-brightened acceleration. 

\subsection{Scalar field: Quantization and near-horizon analysis}
\label{sec:scalar-field-nh}

The Euler-Lagrange equation for 
a scalar field in a generic gravitational background $g_{\mu\nu}$ 
 is given by the classical curved-spacetime Klein-Gordon equation~(\ref{eq:Klein_Gordon_basic}).
Thus, this determines the functional form of the field modes $\phi_{\mathbf{s} }$ in the field expansion of
Eq.~(\ref{eq:field_expansion}), as needed 
for its canonical quantization.
 For the class of metrics~(\ref{eq:RN_metric}), 
 with the set of quantum numbers
 $\mathbf{s}
  = (nl \mathbf{m})$, the modes take the separable form
\begin{equation}
\phi_{  \mathbf{s} } (t,r, \Omega) 
=
R_{ nl}(r) Y_{l \mathbf{m} }(\Omega)e^{-i\omega_{nl}t}
\; ,
\label{eq:separation_of_variables}
\end{equation}
where  $Y_{l \mathbf{m}}(\Omega)$ are ultra-spherical harmonics,
the solutions to the angular part of the Laplacian.
Then, the Klein-Gordon equation can be further reduced to its normal form 
with the Liouville transformation~\cite{forsyth:Liouville}
 $R(r) = \chi (r) u(r)$, where $\chi (r)= [f(r)]^{-1/2} \,r^{-(D-2)/2}$,
such that its radial part becomes
\begin{equation}
u_{nl}''(r) +I_{(D)} (r; \omega_{nl}, \alpha_{l,D}  )\,u_{nl}(r) =0\;  ,
\label{eq:Klein_Gordon_normal_radial}
\end{equation}
where $I_{(D)}$ is an effective potential; e.g., 
an explicit form is given in
Refs.~\cite{nhcamblong,acceler-rad-Schwarzschild}.

In principle, the physics of quantum fields in the gravitational background 
of a black hole can be studied directly from Eq.~(\ref{eq:Klein_Gordon_normal_radial}).
But the intrinsic properties generated by the black hole are essentially driven by 
the presence of the event horizon. 
Thus, it is plausible that some of the most essential features of the relevant physics,
including quantum properties, are captured by the near-horizon behavior.
The possible relevance of the near-horizon physics for black-hole thermodynamics 
was highlighted in several studies starting in the late 1990s~\cite{strominger_nh-BHS, carlip1, carlip2, solodukhin:99, gov:BH_states, gupta:BH1, gupta:BH2, calogero_black_holes}.
These efforts uncovered a form of scale symmetry associated with conformal symmetries,
which appeared to be an important ingredient related to the thermal and quantum properties. More direct evidence for the role played by this symmetry
in black-hole thermodynamics was shown in Refs.~\cite{nhcamblong,nhcamblong-sc}, 
and the same approach was later used to display the related properties of 
 horizon-brightened acceleration radiation~\cite{acceler-rad-Schwarzschild,acceler-rad-Kerr,acceler-rad-Qopt-1,acceler-rad-Qopt-2}. 
 This is the near-horizon CQM approach that we will highlight for the derivation of thermodynamic behavior in most of the remainder of this article. 
 With this purpose in mind, we first begin by defining the near-horizon approximation, 
 with details and associated symmetry to be discussed in the next section.

The near-horizon scheme involves an approximation 
near the outer horizon ${\mathcal H}$, $ r \sim r_{+}$, 
with $r=r_{+}$ being the largest root of the scale-factor equation $f(r)=0$.
  Thus, with the shifted variable $x= r -r_{+}$,
  the Taylor series for the scale factor $f(r)$ 
  starts at first or higher orders. 
 The notation $\stackrel{(\mathcal H)}{\sim}$ will be used to represent this hierarchical expansion.
 Considering the physically relevant {\em nonextremal\/} metrics, which satisfy the condition 
  $f'_{+} \equiv f'(r_{+}) \neq 0$, 
  the function $f(r)$ and its derivatives to second order are given by 
\begin{equation}
f(r)  \stackrel{(\mathcal H)}{\sim}  f'_{+}  \, x \left[ 1 + {O}(x) \right]  \, , \; \; \; 
f'(r)  \stackrel{(\mathcal H)}{\sim}  f'_{+} \left[ 1 + {O}(x) \right]   \, , \; \; \; 
f''(r)  \stackrel{(\mathcal H)}{\sim}  f''_{+} \left[ 1 + {O}(x) \right] \, ,
\label{eq:nh-expansions}
\end{equation}
where $f''_{+} \equiv f''(r_{+}) $.
This near-horizon approximation can be applied to the original Klein-Gordon equation~(\ref{eq:Klein_Gordon_basic}),
leading to
\begin{equation}
\left[\frac{1}{x} \frac{d}{d  x} \left( x  \frac{d}{d x} \right) 
+ \left( \frac{ {\omega}}{f'_{+}} \right)^{2}
\frac{1}{x^2} \right] R(x)
\stackrel{(\mathcal H)}{\sim} 0 \; ,
\label{eq:Kerr_Klein_Gordon_conformal-R}
\end{equation}
followed by the corresponding Liouville transformation  $R(x) \propto x^{-1/2} u(x)$;
or directly to the reduced form of Eq.~(\ref{eq:Klein_Gordon_normal_radial}). Thus,
 the final result, up to leading-order with respect to $x$, 
 is a Schr\"{o}dinger-like equation
\begin{equation}
u''(x)+\frac{ \lambda_{} }{x^{2}}
\,\left[ 1 + {O}(x) \right]u (x)=0
\;  
\label{eq:Klein_Gordon_conformal}
\end{equation}  
[with the replacement of $u(r)$ by $u(x)$], where
 the dominant near-horizon
 physics is driven by the effective Hamiltonian
  \begin{equation}
  \mathscr{H} = {p}_{x}^{2}-  \frac{ \lambda }{ x^{2} }
  \; ,
  \label{eq:CQM-nh-Hamiltonian}
  \end{equation}
 with an inverse square potential of coupling
\begin{equation}
\lambda_{} = \frac{1}{4} + \Theta^{2}\, , \; \; \; \;
 \Theta= \frac{\omega}{  f'_{+} } \equiv  \frac{\omega}{  2 \kappa  }
 \; .
\label{eq:conformal_interaction}
\end{equation}  
In Eq.~(\ref{eq:conformal_interaction}),
$\kappa =  f'_{+}/2$  is the surface gravity of the black hole, 
from Eqs.~(\ref{eq:surface-gravity})--(\ref{eq:surface-gravity_gen-Schwarzschild}).

\begin{figure}[h]
    \centering
    \includegraphics[width=0.4\linewidth]{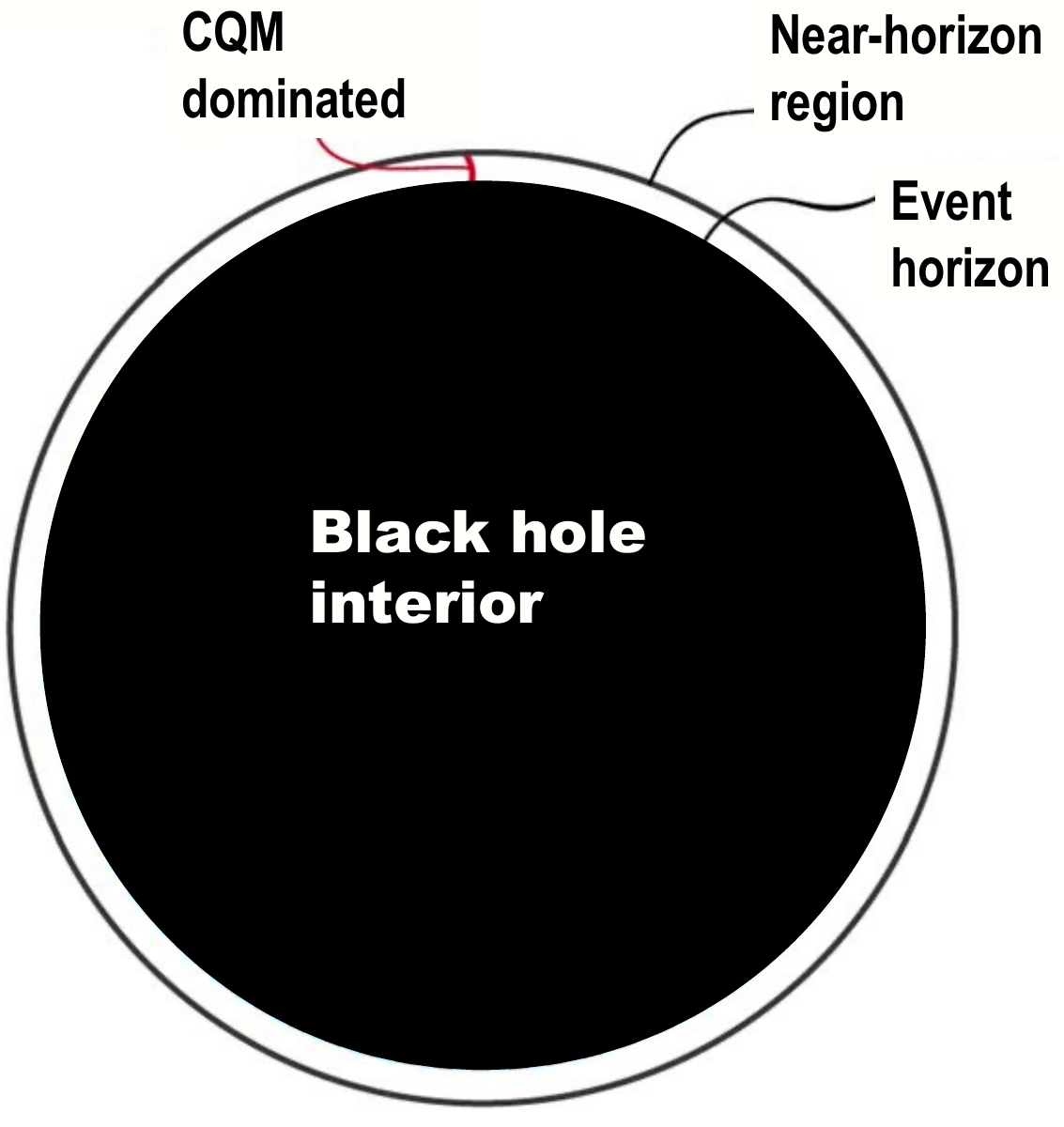}
    \caption{         
    The near-horizon approximation can reveal properties of the black hole near the event horizon,
    including the emergence of conformal quantum mechanics (CQM), and the
    interpretation of quantum fields as providing a thermal atmosphere.}
           \label{fig:BH-atmosphere}
            \vspace*{-0.15in}
       \hspace*{0in}
\end{figure}

Remarkably, even though 
Eqs.~(\ref{eq:Kerr_Klein_Gordon_conformal-R})--(\ref{eq:conformal_interaction}) involve the near-horizon approximation, this simplification does not limit their scope. 
Basically, the near-horizon regime emerges as an effective theory that captures the essence of black-hole thermodynamics 
and the HBAR physics of particles falling into a black hole.
In this view, specifically,
the one-dimensional effective Hamiltonian $\mathscr{H}$
associated with Eq.~(\ref{eq:conformal_interaction}) is a realization of
 the inverse square potential 
 of conformal quantum mechanics~\cite{Qanomaly-molecular-to-BH}.
 The conclusion is that the near-horizon physics exhibits an {\em asymptotic conformal symmetry\/}.
 The near-horizon expansion and emergence of conformal quantum mechanics are
  as depicted in Fig.~\ref{fig:BH-atmosphere}.

\subsection{Conformal symmetry: The unreasonable effectiveness of conformal quantum mechanics (CQM)}
\label{sec:CQM}

{\it Symmetries play a crucial role in most aspects of foundational quantum physics\/}---this has been another
fundamental recurrent theme throughout the first hundred years of quantum mechanics~\cite{Weinberg_QM}.

 A remarkable example that has attracted considerable attention in recent decades
is conformal symmetry~\cite{CFT_DiFrancesco}.
One particular form of this invariance is
the framework known as conformal quantum mechanics (CQM), 
as it was called in the comprehensive presentation of Ref.~\cite{dAFF}.

 \subsubsection{Conformal symmetry and CQM}
 
 \begin{figure}[!h]
	\centering
	\includegraphics[width=0.6\linewidth]{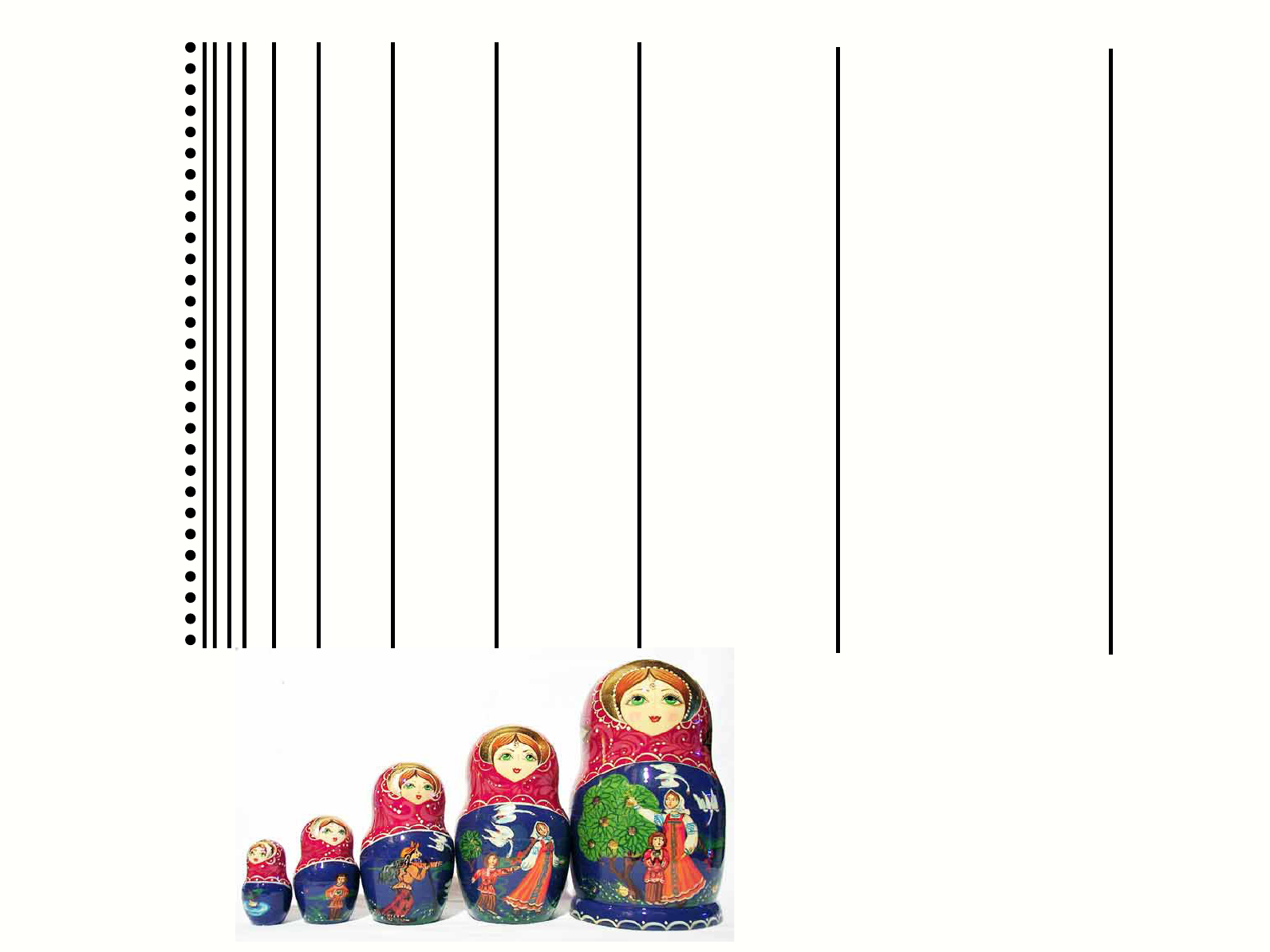}
	\vspace{-0.2in}
	\caption{The near-horizon conformal modes  $\phi (r,t)$ can be pictured with their wavefronts.
 The event horizon is represented by the dotted line. 
 A geometric sequence $x_{(n)} \propto \eta^{n}$ 
 with ratio $\eta = e^{-2\pi/\Theta}$, and $\Theta$ defined via Eq.~(\ref{eq:conformal_interaction}),
 shows the piling up of the modes with an accumulation line at the horizon.
 This manifests scale invariance under arbitrary magnifications, with a
 geometric scaling depicted via the Russian-doll analogy. 
In this graph, $\eta^{-1} = 1.4$.
For the HBAR analysis, the phase of the probability amplitudes,
as shown in Eqs.~(\ref{eq:P_ex_CQM})--(\ref{eq:near-horizon-pex}),
has a similar geometric scaling: $\eta = e^{-2\pi/\sigma} = e^{-\pi/\Theta}$, 
but the frequency scale is twice as big.   }
	\label{fig:Russian-doll-wavefronts}
	\hspace*{0in}
\end{figure}
 
The CQM framework is based on the inverse-square-potential Hamiltonian $\mathscr{H}$ of Eq.~(\ref{eq:CQM-nh-Hamiltonian}) and is manifestly scale invariant at the classical level as it is 
homogeneous 
of degree $-2$~\cite{Cam_DT1,Cam_DT2}.
In addition, this operator is part of an enlarged SO(2,1) symmetry group
which can be interpreted as describing a 
lower-dimensional conformal field theory~\cite{dAFF,jackiw1,jackiw2,jackiw3}.
The algebra of this SO(2,1) group consists of three operators:
$\mathscr{H} $, which is the generator of time displacements (with respect to 
a specified time $t$
conjugate to
$\mathscr{H} $),
 together with
the dilation operator
\begin{equation}
D= t H - \frac{ 
x p + p x
 }{4}
 \label{eq:dilation-operator}
\; 
\end{equation}
(where $p$ is the conjugate momentum),
which performs scale transformations, e.g., as illustrated in Fig.~\ref{fig:Russian-doll-wavefronts});
and the special conformal operator
\begin{equation}
K= t^{2} H -  
\frac{t \, (px + x p) }{2}
 +  \frac{m
x^{2} }{2}
\; .
\end{equation}
which is the generator of inverse time displacements.
These operators generate the 
SO(2,1) {\em conformal algebra\/}
\begin{equation}
[D,H]
= - i \hbar H
  \;  ,
\; \;
[K,H]
= - 2 i \hbar D
\;  ,
\; \;
[D, K]
=  i \hbar K
\;  .
\label{eq:naive_commutators}
\end{equation}
This symmetry algebra has been recently studied in its most general setting using path integral 
methods~\cite{Cam_CQM-PI}, extending earlier results on path integrals 
of the inverse square potential~\cite{Cam_CQM-GF-ISP, Cam_SQM-PI}.

\subsubsection{Physical realizations of CQM}
 
Paraphrasing Wigner~\cite{Wigner_math-phil}, 
the unreasonable effectiveness of CQM is manifested in its broad range of physical applications.
This versatility was first recognized in the pioneering work of Jackiw~\cite{jackiw1, jackiw2, jackiw3}, and led to the discovery of physical realizations
in molecular physics~\cite{Qanomaly-molecular,Qanomaly-molecular-EFT,Qanomaly-molecular-to-BH}, nanophysics~\cite{CQM-renormalization-PLA}, 
nuclear and particle physics~\cite{CQM-renormalization-PLA}, 
and the black-hole thermodynamics and HBAR properties discussed herein.
 In this sense, CQM can be regarded both as a theoretical tool for
 fundamental physics and as a practical tool for physical systems.
 
Many of these realizations involve the emergence of quantum symmetry breaking, i.e.,
a quantum anomaly~\cite{tre:85,don:92}. This is one of the three main classes of
symmetry breaking in physical systems, in addition to explicit and
spontaneous symmetry breaking. Specifically, an anomaly
is the breaking of a classical invariance upon
quantization. The relevance of this phenomenon has been
recognized in high-energy physics since the introduction of the
Adler-Bell-Jackiw anomaly~\cite{a-b-j1,a-b-j2,a-b-j3}.
The CQM anomaly, when its presence is allowed by the relevant physics, is based on the fact that 
the conformal theory is singular and ill-defined for a sufficiently {\em strong coupling\/}:
any bound
state energy could be rescaled to arbitrarily negative energies via the dilation operator $D$ of
Eq.~(\ref{eq:dilation-operator}).
Thus, the Hamiltonian is unbounded from below due to the scale symmetry.
 While this naively appears to be an intractable problem, 
an alternative viewpoint is available:
the singular behavior of the conformal interaction
reveals the existence of additional {\em ultraviolet physics\/}. This is similar to the divergences familiar 
from quantum field theory and leads to a low-energy version of quantum-mechanical symmetry breaking associated with 
the renormalized theory. Moreover, the anomalous regime of the theory has a well-defined extension of the
symmetry algebra~\cite{Qanomaly-molecular-to-BH, Qanomaly-delta-2D, Qanomaly-ISP-2D}.

Using an effective-field-theory approach, the renormalized theory
 typically generates a conformal tower of bound states~\cite{CQM-renormalization-PLA}
 $E_{ n }
=
E_{0}
\,
e^{  - 2 \pi n/ \Theta } $ ($n =0,1, \dots$), where $\Theta$ is the conformal parameter associated with the 
strength of the inverse square potential---this is defined in a manner similar
to Eq.~(\ref{eq:conformal_interaction}) for near-horizon black-hole physics.
Even though the symmetry
is broken, and an energy scale arises associated with a ground state, the spectrum still exhibits a
residual discrete version of the scale symmetry: $E_{n}$ is a geometric sequence with ratio $\eta = e^{-2\pi/\Theta}$.
Thus, the spectrum looks identical from any ``vantage point'' via proportional rescalings.
In energy space, this {\em geometric scaling\/} is identical to the Russian-doll symmetry shown 
in Fig.~\ref{fig:Russian-doll-wavefronts}
for the black-hole near-horizon physics of quantum field modes.
A limitation is typically imposed by additional physics, making the conformal tower be a ``window''
limited between two characteristic ultraviolet and infrared length scales $L_{\rm UV}$
and $L_{\rm IR}$, with a total number of states $N_{\rm conf} \sim (\Theta/\pi) \ln \left( L_{\rm IR}/L_{\rm UV} \right)$.
This general framework~\cite{CQM-renormalization-PLA} can also give additional predictions;
for example, giving corrections
associated with the existence of an infrared cutoff,
as shown in Ref.~\cite{Qanomaly-molecular-EFT}, 
where these techniques give a modified values for critical dipole moments 
in the formation of molecular dipole-bound anions~\cite{Qanomaly-molecular}.

 \subsubsection{CQM in black hole physics}

The relevance of conformal symmetry for the quantum and thermal 
 properties black holes has been a recurrent theme in 
fundamental physics, using a variety of approaches, e.g.,
 Refs.~\cite{strominger-vafa_BHS,strominger_nh-BHS},
\cite{carlip1,carlip2},
 \cite{ryu-takyanagi},
and \cite{strominger-KerrCFT}--\cite{strominger_hiddenKerr}. 
More specifically,
conformal quantum mechanics (CQM)~\cite{dAFF},
as a model based upon the Hamiltonian $\mathscr{H}$ 
 of Eq.~(\ref{eq:CQM-nh-Hamiltonian}),
{\it can be used as a probe of black hole thermodynamics\/} and related phenomena. 
Indeed,
the CQM approach to black holes provides a universal model of black hole entropy from near-horizon 
physics~\cite{nhcamblong, nhcamblong-sc, nhcamblong-heat-kernel, nhcamblong-conformal-tightness}, 
and can be used for the framework of HBAR 
radiation we are reviewing in this 
article~\cite{acceler-rad-Schwarzschild, acceler-rad-Kerr,acceler-rad-Qopt-1, acceler-rad-Qopt-2}.
In the current context,
this conformal symmetry is revealed by
the disappearance of all characteristic field scales; in particular, the 
field parameters $\mu_{\Phi}$  and $\xi$ in Eqs.~(\ref{eq:scalar_action}) and (\ref{eq:Klein_Gordon_basic})
play no role in the near-horizon physics. 
Moreover, this invariance, as represented in Fig.~\ref{fig:Russian-doll-wavefronts},
 can be pictured as arising from a gravitational blueshift 
that grows infinitely, with an accumulation point towards the event horizon; as a result,
any other physical scales are asymptotically erased~\cite{parkerprl}.

 For near-horizon CQM, the leading form of the field modes 
 can be obtained from a pair of independent solutions of the CQM equation (\ref{eq:Klein_Gordon_conformal}), 
\begin{equation}
    u_{\pm}(x) = x^{\frac{1}{2}\pm\sqrt{\frac{1}{4}-\lambda}} = \sqrt{x} \, x^{\pm i\Theta} 
    \label{eq:u-of-x-nh}
\end{equation}
where $\Theta = {\omega}/{f_+'}$ as defined in Eq.~(\ref{eq:conformal_interaction}).
These are outgoing/ingoing CQM modes that are 
normalized as asymptotically exact WKB local waves~\cite{nhcamblong-sc}
and display a logarithmic-phase singular behavior associated with scale invariance,
as $x^{i\Theta} = e^{i \Theta \ln x}$.
The logarithm $\ln x$ itself corresponds to the
familiar tortoise coordinate of generalized Schwarzschild 
geometries~\cite{GR_Carroll-2003, MTW-gravitation}; 
but writing it in the form of the solution~(\ref{eq:u-of-x-nh}) to the
corresponding near-horizon, Schr\"{o}dinger-like 
Eq.~(\ref{eq:Klein_Gordon_conformal}),
makes the CQM scale invariance more manifest,
as displayed in Fig.~\ref{fig:Russian-doll-wavefronts}.
In addition, when their time dependence is made explicit, 
these solutions, in the form of Eq.~(\ref{eq:separation_of_variables}), 
 give the outgoing and ingoing CQM modes
\begin{equation}
    \phi_{\boldsymbol{s}} (r,\Omega, t) \stackrel{(\mathcal H)}{\sim}
     \Phi^{\pm {\rm \scriptscriptstyle (CQM)}}_{\boldsymbol{s}} 
     \stackrel{(\mathcal H)}{\propto}
     x^{\pm i\Theta} 
     Y_{l \mathbf{m}}(\Omega)
      e^{-i\omega t} 
     \label{eq:CQM_modes}
     \; ,
\end{equation}
where $ \stackrel{(\mathcal H)}{\propto}$ 
denotes the hierarchical near-horizon expansion.

 
A closely related conformal ingredient revealed by the gravitational background consists of 
 the near-horizon geodesic equations~\cite{acceler-rad-Schwarzschild}; 
 see generalizations in Appendix~\ref{app:spacetime}.
The symmetries generated by the Killing vectors~(\ref{eq:Killing-vectors})
lead to geodesic planar orbits with: (i)
 spacetime velocity $ {\bf u}$, which is normalized (as all spacetime trajectories) with
   \begin{equation}
{\bf u}  \cdot {\bf u} = -1
\; ;
\label{eq:normalized-velocity}
\end{equation}
  and (ii)
 conserved energy and angular momentum components (per unit mass)
  \begin{equation}
e = - \boldsymbol{\xi}_{(t)}  \cdot {\bf u} = f (r) \frac{dt}{d\tau}
\; \; \;  , \; \; \; 
 \ell = \boldsymbol{\xi}_{(\phi)}  \cdot {\bf u} = r^2\frac{d\phi}{d\tau} \; ,
\label{eq:conserved-quantities} 
\end{equation}
choosing the azimuthal angle $\phi$ of the orbital plane. Use of Eq.~(\ref{eq:normalized-velocity}) 
amounts to the assignment of
an invariant particle mass $\mu$ in addition to the conserved quantities of Eq.~(\ref{eq:conserved-quantities}).

Straightforward application of the expressions in Eqs.~(\ref{eq:normalized-velocity}) and (\ref{eq:conserved-quantities})
gives an integrated form of the geodesic equations,
leading directly to the near-horizon geodesics 
as functional relationships $\tau=\tau(x)$ and $t=t(x)$, with the explicit dependence
\begin{align}
    \tau & \stackrel{(\mathcal H)}{\sim}  - k x +\mathrm{ const.} +  {O}(x^2)\;,
    \label{eq:tau_in_x} 
    \\
    t &  \stackrel{(\mathcal H)}{\sim}  -\frac{1}{2 \kappa}\ln x - C \, x
    + \mathrm{ const.} + {O}(x^2)
    \; . 
    \label{eq:t_in_x} 
\end{align}
As displayed in Eqs.~(\ref{eq:tau_in_x}), and (\ref{eq:t_in_x}),
 there are two constants,
  $k=1/e$ and $C$, that govern the linear terms in $x$; while 
  $k$ only depends on the specific energy $e$, the constant
  $C$ depends on both $e$ and the specific angular momentum $\ell$ of the atom, as well as the black hole parameters
 (see details in Ref.~\cite{acceler-rad-Schwarzschild}). 
However, these constants, as shown in Sec.~\ref{sec:HBAR}, do not play a direct role in the 
dominant part of the HBAR acceleration radiation formula; in effect,
unlike the governing scale symmetry of CQM,
the metric spacetime symmetries only act to simplify the geodesic initial value problem but 
do not drive the underlying physics of HBAR.
The final results of Eqs.~(\ref{eq:t_in_x}) 
are the geodesic equivalent of the hierarchical near-horizon expansion of the metric 
shown in Eq.~(\ref{eq:nh-expansions}).
The logarithmic term is obviously the dominant near-horizon part as $x \rightarrow 0$, revealing the existence
of scale invariance, which corresponds to
the familiar gravitational frequency shift~\cite{parkerprl}
and is a direct manifestation of the same conformal invariance that yields the logarithmic phase of the field modes.

This concludes the basic near-horizon analysis, which shows that both the field modes and the geodesics are controlled
by the scale-invariant features of CQM.

\section{Quantum and thermal nature of horizon-brightened acceleration radiation (HBAR)}
\label{sec:HBAR} 

With the results of the analysis of the effects of the gravitational field from the previous section, 
the HBAR radiation can be thoroughly computed with the aid of
the transition probabilities~(\ref{eq:P_ex_explicit}) and (\ref{eq:P_ab_explicit}).
 These results apply to the
 the HBAR thought experiment of 
Fig.~\ref{fig:HBAR-setup},
 where atoms are freely falling into the black hole.
With these probabilities, we will derive the emission of radiation fields due to the fall of particles through the event
horizon, as well the associated black hole thermodynamics. 
 To highlight the role played by conformal quantum mechanics,
 Fig.~\ref{fig:HBAR-setup-2} displays the 
  near-horizon region as generating the relevant physics.
 Once this near-horizon analysis is enforced for the HBAR field, a remarkable
 set of properties and similarities with black hole thermodynamics emerge, as we will discuss in this section.

\subsection{HBAR transition probabilities}

\begin{figure}[h]
    \centering
    \includegraphics[width=0.85\linewidth]{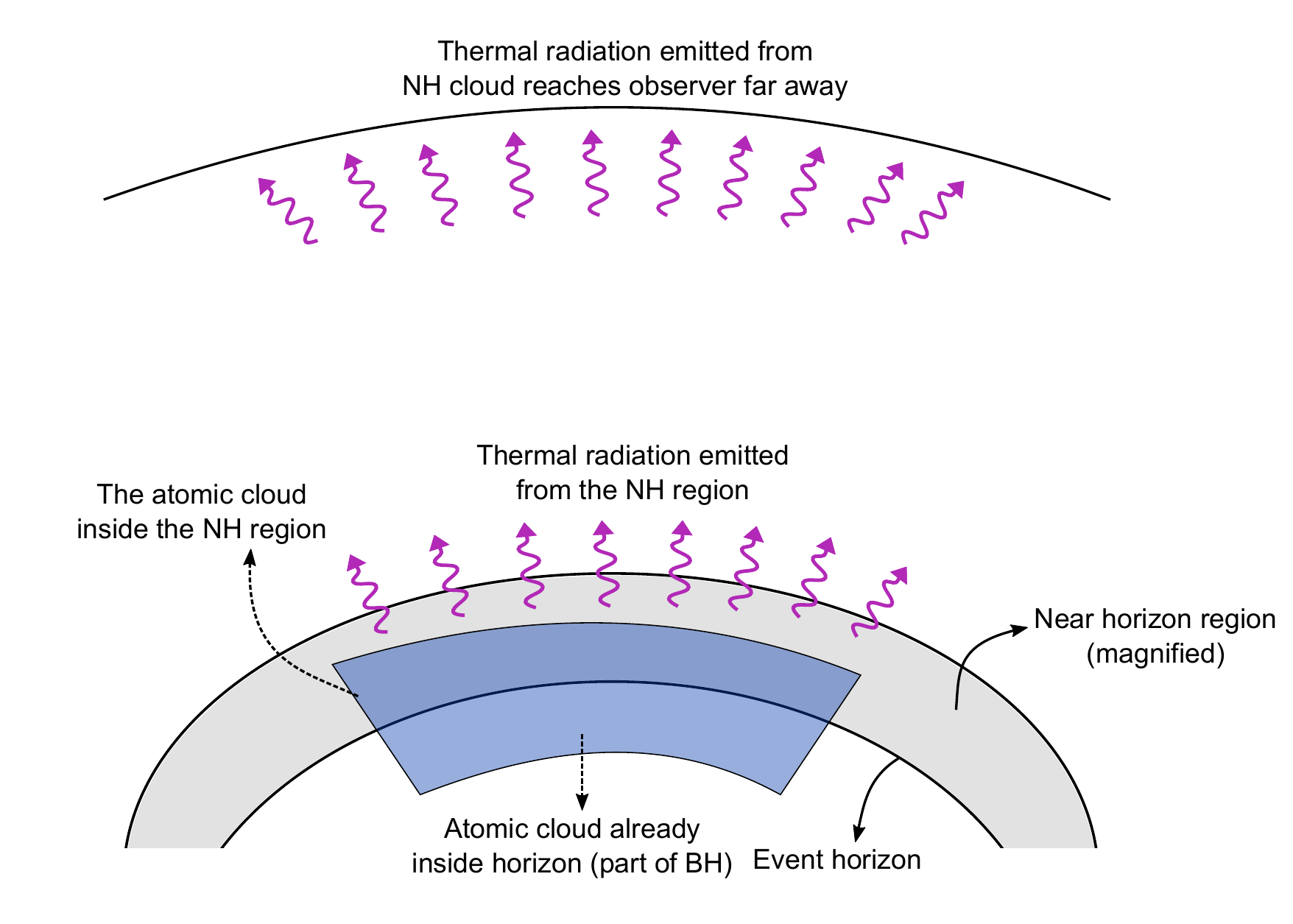}
    \caption{The thought experiment for the HBAR model displays the 
    relevance of the near-horizon analysis leading to conformal quantum mechanics (CQM).}
       \label{fig:HBAR-setup-2}
       \hspace*{0in}
\end{figure}

The two direct consequences of the gravitational field needed for the calculations of the HBAR radiation field are
the functional forms of the scalar field modes and the geodesic equations.
In their near-horizon CQM forms, these are Eqs.~(\ref{eq:CQM_modes}) and (\ref{eq:tau_in_x})--(\ref{eq:t_in_x}).
These results can be applied to 
atoms freely falling into the black hole, in the HBAR thought experiment of 
Figs.~\ref{fig:HBAR-setup} and ~\ref{fig:HBAR-setup-2}.
where 
the possible emission of outgoing radiation corresponds to 
the purely outgoing CQM modes of Eq.~(\ref{eq:CQM_modes}). Then, the emission and absorption rates 
under free fall, using Eqs.~(\ref{eq:P_ex_explicit}) and (\ref{eq:P_ab_explicit}), become
    \begin{equation}
            P_{{\mathrm e}, {\boldsymbol{s}} } 
            \stackrel{(\mathcal H)}{\sim}\; 
            g^2k^2 \left|\int_0^{x_f} \; dx\; x^{-i\Theta} 
            e^{i{\omega} t(x)} e^{i\nu\tau(x)} \right|^2
        \; 
        \label{eq:P_ex_CQM}
    \end{equation}
    and
    \begin{equation}
            P_{{\mathrm a}, {\boldsymbol{s}} }  
            \stackrel{(\mathcal H)}{\sim}\; 
            g^2k^2 \left|\int_0^{x_f} \; dx\; x^{i\Theta}  e^{-i{\omega} t(x)} e^{i\nu\tau(x)} \right|^2
        \; , 
        \label{eq:P_ab_CQM} 
    \end{equation}
where $k=1/e$ and $x_f$ is an approximate scale setting the upper 
 boundary of the region of dominance of the near-horizon CQM behavior.
In addition, from Eqs.~(\ref{eq:tau_in_x}) and (\ref{eq:t_in_x}), the emission rate becomes
\begin{equation}
        R_{{\mathrm e}, {\boldsymbol{s}} }   =  \mathfrak{r} \, g^2 k^2 \left|\int_0^{x_f} dx 
        \;
        x^{-i{\omega} /\kappa} e^{-i s x} \right|^2
        \; ,
 \label{eq:near-horizon-pex}
\end{equation}
where $s=C{\omega} + \nu/e $.
The integral of Eq.~(\ref{eq:near-horizon-pex}) is essentially the probability amplitude, where 
the field-mode and atom contributions yield phases leading to a combined competition of 
 two oscillatory factors: $x^{-i{\omega} /\kappa}$ and $e^{-isx}$. 
Clearly, the factor $x^{-i{\omega} /\kappa}= x^{-2 i \Theta}$ 
 is the one that governs the final outcome, with 
 the CQM scale invariance and a conformal parameter  ${\omega} /\kappa = 2\Theta$. Notice that this 
 parameter includes two contributions of 
$\Theta$: one from the near-horizon field modes
and the one from the geodesic motion of the atom (relative to the fields).
This is the factor that exhibits the Russian-doll behavior depicted in
 Fig.~\ref{fig:Russian-doll-wavefronts}).
On the other hand,
 the $e^{isx}$ factor makes the integral average out to essentially zero away from the horizon
 because of its highly oscillatory nature in the limit $\nu\gg {\omega} $.
  Consequently, this behavior generates the leading value of the integral 
  almost exclusively from the near-horizon region. Therefore,
  the upper limit $x_f$ appears as effectively infinite: it can displaced to infinity 
  as a result of scale invariance and cancellations,
   leading to
\begin{equation}
    R_{{\mathrm e}, {\boldsymbol{s}} } 
     =   \mathfrak{r}  \, g^2 k^2  \left|\int_0^{\infty} \; dx\;  x^{-i{\omega} /\kappa} e^{-i{s} x} \right|^2 
     = \frac{2\pi  \mathfrak{r} \,  g^2 {\omega} }{\kappa\,\nu^2}\;\frac{1}{e^{2\pi{\omega} /\kappa}-1}
     \; .
    \label{eq:R_ex_steps} 
\end{equation}
Interestingly, as anticipated in the previous section,
the probability is independent of
 the constants $k$ and $C$; thus, it is independent of the initial conditions.

In conclusion, Eq.~(\ref{eq:R_ex_steps})
shows the existence
of acceleration radiation with a Planckian distribution, provided 
we use the Boulware state $\left| B \right\rangle$ (with the field modes associated with stationary, 
generalized Schwarzschild coordinates).
This points to the thermal nature of the emission rate. 
However, this result needs to be examined in greater detail to show it 
is fully endowed with all the properties of a thermal state. 

 \subsection{Thermal behavior: Detailed balance via the Boltzmann factor}
 \label{sec:thermal_Boltzmann}

We can start by exploring the thermal behavior
 through the ratio of the emission and absorption rates. 
These rates could be computed separately, but 
direct inspection of Eq.~(\ref{eq:P_ab_CQM}) 
shows that the absorption rate follows from the emission rate 
via the replacement ${\omega}  \rightarrow -{\omega} $. Then,
\begin{equation}
    R_{{\mathrm a}, {\boldsymbol{s}} } 
     = \frac{2\pi r g^2 {\omega} }{\kappa\,\nu^2}\;\frac{1}{1-e^{-2\pi{\omega} /\kappa}}\;,
       \label{eq:em_ab_probabilities}
\end{equation}
leading to the probability ratio
\begin{equation}
    \frac{R_{{\mathrm e}, {\boldsymbol{s}} } }{R_{{\mathrm a}, {\boldsymbol{s}} } } 
    = e^{-2\pi{\omega} /\kappa}
    \,.
     \label{eq:ratio_em_ab}
\end{equation}
This ratio has a straightforward interpretation: it
corresponds to a thermal state with an effective temperature 
$T= \beta^{-1}$ by 
 detailed-balance,
  governed by the Boltzmann factor
\begin{equation}
    \frac{R_{{\mathrm e}, {\boldsymbol{s}} } }{R_{{\mathrm a}, {\boldsymbol{s}} } } = e^{-\beta {\omega} }
    \, . \label{eq:ratio_em_ab_Boltzmann}
\end{equation}
Compatibility of  Eqs~(\ref{eq:ratio_em_ab}) and (\ref{eq:ratio_em_ab_Boltzmann}) 
 implies that the field is a thermal state with temperature
\begin{equation}
T= (k_{B} \beta)^{-1}  = \frac{\hbar \kappa}{2 \pi k_{B} c }  \equiv (k_{B} \beta_{H})^{-1} = T_H
\; ,
\label{eq:temperature=Hawking}
\end{equation}
which is identical to the black-hole Hawking temperature~(\ref{eq:Hawking-temperature}), 
when all the units are restored by dimensional analysis.
In addition, this temperature coincides with that of the Planck distribution~(\ref{eq:R_ex_steps}). 

 It should be noted that 
  Eqs.~(\ref{eq:ratio_em_ab})--(\ref{eq:temperature=Hawking})
  do show the existence of a {\it unique temperature $T$ defined by detailed balance 
 for all the field modes\/}.
    In the absence of such unique-temperature condition, the ``temperature'' would not 
   satisfy all the features of thermality, but would instead be 
   an effective, mode-dependent parameter. This shows that
$
T = T_{H}
$
is a candidate for a {\it genuine thermodynamic temperature associated with a thermal state\/}.

In conclusion,
this proof also shows the governing role of near-horizon CQM,
through the logarithmic singular nature of its modes and geodesics. This leads
 directly to the Boltzmann factor
in the ratio of the probabilities $P_{{\mathrm e}, {\boldsymbol{s}} } $ and $P_{{\mathrm a}, {\boldsymbol{s}} } $,  Eq.~(\ref{eq:ratio_em_ab_Boltzmann}).
A more thorough thermal characterization of the state of the field can be obtained 
via the steady-state field density matrix, as discussed in the next section.

 \subsection{Thermal behavior: Generalizations and steady-state field density matrix}
 \label{sec:conformal_steady_state}
 
We now extend the results of the previous section
 to a complete derivation of thermality under the most general conditions. For this discussion,
 we will use relevant parts of Appendix~\ref{app:spacetime} to accommodate
  a black hole rotational degree of freedom via its angular momentum $J$ and angular velocity $\Omega_{H}$.

The basic framework described so far remains formally identical, and the 
results do apply without further changes
 provided that the frequency $\omega$ is replaced by the ``corotating frequency'' 
  ${\tilde{\omega}} = \omega - \Omega_{H} m $,
which involves the energy $\omega$ (measured by an asymptotic observer) 
and the axial component $m$ of the field angular momentum.
 Thus, Eqs.~(\ref{eq:ratio_em_ab}) and
 (\ref{eq:ratio_em_ab_Boltzmann})  generalize to the form
 \begin{equation}
    \frac{R_{e, {\boldsymbol{s}} } }{R_{a, {\boldsymbol{s}} } } = e^{-2\pi\tilde{\omega}/\kappa} \,
 = e^{-\beta \tilde{\omega}}
    \label{eq:ratio_em_ab_Kerr-gen}
    \; .
\end{equation}
The final conclusion remains the same:
\begin{quotation}
\noindent
 The value of the effective temperature defined by this procedure in 
Eq.~(\ref{eq:temperature=Hawking}) is the same as the Hawking temperature of the black hole.
\end{quotation}
A more detailed analysis
of the thermal nature of the state of the HBAR radiation field
can be fully established with the master equation for the field density matrix:
  For a cloud of freely falling atoms, with random injection times,
  the diagonal part of the equation has the form~(\ref{eq:master_equation_final_multimode}).
Thus, the properties are essentially geometry-independent and apply equally well to the
 generalized Schwarzschild and Kerr geometries.
The steady-state density matrix 
$  {{\rho}}^{\mathrm (SS)}_{\rm diag}(  \boldsymbol{ \left\{  \right. } n  \boldsymbol{\left. \right\}  } )$
is obtained when the time derivative is zero: $\dot{\rho}_{\rm diag}=0$,
and using Eq.~(\ref{eq:ratio_em_ab_Boltzmann});
 the equation is easily solved for each mode, and the combined operator is a product of independent single-mode pieces, as follows:
 \begin{equation}
    {{\rho}}^{\mathrm (SS)}_{\rm diag}(  \boldsymbol{ \left\{  \right. } n  \boldsymbol{\left. \right\}  } )
= N \, \prod_{j}  \left(  \frac{R_{e, j } }{ R_{a, j} } \right)^{n_{j}} 
  = \frac{1}{Z}
   \prod_{j}  
    e^{-n_{j} \beta \tilde{\omega}_{j}}
    \; ,
    \label{eq:steady_state_multimode_combined}
\end{equation}
where $Z= N^{-1}= 
\prod_{j} Z_{j}
= \prod_{j} 
\left[ 1- e^{- \beta \tilde{\omega}_{j}}\right]^{-1}$ is the partition function.
Finally, along with Eq.~(\ref{eq:ratio_em_ab_Boltzmann}),
this verifies that:
(i) it is a thermal distribution at the Hawking temperature;
(ii) the steady-state average occupation numbers
are $ \left\langle n_{j} \right\rangle \! ^{\! \! ^{\mathrm (SS)}}  = \left( e^{\beta \tilde{\omega}_{j}} -1 \right)^{-1}$.

In conclusion, the density matrix of Eq.~(\ref{eq:steady_state_multimode_combined})
defines a steady-state thermal state for HBAR radiation, 
which includes the primary thermal properties of
Eqs.~(\ref{eq:ratio_em_ab_Boltzmann}) and (\ref{eq:temperature=Hawking}): 
the Boltzmann factor and the Hawking temperature.
Moreover, these results apply to a large class of black holes, both 
of generalized Schwarzschild and Kerr types.
Remarkably, these properties emerge from near-horizon CQM for
all field modes.
Finally, the outcome of this problem is closely related to the intrinsic thermodynamics of the black hole~\cite{Scully_2018_HBAR}. 
Indeed, in the next section,
we turn our attention to this general thermodynamic problem, with fundamental origins in quantum physics:
a comprehensive analysis of HBAR thermodynamics.

\section{Quantum information and quantum thermodynamics: 
HBAR-black hole correspondence}
\label{sec:HBAR-thermo}

In this section, we probe deeper into various quantum aspects of spacetime
by extending the results from the density matrix via its quantum-information measure given by the 
 von Neumann entropy~(\ref{eq:vonNeumann-entropy}):
$S = - 
\mathrm{Tr} \left[  {\rho} \ln {\rho} \right] $.
To simplify the analysis of structural analogies between HBAR thermodynamics and black hole thermodynamics,
we will continue using Planck natural units 
($\hbar =1$, $c=1$, $k_{B} =1$, $G=1$) 
throughout the remainder of the paper.

Our prime example of HBAR radiation displays several features of the questions being formulated within one of the most
recent outgrowths of quantum theory: 
{\it quantum thermodynamics\/}~\cite{QThermo_Gemmer-et-al, QThermo_Deffner-Campbell}.
This is an interdisciplinary field
that focuses on the relations between quantum physics and thermodynamics, 
using tools from information theory and open quantum systems. 
The emergence of thermodynamic behavior from quantum principles, and 
the description of systems out of equilibrium are some of the signatures of this novel emphasis,
bridging the gap with nonequilibrium statistical mechanics~\cite{Zwanzig_noneq-SM}.
For our purposes, starting from the density matrix~(\ref{eq:steady_state_multimode_combined}),
which we have derived within an open systems approach with standard quantum-optics techniques, a complete thermodynamic
analysis can be carried out, including the time evolution of the thermodynamic states using entropy flux.
In that sense, HBAR thermodynamics provides a thought experiment that illustrates techniques from open quantum systems
theory and quantum thermodynamics.

An important comparative remark on vacuum states in gravitational fields is relevant to 
the correspondence between HBAR and ordinary black hole thermodynamics.
The final outcome is governed by the properties of the stationary configuration achieved by the black hole,
according to the no-hair theorem~\cite{frolov}:
this is the ultimate reason for the remarkable correspondence.
However, the thermodynamics and the radiation fields of these two 
systems are of very different origin in terms of the initial setup.
 Hawking radiation assumes that the outgoing state of the field is the Unruh vacuum rather than the Boulware
 vacuum of the HBAR field.
 The Boulware vacuum is defined with normal modes of positive frequency
with respect to the Killing vector $\partial_{t}$.
Instead, the Unruh vacuum is defined with
modes incoming from past infinity to be of positive
frequency with respect to $\partial_{t}$;
and those that emerge from the past horizon of
positive frequency with respect to the coordinate $U$.
The coordinate $U$ is the affine parameter along the past horizon, and is
associated with the temporal ($T$) and radial ($R$)
Kruskal-Szekeres coordinates~\cite{GR_Carroll-2003, MTW-gravitation, GR_Wald-1984, frolov}:
$U=T-R$, i.e., the Kruskal-Szekeres outgoing null coordinate.
From the way it is set up, 
the Unruh vacuum is expected to approximate the field configuration
associated with a real gravitational collapse leading to the final formation of a black hole. Evidently, this
corresponds to a configuration where the black hole did not exist in the distant past and has achieved a final
stationary configuration in the future.
By contrast, the Boulware state fails to model this expected physical outcome---this is the reason why
it has only been proposed via a thought experiment for HBAR radiation.

\subsection{Radiation field entropy: HBAR entropy flux}
\label{sec:radiation-HBAR-entropy}

One of the straightforward consequences of the von Neumann entropy definition~(\ref{eq:vonNeumann-entropy})
 is its rate of change, known as the entropy flux, which is given by 
\begin{equation}
\dot{S} = - \mathrm{Tr} \left[  \dot{{\rho}} \ln {{\rho}} \right] 
\; .
\label{eq:entropy-flux}
\end{equation}
This rate allows the extension of steady-state properties to nonequilibrium states, and their information-theoretical measures.
Specifically, Eq.~(\ref{eq:entropy-flux}) directly applies to our case study:
particles falling into a black hole, and their concomitant HBAR 
field, i.e., it gives the entropy flux of their ``photons'' or field quanta.
As in Sec.~\ref{sec:conformal_steady_state}, 
a thermal steady-state density matrix can be directly obtained from near-horizon CQM, under 
the most general initial conditions and types of black holes.
This is the horizon brightened acceleration radiation (HBAR) entropy in Ref.~\cite{Scully_2018_HBAR}.

The HBAR entropy flux, from the general von Neumann expression of Eq.~(\ref{eq:entropy-flux}), 
can be computed by evaluating the operator trace via 
\begin{equation}
 \dot{S}_{\mathcal P} =  -  \sum_{\boldsymbol{ \left\{  \right. } n  \boldsymbol{\left. \right\}  } } 
 \dot{{\rho}}_{\rm diag}(  \boldsymbol{ \left\{  \right. } n  \boldsymbol{\left. \right\}  } )
 \ln  \left[   {\rho}_{\rm diag}(  \boldsymbol{ \left\{  \right. } n  \boldsymbol{\left. \right\}  } ) \right]
 =
- \sum_{  n_1,n_2, \ldots }
  \dot{\rho}_{ n_1,n_2, \ldots ;   n_1,n_2, \ldots }
 \ln   {\rho}_{ n_1,n_2, \ldots ;   n_1,n_2, \ldots } 
    \; ,
    \label{eq:entropy_rate-change}
\end{equation}
which is a diagonal sum over all the states
${\boldsymbol{ \left\{  \right. } n  \boldsymbol{\left. \right\}  } }$ for all the field modes.
If, in addition, the 
system is near a steady-state configuration, the leading order
 of Eq.~(\ref{eq:entropy_rate-change}) 
is given by approximating the logarithm of the 
density matrix, with the value
 ${\rho}^{\mathrm (SS)}_{\rm diag}$ from
Eq.~(\ref{eq:steady_state_multimode_combined}):
\begin{equation}
    \dot{S}_{\mathcal P} \approx 
 - \sum_{\boldsymbol{ \left\{  \right. } n  \boldsymbol{\left. \right\}  } } 
  \dot{{\rho}}_{\rm diag}(  \boldsymbol{ \left\{  \right. } n  \boldsymbol{\left. \right\}  } ) 
 \ln  \left[   {\rho}^{\mathrm (SS)}_{\rm diag}(  \boldsymbol{ \left\{  \right. } n  \boldsymbol{\left. \right\}  } ) \right]
 =
 - \sum_{j}  \sum_{\boldsymbol{ \left\{  \right. } n  \boldsymbol{\left. \right\}  } } 
  \dot{{\rho}}_{\rm diag}(  \boldsymbol{ \left\{  \right. } n  \boldsymbol{\left. \right\}  } ) 
\ln  {\rho}_{n_{j},n_{j}}^{\mathrm (SS)}
    \; ,
    \label{eq:entropy_rate-change_SS}
\end{equation}
which involves a reversal in the order of the sums and 
the explicitly factorized form of the HBAR density matrix~(\ref{eq:steady_state_multimode_combined}).
Then, the thermal nature of the steady-state density matrix
${\rho}^{\mathrm (SS)}_{\rm diag}$ converts  Eq.~(\ref{eq:entropy_rate-change_SS}) into
\begin{equation}
    \dot{S}_{\mathcal P} \approx 
 \sum_{j}  \sum_{\boldsymbol{ \left\{  \right. } n  \boldsymbol{\left. \right\}  } } 
  \dot{{\rho}}_{\rm diag}(  \boldsymbol{ \left\{  \right. } n  \boldsymbol{\left. \right\}  } ) 
  \left[
  n_{j} \beta \tilde{\omega}_{j} - \ln (1-e^{- \beta \tilde{\omega}_{j}}) 
  \right]
    \; .
  \label{eq:entropy_rate-change_SS_2}
\end{equation}
The sums are constrained by two conditions:
the trace normalization
$ 
\mathrm{Tr} \left[  
{\rho} \right] = 1$ and the dynamic generalization of the occupation-number averages
\begin{equation}
  \left\langle n_{j} \right\rangle \equiv 
\sum_{
 \boldsymbol{ \left\{  \right. } n  \boldsymbol{\left. \right\} }
}
n_{j} \,   {{\rho}}_{\rm diag}(  \boldsymbol{ \left\{  \right. } n  \boldsymbol{\left. \right\}  } ) 
    \; ,
    \label{eq:average-occupation_dynamical}
\end{equation}
where these quantities $\left\langle n_{j} \right\rangle \equiv  \left\langle n_{ \boldsymbol{s} } \right\rangle $
are defined for each set of field-mode numbers
${\boldsymbol{s}} =  \{\tilde{\omega},l,m\}$. 
In
Eq.~(\ref{eq:average-occupation_dynamical}), 
the averages $  \left\langle n_{j} \right\rangle$ are not simply given by
the steady-state average occupation numbers
 $ \left\langle n_{j} \right\rangle \! ^{\! \! ^{\mathrm (SS)}}  = \left( e^{\beta \tilde{\omega}_{j}} -1 \right)^{-1}$ or 
 the exact Planck distribution, but they include a modification that guarantees
  a nonzero flux through $ \dot{\, \, \left\langle n_{ \boldsymbol{s} } \right\rangle } \neq 0$.
  In addition, to the same order,
the constant trace normalization
yields the vanishing of
 the second-term in Eq.~(\ref{eq:entropy_rate-change_SS_2}). 
Thus, the entropy flux  is given by
\begin{equation}
    \dot{S}_{\mathcal P} \approx \beta_{H}
 \sum_{{\boldsymbol{s}} = \{\tilde{\omega},l,m\} } 
  \! \!   \! \! 
   \dot{\, \, \left\langle n_{ \boldsymbol{s} } \right\rangle }  \, \tilde{\omega} 
   =
   \frac{2\pi}{\kappa}
 \sum_{{\boldsymbol{s}} = \{\tilde{\omega},l,m\} } 
  \! \!   \! \! 
   \dot{\, \, \left\langle n_{ \boldsymbol{s} } \right\rangle }  \, \tilde{\omega} 
    \; ,
    \label{eq:Sp_1}
\end{equation}
which can be written in terms of the unique Hawking temperature~(\ref{eq:temperature=Hawking}) as 
geometrical surface gravity; in geometrized units
\begin{equation}
T_{H}= \beta_{H}^{-1}  = \frac{ \kappa}{2 \pi } 
\; .
\label{eq:temperature=Hawking_geom-units}
\end{equation}

 Equation~(\ref{eq:Sp_1}) can be interpreted physically in terms of the photons of the acceleration radiation:
the product $ \dot{\left\langle n_{ \boldsymbol{s} } \right\rangle }  \, \tilde{\omega} $ 
is the energy flux of the field quanta at a given frequency.
In the more general type of black holes with rotation (Kerr geometry), the photon energies 
in Eq.~(\ref{eq:Sp_1}) are measured in the corotating frame:
${\tilde{\omega}} = \omega - \Omega_{H} m $,
which involve the energy $\omega$ (measured by an asymptotic observer) 
and the axial component $m$ of the field angular momentum (along the black hole's rotational axis) 
coupled to the
black hole's angular momentum $\Omega_{H}$.
Thus,
the net corotating energy flux is
\begin{equation}
\dot{\tilde{E}}_{\mathcal P}
=
 \sum_{{\boldsymbol{s}} = \{\tilde{\omega},l,m\} } 
  \! \!   \! \!  
   \dot{\, \, \left\langle n_{ \boldsymbol{s} } \right\rangle }      \, \tilde{\omega} 
   =
    \sum_{{\boldsymbol{s}} = \{\tilde{\omega},l,m\} } 
  \! \!   \! \!   
   \dot{\, \, \left\langle n_{ \boldsymbol{s} } \right\rangle }
\,  \bigl( \omega - \Omega_{H} m \bigr)
= \dot{E}_{\mathcal P} - \Omega_{H} \dot{J}_{{\mathcal P},z} 
\; ,
\label{eq:tilde-E_P-change}
 \end{equation} 
 which consists of  a combination of the change in the total energy $E_{\mathcal P} $ 
 and axial angular momentum $J_{{\mathcal P},z} $ of the photons. 
This sequence of quantum conditions leads to a compact 
formula for the HBAR von Neumann entropy flux,
\begin{equation}
    \dot{S}_{\mathcal P} = \beta_{H}
    \bigl( \dot{E}_{\mathcal P} -\Omega_{H} \dot{J}_{{\mathcal P},z}   \bigr)
    =  \beta_{H} \dot{\tilde{E}}_{\mathcal P}
    \; .
    \label{eq:Sp_2}
\end{equation}
Or, in terms of thermodynamic changes,
\begin{equation}
   \delta {S}_{\mathcal P} = \beta_{H} 
  \bigl(  \delta {E}_{\mathcal P} -\Omega_{H} \, \delta {J}_{{\mathcal P},z}  \bigr)
   \equiv
   \delta {S}_{\mathcal P}^{\, \rm (th)}
    \; ,
    \label{eq:Sp_3}
\end{equation}
which is identical to the changes in 
the usual thermodynamic entropy $ {S}_{\mathcal P}^{\, \rm (th)}$.

 In conclusion, 
  fluxes and changes in the HBAR von Neumann and thermodynamic entropies of the radiation field coincide,
  governed by Eq.~(\ref{eq:Sp_3}).
Incidentally,
the same conclusions can be derived 
directly from the von Neumann entropy, 
$S = - \mathrm{Tr} \left[  {\rho} \ln {\rho} \right] $,
with the replacement of 
$\dot{{\rho}}_{\rm diag}(  \boldsymbol{ \left\{  \right. } n  \boldsymbol{\left. \right\}  } ) $
by
${{\rho}}_{\rm diag}(  \boldsymbol{ \left\{  \right. } n  \boldsymbol{\left. \right\}  } ) $
in Eqs.~(\ref{eq:entropy_rate-change})--(\ref{eq:entropy_rate-change_SS_2}),
leading to 
$S_{\mathcal P} = \beta (E_{\mathcal P} -F_{\mathcal P})$,
in terms of the Helmholtz free energy $F_{\mathcal P}$ that satisfies 
$\beta F_{\mathcal P}=
- \ln Z = \sum_{j}\ln \left( 1 - e^{-\beta \tilde{\omega}_{j} } \right)
$ from 
Eq.~(\ref{eq:steady_state_multimode_combined}).
The equivalence of entropies and entropy changes described by 
Eq.~(\ref{eq:Sp_3}) is ultimately due 
to the universal behavior of the black hole as a temperature reservoir
that has a {\em unique Hawking temperature\/}~(\ref{eq:temperature=Hawking_geom-units}).

\subsection{HBAR-black-hole thermodynamic correspondence}
\label{sec:HBAR-BH_correspondence}

The HBAR entropy flux described by Eqs.~(\ref{eq:Sp_2}) and (\ref{eq:Sp_3}) has a form structurally identical to
the thermodynamic changes of the black hole itself,
\begin{equation}
\delta S_{\mathrm{BH}}
= \beta_{H} \bigl( \delta M -  \Omega_H \delta J \bigr)
\; ,
  \label{eq:BH-entropy-changes}
\end{equation}
which are expressed in terms of the black hole entropy  $S_{\mathrm{BH}}$,
its mass $M$ and its angular momentum $J$. 
Equation~(\ref{eq:BH-entropy-changes}) relates the entropy and energy changes with the black hole rotational work as 
 required by general relativistic and thermodynamic arguments only; 
 in particular, the combination
 \begin{equation}
 \delta \tilde{M}
=  \delta M -  \Omega_H \delta J
\; 
\label{eq:tilde-M-change}
\end{equation}
is the black hole corotating energy change.

On the other hand, the celebrated black-hole Bekenstein-Hawking entropy~(\ref{eq:BH-entropy}),
as mentioned in Sec.~\ref{sec:introduction},
involves the geometrical area of the event horizon~\cite{bekenstein-S_1972, bekenstein-S_1973}; in
Planck units, it takes the form 
 \begin{equation}
  S_{\mathrm{BH}} = \frac{1}{4} A
    \; , 
\label{eq:BH-entropy-geometrized}
  \end{equation}
  so that
 \begin{equation}
\delta S_{\mathrm{BH}} = \frac{1}{4}   \delta A
\; . 
 \label{eq:BH-entropy-area-changes} 
\end{equation}
  It is noteworthy that, while various thermodynamic and quantum 
  calculations indicate that there 
  exists a proportionality between entropy and area, 
 the correct value $1/4$ of the proportionality constant cannot be obtained by simple arguments.
 However, 
   the specific value of the Hawking temperature~(\ref{eq:temperature=Hawking}):
   $T_{H}=\beta_{H}^{-1}=\kappa/2 \pi$,
   driven by quantum-mechanical radiation effects,   
    does uniquely fix the proportionality prefactor~\cite{hawking74,hawking75}.
Moreover, these results are known to be valid for all black holes, including their angular momentum.
As summarized in Appendix~\ref{app:spacetime},
the black-hole area changes,
accounting for the Hawking temperature~(\ref{eq:temperature=Hawking}), 
 are geometrically determined to be~\cite{frolov}
\begin{equation}
  \delta A  = 4 \beta_{H} \left( \delta M -  \Omega_H \delta J\right)
\; .
  \label{eq:BH-area-changes}
\end{equation}
In effect, Eq.~(\ref{eq:BH-area-changes})
follows from the geometric area change of Eq.~(\ref{eq:BH-area-changes-geom_Kerr}) combined with
the Hawking temperature~(\ref{eq:temperature=Hawking}), 
and the substitution~(\ref{eq:tilde-M-change}). Therefore,
 the required generic black hole entropy changes~(\ref{eq:BH-entropy-changes})
are indeed enforced with the Bekenstein-Hawking entropy~(\ref{eq:BH-entropy-geometrized}) with a numerical
factor $1/4$.
 
These remarkable results reveal an {\em HBAR-black-hole correspondence\/}
that extends the familiar {\it universal thermodynamic relations\/} inherent to the black hole:
the Hawking temperature and the Bekenstein-Hawking entropy.
The following properties
capture the essence of this correspondence, including the rationale for its existence.
  
$\bullet$ First,
the intrinsic thermodynamic nature of the temperature and its role in thermal equilibrium lead to
the {\em unique Hawking temperature\/}~(\ref{eq:temperature=Hawking_geom-units}),
which is both geometrical and quantum-mechanical,
and is common to both the black hole itself and the HBAR radiation field.

$\bullet$ Second,
for the entropy, energy, and angular momentum variables,
the {\em analog fundamental thermodynamic relations\/},
 Eqs.~(\ref{eq:Sp_3}) and (\ref{eq:BH-entropy-changes}),
  provide a rigorous thermodynamic correspondence
 \begin{equation}
\big( 
{S}_{\mathcal P} , {E}_{\mathcal P} ,  {J}_{{\mathcal P},z} 
\bigr) 
\xlongleftrightarrow{\beta = \beta_{H}}
\bigl( 
S_{\mathrm{BH}} , M , J 
\bigr) 
\; \; \;
\; .
\label{eq:HBAR-BH-correspondence}
\end{equation}
Moreover, this correspondence can be generalized to subsume
any other relevant degrees of freedom consistent with no-hair theorems~\cite{frolov}; in particular, this includes
charged black holes (Kerr-Newman geometry in 4D), with an additional charge variable.  

$\bullet$ Third,
  the HBAR-field and the black hole entropies are governed by the common {\em near-horizon conformal symmetry  
  of CQM\/},  which determines the characteristic temperature, as seen in the
  steps leading to Eq.~(\ref{eq:temperature=Hawking}).
This is the ultimate reason why the quantum field appears to mirror 
  the black hole degrees of freedom in thermal equilibrium. 
  
  $\bullet$ Fourth,
  the HBAR-black-hole thermodynamic correspondence~(\ref{eq:HBAR-BH-correspondence})
and the Bekenstein-Hawking entropy-area relation 
of Eq.~(\ref{eq:BH-entropy-area-changes}) imply the existence of an analog 
{\em entropy-area relation for the HBAR entropy\/}, 
\begin{equation}
    \dot{S}_{\mathcal P} = \frac{1}{4} \bigl| \dot{A}_{\mathcal P} \bigr|
     \label{eq:HBAR_final}
     \; ,
\end{equation}
where $\bigl| \dot{A}_{\mathcal P} \bigr|$ is the absolute value of the
change in the event-horizon area due to the emission of acceleration radiation.
\\
This shows, {\em inter alia\/}, that,
once the temperature is fixed, there is a unique entropy-area relation, with a proportionality prefactor equal to $1/4$.

   $\bullet$ Finally,
even though the HBAR field is not the better known Hawking radiation, 
as seen far from the black hole, they both have identical properties:
thermal nature and characterized by the same Hawking temperature 
$T_{H}= \kappa/2\pi$.
As a result, the HBAR-black-hole correspondence enlarges 
 Eq.~(\ref{eq:HBAR-BH-correspondence})
with the additional mapping
\begin{equation}
\text{(HBAR field)}
\xlongleftrightarrow{\beta = \beta_{H}}
\text{(Hawking radiation)}
\; \; \;
\; .
\label{eq:HBAR-BH-correspondence_radiation}
\end{equation}
 
Some final remarks are in order for context.
Regarding the proof of the HBAR area-entropy-flux relation~(\ref{eq:HBAR_final}),
first proposed in Ref.~\cite{Scully_2018_HBAR},
it is structurally mandated by the
HBAR-black-hole thermodynamic correspondence~(\ref{eq:HBAR-BH-correspondence})
and the Bekenstein-Hawking relation~(\ref{eq:BH-entropy-area-changes}).
However,
a more direct proof follows by evaluating the 
 area changes~(\ref{eq:BH-area-changes}) leading to Eq.~(\ref{eq:BH-entropy-area-changes}), which
 have an absolute value
\begin{equation}
|\dot{A}|
=
 4 \beta_{H}
 |\dot{\tilde{M}}|
 \; ,
 \end{equation}
with the corotating energy change $\dot{ \tilde{M} } = \dot{ {M} } - \Omega_H \dot{J}$ as in Eq.~(\ref{eq:tilde-M-change}).
Now, the contribution to $ |\dot{\tilde{M}}|$ due to photon emission is the corotating energy flux~(\ref{eq:tilde-E_P-change}):
$| \dot{\tilde{M}}| = \dot{\tilde{E}}_{\mathcal P} $, 
which leads to the change associated with acceleration radiation:    
\begin{equation}
 |\dot{A}_{\mathcal P}|
= 4 \beta_{H}
 \dot{\tilde{E}}_{\mathcal P}
 = 4  \dot{S}_{\mathcal P}
     \label{eq:HBAR_final_derivation}
     \; ,
\end{equation}
where Eq.~(\ref{eq:Sp_2}) is used as a last step.
Equation~(\ref{eq:HBAR_final_derivation}) amounts to the area-entropy-flux relation~(\ref{eq:HBAR_final}).

One important qualification of the statements above is that the HBAR radiation 
  is mediated by the interaction of the field with the atoms. 
 The energy and angular momentum transfers between the black hole, atoms, and field
 satisfy conservation laws,
 which can be expressed as a single energy conservation statement 
    \begin{equation}
      ( \delta M -  \Omega_H \delta J )
      +   \delta \tilde{E}_{\mathcal P} 
        + \delta \tilde{E}_{\mathcal A} = 0 
        \; ,
         \label{eq:conservation}
    \end{equation}
 where  $\delta \tilde{E}_{\mathcal P} $ and $\delta \tilde{E}_{\mathcal A} $  are the field and atom corotating energy changes, 
 i.e., their values in the frame corotating with the black hole.
 Thus, the black hole area change can be regarded as arising from two distinct contributions,
        \begin{equation}
 \dot{A} =  \dot{A}_{\mathcal P}
 +  \dot{A}_{\mathcal A}
   \; ,
    \label{eq:area-changes}
\end{equation}
  including one part associated with the atoms,
 $\dot{A}_{\mathcal A} = - 4 \beta_{H} \dot{\tilde{E}}_{\mathcal A}$.
 As a result,
the area change $ \dot{A}_{\mathcal P}$
 associated with the HBAR radiation field
  in the area-entropy-flux relation~(\ref{eq:HBAR_final}):
 (i)
 is only a fraction of the total change in the black hole area;
 (ii) has a 
 sign reversal
(a positive corotating energy corresponds to a decrease in the area of the black hole).
The corresponding increase 
 in the HBAR entropy is still consistent 
 with the generalized second law of thermodynamics (GSL), 
 which involves the total sum of the entropies and not to the entropy of the black hole or of the radiation field alone.
The atoms falling into the black hole have $\delta \tilde{E}_{\mathcal A} <0$
with a corresponding area increase $\delta A_{\mathcal A}>0$.
For nonrelativistic atoms 
the overall area of the black hole does increase. 
Finally, the generalized second law of thermodynamics,
$\delta S_{\rm total} =
\delta {S}_{\mathrm{BH}}
+
\delta {S}_{\mathcal A}
+
\delta {S}_{\mathcal P}
 \geq 0$, 
 leads to
 \begin{equation}
 \delta S_{\mathcal A} \geq \beta_{H}
 \bigl(  \delta {E}_{\mathcal A} -\Omega_{H} \, \delta {J}_{{\mathcal A},z}  \bigr)
\; ,
\end{equation} 
and all the previous statements are compatible with this condition.

In short,
the HBAR and Hawking radiation fields have many properties of quantum nature in common. 
In some sense, they are analogue systems that
require a black hole, even though their origin is different, with HBAR being fed by 
 an atomic cloud mediating the radiation process.
For fairly generic random-injection conditions, both phenomena, when probed far from the black hole, have 
formally identical properties. 
Finally, the HBAR area-entropy-flux relation~(\ref{eq:HBAR_final}), along with the  broader 
 HBAR-black hole correspondence defined 
by Eqs.~(\ref{eq:HBAR-BH-correspondence}) and (\ref{eq:HBAR-BH-correspondence_radiation}) are
insightful result results that 
point to a {\em deep connection between the radiation field and the black hole itself\/}, and they both appear to be 
{\em governed by near-horizon conformal symmetry.\/}

\section{Conclusions: Frontiers of quantum knowledge and HBAR proposals}
\label{sec:conclusions}

The emergence of a quantum-thermodynamic framework for black holes is
a remarkable development that has made explicit the subtle interplay between quantum physics and 
relativistic spacetime following the seminal discoveries of 
Refs.~\cite{bekenstein-S_1972, bekenstein-S_1973,bardeen-carter-hawking1973, hawking74,hawking75,hawking76}, 
and with similar results for accelerated systems~\cite{Fulling-1973_CFQ,Davies-1975_Unruh,unruh-notes,ufd}.
As a result, one cannot overstate the central role played by black hole thermodynamics 
and the Hawking effect in the frontiers of contemporary theoretical physics. 
The fact that quantum physics is compatible with
gravity in a semiclassical limit is a nontrivial test of robustness that points to a possible theory of quantum gravity. Most
importantly, the paradoxes involved by the standard Hawking effect and the black hole information paradox are topics of current interest~\cite{BH-InfoLoss_review-Mathur-2009, BH-InfoLoss_review-Perez-2017, BH-InfoLoss_review-Almheiri-2021, BH-InfoLoss_review-Raju-2021}
 that suggest further compatibility via quantum information 
theory---but this still remains an open problem.

\subsection{Network of quantum connections}

In this review article, we offered a comprehensive overview of various
aspects of such framework of 
{\it quantum thermodynamics\/}~\cite{QThermo_Gemmer-et-al, QThermo_Deffner-Campbell}
and {\it quantum information\/}~\cite{nielsen00, Qinfo-review_LewisSwan,peres2004quantum,Mann_RQI-review},
in connection with 
the acceleration radiation emitted by particles 
in a near-horizon black hole background.
Some of the highlights of this review article involve the following conceptual points.
\begin{itemize}
\item
There exists a tight network of connections
between quantum field theory, spacetime physics, and quantum optics. 

\item
The complete framework involves thermodynamic and quantum-information features that generalize the more familiar 
flat-spacetime relations.
Thus, this is an ideal laboratory to explore subtle conceptual and technical features of quantum physics---making this a
suitable topic to highlight a century of stunning achievements of quantum mechanics.

\item
In this relational setting, we reviewed
a variety of successful techniques that have provided conceptual and computational tools for the physics
of accelerated systems and gravitational backgrounds, including black holes.
These techniques include both standard quantum field theory tools and an arsenal of related methods 
developed in the context of quantum optics.

\item
The theory of open quantum systems, with the density matrix as central object, plays a dominant role in describing the ensuing thermodynamic and information-theoretic framework.

\item
 Specifically, the acceleration radiation due to the fall of particles in a Boulware vacuum can be studied as a thought experiment
with a field vacuum configuration enforced by boundary setups similar to laser cavities around
the black hole.  
Such Boulware-vacuum configurations amount to a cavity that is accelerated in a relativistic sense near the 
event horizon
of black hole, generating a relative acceleration between the field and falling particles.
The ensuing radiation is known as horizon-brightened acceleration radiation (HBAR).

\item
The HBAR field 
 leads to a full-fledged form of
HBAR thermodynamics that
is in one-to-one correspondence with black hole thermodynamics.

\item
Specifically, HBAR thermodynamics
 includes a characterization of both temperature and entropy, 
 as well as the creation of radiation with formal 
 analogies with Hawking radiation.
The conclusion of this analysis is that the black hole acts as a reservoir with the same manifestations 
as in its development of standard black hole thermodynamics.  
\end{itemize}

\subsection{Experimental realizations}

It should be stressed that
the relevant experiment realizing an HBAR field with astronomical black holes involves mirrors and cavities in black hole backgrounds. This may not be experimentally feasible in the foreseeable future. However,
{\em in the best tradition of a gedanken experiment\/},
it does provide deep insights into and theoretical tests of black hole thermodynamics.
On the other hand, one may evaluate possible experimental realizations of this thought experiment.
Such realizations can be classified into two categories.
\begin{itemize}

\item
{\em Black-hole analog systems.\/}
As the study of black-hole analog systems in the lab continues to 
evolve~\cite{BH-analogues_review-1, BH-analogues_review-2}, 
it might be possible to test some of these ideas in Earth-based labs with such black-hole analogs 
in the not-too-distant future.
However, at present, no specific realizations of this kind, i.e., simulating a black hole, have been offered.

\item
{\em Equivalent systems that simulate relative acceleration.\/}
As discussed in the introduction, there is vast literature dealing with 
mirrors and cavities involving various forms of acceleration~\cite{Fulling-Davies_mirror-1, Davies-Fulling_mirror-2, Anderson-etal_mirror-3, Good-etal_mirror-4,dynamical-Casimir-curved,Fulling-Wilson_2018_EP,Ben-Benjamin-etal_2019_Unruh-rev}.
A specific proposal is discussed below, in which accelerated mirrors 
can simulate physics equivalent to the HBAR field emission.

\end{itemize}

\begin{figure}[h]
    \centering
    \includegraphics[width=0.675\linewidth]{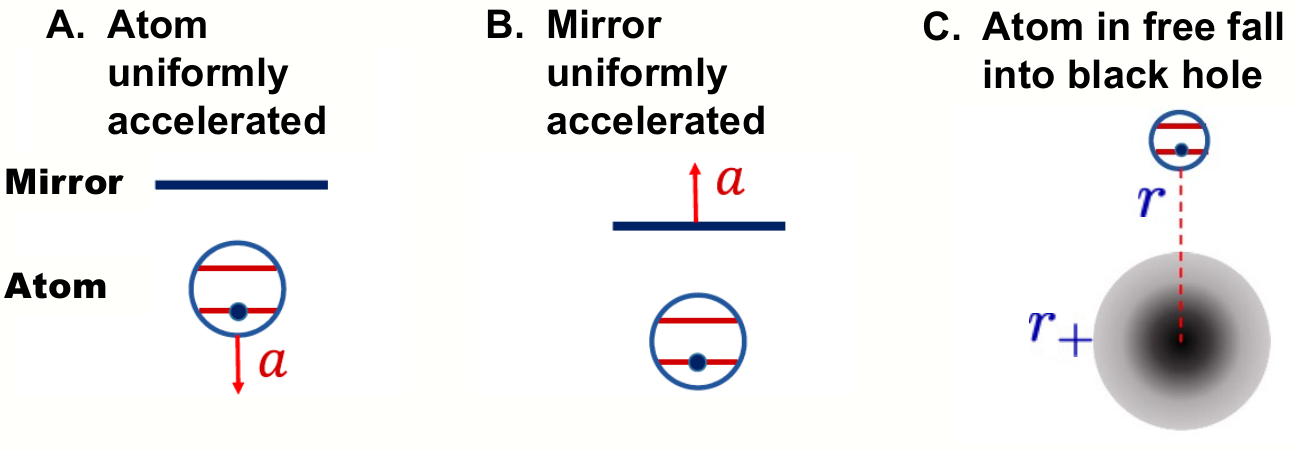}
    \caption{In the relative equivalence principle, configurations B and C 
    are equivalent as they both involve an accelerated background field and a stationary atom. Configuration C is the standard HBAR setup, where these are relativistic covariant concepts and the field is arranged as a Boulware vacuum.}
       \label{fig:acceleration-configurations}
       \hspace*{0in}
       \vspace*{-0.05in}
\end{figure}

The most basic equivalent system that simulates relative acceleration was proposed in the original seminal paper
of Ref.~\cite{Scully_2018_HBAR}.
This is illustrated by the comparison of three configurations in Fig.~\ref{fig:acceleration-configurations}.
The uniformly accelerated mirror (configuration B) produces an equivalent radiation spectrum to that of standard HBAR (configuration C); and these are distinctly different from a system
where the atoms are accelerated and the field is stationary (configuration A).
This equivalence is an example of the so-called relative
equivalence principle discussed in~\cite{Fulling-Wilson_2018_EP,Ben-Benjamin-etal_2019_Unruh-rev}.
As pointed out in Ref.~\cite{Svidzinsky_2018_mirrors-virtual},
there is a specific laboratory dynamical-Casimir setup that can ideally reproduce the radiation field of HBAR:
 a superconducting transmission line
microwave cavity terminated by a SQUID and coupled to
an ensemble of polar molecules.
This is a third equivalent system that would simulate the physics of
 an accelerating mirror~\cite{Wilson_dynamical-Casimir_2011, Lahteenmaki_dynamical-Casimir_2013, Johansson_dynamical-Casimir_2009}
and a two-level atom~\cite{Wallquist_quantum-devices_2009}. Effectively, the SQUID simulates the same boundary condition as the moving mirror~\cite{Johansson_dynamical-Casimir_2009},
with a large effective acceleration $a \gg c \omega$, where $\omega$ is the microwave-photon frequency. In this setup,
a two-level atom can be replaced by a large ensemble of polar molecules trapped closed to the transmission line
surface; the end result is an excitation probability of the order of $P \sim 10^{-2}$ under a reasonable choice 
of parameters. It is conceivable that similar proposals may lead to empirical verifications of the main physics 
associated with horizon-brightened acceleration radiation.

\subsection{Further developments}

It is also noteworthy that progress has been made in 
the study of  acceleration radiation 
and HBAR entropy in a variety of black hole backgrounds.
This includes an early study of HBAR radiation for detectors moving 
along null geodesics~\cite{HBAR-null-geod_Majhi-2019};
and a variety of extensions to other stringy and braneworld black 
holes~\cite{NH-BH-HBAR_Sen-2022, HBAR-branesBH_Das-2024},
 in quantum-corrected black hole 
geometries~\cite{QcorrBH-HBAR_Sen-2022, QcorrBH-HBAR_Jana-2024, QcorrBH-HBAR_Jana-2025},
and with a derivative coupling~\cite{HBAR-derivative-coupling_Das-2025}.

Finally, the techniques discussed in this review, from the generic quantum optics tools to the specific applications of HBAR, can be used in various extensions of the theory for a variety of gravitational backgrounds and arbitrary distributions of detectors and mirrors, as well as other possible applications.
As in some of the recent 
references~\cite{Dodonov_dynamical-Casimir-effect,Crispino-et-al_2008_Unruh-effect,Pasante_2018_dispersion-vacuum, Masood-etal_2024_atom-field-rev}, 
these analyses help clarify deep 
conceptual problems and point to the resolution of apparent paradoxes in relativistic systems
 and relativistic quantum information.

\bigskip

\acknowledgments{
H.E.C.  thanks Prof.\ C. A. Garc\'{i}a Canal for insightful comments.}

\bigskip

\bigskip

{\bf FUNDING} 

This material is based upon work supported by the Air Force Office of Scientific Research under Grant No. FA9550-21-1-0017 (C.R.O. and A.C.). C.R.O. was partially supported by the Army Research Office (ARO), grant W911NF-23-1-0202. H.E.C. acknowledges support by the University of San Francisco Faculty Development Fund. M.S. work was supported by U.S. Department of Energy (DE-SC-0023103, FWP-ERW7011, DE-SC0024882); Welch Foundation (A-1261); National Science Foundation (PHY-2013771); Air Force Office of Scientific Research (FA9550-20-1-0366). W.U. thanks the Natural Sciences and Engineering Research Council of Canada (NSERC) (Grant No. 5 80441), and the TAMU Hagler Institute for Advanced Studies for their support.


\bigskip

\bigskip

\bigskip

\bigskip

{\bf APPENDICES} 

\begin{appendix}

\section{BASIC ELEMENTS OF QUANTUM OPTICS}
\label{app:QOptics}

In this appendix, we review the basic foundational elements of quantum optics needed for the main body of this article.
These ingredients involve interacting atomic systems and electromagnetic fields, as described below
and in Sec.~\ref{sec:QOptics-interactions}.
Throughout the article, we have used a scalar field to describe the relevant physics; instead, here
we show how these elements are defined with the full-fledged electromagnetic field.

 Quantum optics, by itself, is a huge interdisciplinary field. It is also the underlying theory that led to the discovery and further development of the laser~\cite{Maiman-laser_1960,Bertolotti_masers-lasers},
 which is used in a variety of setups in experiments in physics, biology, chemistry, and other fields of science and engineering. 
 This broad field was originally introduced to model the interaction between electromagnetic fields as photons and 
 ordinary matter. 
  For our purposes, we provide an outline of
  the quantization of the electromagnetic fields and the atom-field dipole interaction. 
  The atom-field interaction in quantum optics 
 is of widespread use in a variety of applications; in particular, for systems involving spacetime backgrounds 
 (gravitational fields and noninertial systems), it leads to
 a definition of the Unruh-DeWitt detector, 
 which is a standard probe of nontrivial spacetime effects, 
 as seen in Secs.~\ref{sec:QOptics-interactions_Hamiltonian} and \ref{sec:QOptics-interactions_UDW}.
 More detailed descriptions and derivations of various theoretical and experimental
 aspects of quantum optics can be found 
in the standard Refs.~\cite{scullybook,meystrebook}.
 
\subsection{Quantization of the electromagnetic fields}
\label{sec:EM-field-quantization}

From the free Maxwell's equations in classical electrodynamics, the electric field $\mathbf{E}$ 
is shown to obey the wave equation 
\begin{equation}
    \nabla^2 \mathbf{E} - \frac{\partial^2\mathbf{E}}{\partial t^2} = 0
    \;
\end{equation}
(with the speed of light $c =1$).
Now,
 plane waves constitute a basis set for the solutions of the wave equation 
 in flat spacetime. Thus, for ordinary laboratory experiments where quantum electrodynamics and
 quantum optics were developed, one can expand the electric fields in terms of this set of plane waves
\begin{equation}
    \mathbf{E}(\mathbf{r},t) \
    = \sum_{\mathbf{k}} 
    \boldsymbol{\varepsilon}_{\mathbf{k}} 
     \mathcal{E}_{\mathbf{k}}  \alpha_{\mathbf{k}} 
      e^{i \mathbf{k} \cdot \mathbf{r}  - i\omega_k t }
    + \mathrm{H.c.}
    \; , 
    \label{eq:classical-EM-mode-expansion}
\end{equation}
as a particular case of the general expansion of the form~(\ref{eq:field_expansion}),
in which the labeling $\mathbf{s}$ of the states is in terms of the wave number $\mathbf{k}$.
In this standard treatment, the field is confined 
in a large but finite cubic cavity of length $L$ and volume $V=L^3$, so that the field momentum is discrete instead of continuous. 
To change the distribution from discrete to continuous, one has to replace the sum over momentum with an integral: 
\begin{displaymath}
\sum_{\mathbf{k}} \longrightarrow 2\left(\frac{L}{2\pi}\right)^3 \int d^3k
\; .
\end{displaymath} 
In Eq.~(\ref{eq:classical-EM-mode-expansion}), $\boldsymbol{\varepsilon}_k$ is the electric field polarization unit vector, 
$\mathbf{k}$ is the momentum of the field, $\omega_{\mathbf{k}}  = |\mathbf{k}|$ is the frequency of the field, $\alpha_{\mathbf{k}} $ is the dimensionless amplitude of each mode, $\mathcal{E}_{\mathbf{k}} $ is the electric field strength
\begin{equation}
    \mathcal{E}_{\mathbf{k}} 
     = \left(\frac{\omega_{\mathbf{k}}  }{2\epsilon_0 V} \right)^{1/2}\;,
\end{equation}
in naturalized SI units.
Due to the periodic boundary conditions, the momentum components take the values
 $k_i = 2\pi n_i/L$, where $n_i$ are non-negative integers. Hence, the set of numbers
  $(n_x,n_y,n_z)$ defines a mode of the field. Following Maxwell's equations, the momentum vector and the polarization vector obey the constraint 
\begin{equation}
    \mathbf{k}\cdot \boldsymbol{\varepsilon}_{\mathbf{k}} = 0\;,
\end{equation}
which means that the fields are transverse (orthogonal to the propagation direction), and the polarization vector only has two independent directions. 
In the summary that follows, we consider the electric field polarized in a certain direction (for example, the $x$ axis).

The canonical quantization of the electric field is implemented 
as in Eqs.~(\ref{eq:field_expansion}) and (\ref{eq:field-operator-algebra})
by promoting the mode amplitudes 
$\alpha_{\mathbf{k}} $ to quantum annihilation operators. Thus, the quantized electric field takes the form
\begin{equation}
    \hat{\mathbf{E}}(\mathbf{r},t) = \sum_{\mathbf{k}} 
     \boldsymbol{\varepsilon}_{\mathbf{k}} 
     \mathcal{E}_{\mathbf{k}} 
     \hat{a}_{\mathbf{k}} 
      e^{i \mathbf{k} \cdot \mathbf{r}  - i\omega_k t }
    +  \mathrm{H.c.}
    \; , 
\end{equation}
where $\hat{a}_{\mathbf{k}} $ is the corresponding annihilation operator, 
and its Hermitian adjoint is the creation operator $\hat{a}_{\mathbf{k}} ^\dagger$.
 The creation and annihilation operators satisfy a particular form of the 
 general canonical commutation relations~(\ref{eq:field-operator-algebra}) 
 that are a consequence of the basic canonical commutators of conjugate field variables,
 e.g., 
$[\ha_{\mathbf{k}} ,\ha_{\mathbf{k}'}^\dagger] = \delta_{{\mathbf{k}} {\mathbf{k}}'}$, with the other commutators being zero.
Then,
 the Hamiltonian of the electromagnetic fields can be cast in the form
\begin{equation}
    \mathcal{H} = \sum_k \omega_{\mathbf{k}} \!
    \left(\ha^\dagger_{\mathbf{k}}  \ha_{\mathbf{k}}  + \frac{1}{2} \right)\;,
\end{equation}
which, with the removal of the zero-point energy
as in Eq.~(\ref{eq:H_field}), yields
\begin{equation}
    \mathcal{H} = \sum_k \omega_{\mathbf{k}} 
   \,
    \ha^\dagger_{\mathbf{k}} 
     \ha_{\mathbf{k}} 
      \;. \label{eq:QO-Hamiltonian}
\end{equation}
The energy eigenstates build up the Fock space, where the states of the system have any number of particles,
 with $\ket{0}$ being the vacuum state and $\ket{n_1,n_2,\cdots,n_j,\cdots}$ being an excited state with 
 the occupation number $n_i$ referring to the number of photons with momentum $\mathbf{k}_i$. 
The action of the creation and annihilation operator in Fock space is given by
\begin{align}
    \ha_{k_j} \ket{n_1,n_2,\cdots,n_j,\cdots} &= \sqrt{n_{j}} \ket{n_1,n_2,\cdots,n_j - 1,\cdots}\;,\\
    \ha_{k_j}^\dagger \ket{n_1,n_2,\cdots,n_j,\cdots} &= \sqrt{n_{j}+1} \ket{n_1,n_2,\cdots,n_j + 1,\cdots}\;.
\end{align}
Additional details of this field-theory construction are given in Sec.~\ref{sec:scalar-field_curved-ST}.

A common model in quantum optics involves
 the use of single field mode inside a cavity interacting with a two-state atom. 
 In that case, the sum over the momentum modes in Eq.~(\ref{eq:QO-Hamiltonian}) is not enforced
  and the system is described in a Hilbert space corresponding to one photon.

\subsection{Atom-field interaction}
\label{sec:Atom-Field-interaction}

A typical setup consists of an uncharged two-state atom interacting with the electromagnetic field. The Hilbert space of the two-state atom is the same as a spin-1/2 particle, and hence the Hamiltonian can be written as the spin-1/2 Hamiltonian with the spin oriented along the $z$-direction without any loss of generality. The interaction between the atom and the field is modeled by a dipole interaction. Therefore, the total Hamiltonian of the field-atom system can be written as in Eq.~(\ref{eq:total-hamiltonian}):
$
H = H_{\rm at} + H_{\rm em} + H_{\rm int}$,
where we more generally have
\begin{align}
\setlength{\abovedisplayskip}{5pt}
\setlength{\belowdisplayskip}{5pt}
	H_{\rm at} &= E_a \ket{a} \bra{a} + E_b \ket{b} \bra{b}\;, \\
	H_{\rm em} &= \sum_{k}^{}\omega_{\mathbf{k}} \,  \ha_{\mathbf{k}} ^{\dagger} \ha_{\mathbf{k}}  \;,\\
	H_{\rm int} &= \hat{\mathbf{P}}(t)\cdot \hat{\mathbf{E}}(t,\mathbf{r})\;. 
\end{align}
Here, $\ket{a}$ and $\ket{b}$ are the excited and ground state of the atom with energies $E_a$ and $E_b$. The frequency of transition of the atom is then defined as $\nu = E_a - E_b$. The dipole moment operator $\mathbf{P}$ can be written as
	\begin{equation}
	\mathbf{P}(t) = \sum_{i,j} \ket{i}\bra{i}\hat{\mathbf{P}}(t)\ket{j}\bra{j} = \sum_{i,j} \hat{\sigma}_{ij} \boldsymbol{\mathcal{P}}_{ij}\; e^{i(E_i -E_j)t} \label{eq:dipole-quantized}\;,
	\end{equation}
where $\hat{\sigma}_{ij} = \ket{i}\bra{j}$ with $i,j\in\{a,b\}$ and $\boldsymbol{\mathcal{P}}_{ij} = \bra{i}\hat{\mathbf{P}}(0)\ket{j}$ are the dipole moment matrix elements in the atomic basis. Now, the dipole moment matrix has zero diagonal and non-zero off-diagonal terms due to the parity of the dipole moment operator:
\vspace{-0.5em}
	\begin{equation}
	\boldsymbol{\mathcal{P}}_{ij} = e \bra{i}\mathbf{r}\ket{j} = \begin{cases}
	0 \hskip 2.8em \text{if } i=j\\
	\boldsymbol{\mathcal{P}}_{ab} \hskip 2em \text{if } i\neq j
	\end{cases}\;,
	\end{equation}
where $\boldsymbol{\mathcal{P}}_{ab}$ is considered real and $\boldsymbol{\mathcal{P}}_{ab} = \boldsymbol{\mathcal{P}}_{ba}$.

This is the standard setup that justifies the use of a model with a scalar field and an analogue monopole coupling, as in 
Sec.~\ref{sec:QOptics-interactions}.

\section{SPACETIME PHYSICS AND BLACK HOLES}
\label{app:spacetime}
 
Given their observational and theoretical relevance, 
this appendix summarizes the main definitions and properties of the 
geometry of nonextremal Kerr black holes~\cite{GR_Carroll-2003, MTW-gravitation, GR_Wald-1984, frolov}.
As this overview shows, leaving aside some technicalities, the basic properties of the near-horizon region, black hole thermodynamics, HBAR physics, and related CQM are the same as for generalized Schwarzschild black holes
of Sec.~\ref{sec:spacetime_CQM}.

\subsection{Kerr metric in Boyer-Lindquist coordinates}

\begin{figure}[h]
    \centering
    \includegraphics[width=0.5\linewidth]{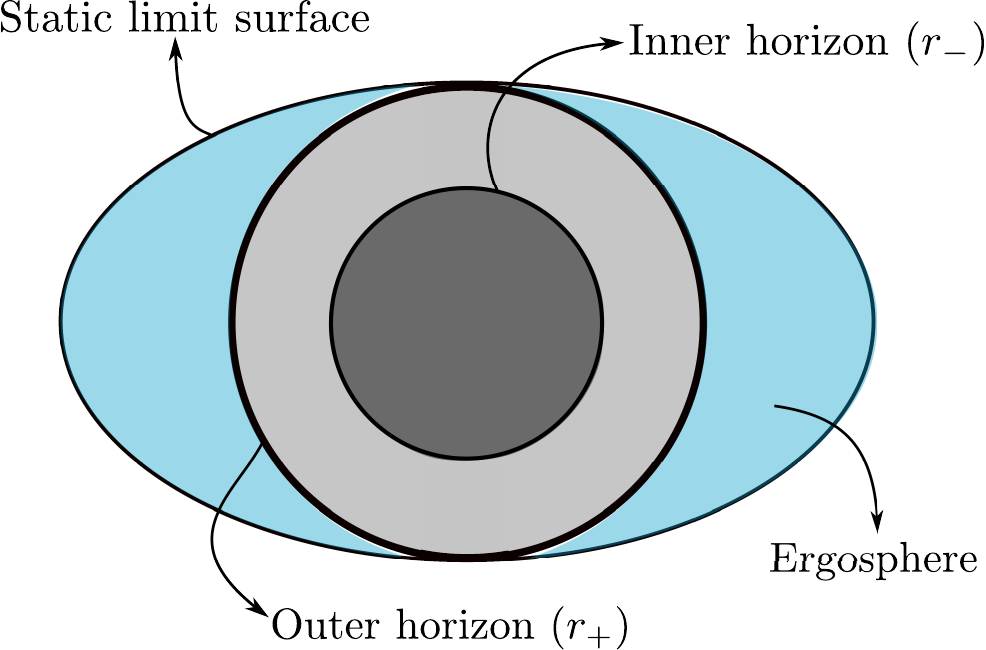}
    \caption{A rotating black hole represented by the Kerr metric~(\ref{eq:Kerr1}). 
    The horizons (outer and innner) and the static limit hypersurfaces
    are characterized by the conditions $g_{\tilde{t} \tilde{t}} \propto \Delta =0$, i.e., Eq.~(\ref{eq:horizon-r_pm});
     and $g_{tt}  \propto \Delta -a^2 \sin^2 \theta = 0$, respectively.
     In the diagram, we are not displaying the structure inside the inner horizon (including the inner static limit),
      which is thought to be unrealizable for astrophysical black holes.}
       \label{fig:Kerr-BHs}
       \hspace*{0in}
\end{figure}

Kerr black holes are the four-dimensional rotating black holes 
that are of current interest in astronomical realizations~\cite{Kerr-astrophysics};
see Fig.~\ref{fig:Kerr-BHs}.
They are described by the Kerr metric~\cite{Kerr-metric_Teukolsky}
as the unique vacuum solution of the Einstein
 field equations in 4D in the presence of a black hole of mass $M$ and angular momentum $J$. 
In Boyer-Lindquist coordinates $(t,r,\theta,\phi)$, the Kerr metric can be written
in a variety of forms. 
Using geometrized natural units with $c=1$ and $G = 1$, in terms of its coordinate components, it reads
\begin{equation}
ds^2 =
-\frac{(\Delta - a^2 \s^2 \T)}{\rho^2} dt^2
- \frac{4 M r}{\rho^2} a \s^2 \T dt d \phi
+\frac{\rho^2}{\Delta}dr^2 +\rho^2d\T^2
+ \frac{\Sigma^2}{\rho^2} \s^2 \T d \phi^2 
\, 
\label{eq:Kerr1}
\end{equation}
where $a=J/M$, called the rotational Kerr parameter, is the angular momentum per unit mass.
 \medskip
In Eq.~(\ref{eq:Kerr1}), the auxiliary quantities $\D$, $\rho$, and $\Sigma$ are given as
\begin{equation}
\begin{aligned}
\D&= r^2 -2Mr + a^2
\;  \;  , \; \; \; \; \; 
\rho^2 = r^2 + a^2 \C^2\T
\; , \\
\Sigma^2 &= \left( r^2 +a^2 \right) \rho^2 + 2 M r a^2 \, \s^2 \T 
= (r^2 +a^2)^2 - \D a^2 \s^2 \T
\; .
\end{aligned}
\end{equation}
It should be noted that all the results and implications of this appendix and in the main text, 
including the HBAR results, are similarly valid with the addition of a black hole electric charge $Q$, 
if the replacement $2Mr \longrightarrow 2Mr - Q^2$ is made 
(which amounts to $a^2 \rightarrow a^2 + Q^2$ within the function $\Delta$,
when all the terms involving $M$ are rewritten in terms of $\Delta$).

A second form of the Kerr metric uses a shift of the angular coordinate $\phi$
to remove the off-diagonal metric term $g_{t\phi}$, by completing the square in Eq.~(\ref{eq:Kerr1}).
With the auxiliary quantity
 \begin{equation}
 \varpi = - \frac{ g_{t\phi} }{ g_{\phi \phi} }
 \; ,
 \label{eq:Kerr-angular-velocity}
 \end{equation}
this insightful form of the metric reads
\begin{equation}
ds^2  = -\frac{\D \rho^2}{\Sigma^2} dt^2 
 + \frac{\rho^2}{\Delta}dr^2 + \rho^2d\T^2    
+ \frac{\Sigma^2}{\rho^2} \s^2 \T \left( d \phi - \varpi dt \right)^2
\; ,
\label{eq:Kerr3} 
\end{equation}
showing that $\varpi$ can be 
 interpreted as a position-dependent angular velocity~\cite{Landau-Lifshitz_CTFs} 
 for the frame dragging of spacetime around the black hole.
 This is a form of the metric that is ideally suited for the near-horizon analysis,
 as it displays a structure similar to the generalized Schwarzschild metrics~(\ref{eq:RN_metric}),
 leading directly to the CQM dominance and ensuing thermodynamic implications.

\subsection{Kerr spacetime symmetries and structure}

 The symmetries of Kerr spacetime
can be analyzed in a manner similar to the Schwarzschild geometry~(\ref{eq:RN_metric}), 
with the Killing vectors of Eq.~(\ref{eq:Killing-vectors}). These are associated with
independence with respect to time $t$ and azimuthal angle $\phi$, as arising from the stationary 
and axisymmetric invariances of the geometry.
In addition, similar techniques (as in Sec.~\ref{sec:BH-geom_horizons}),
can be used for the analysis of the structure of Kerr spacetime, 
including the horizons and details implied by the black hole's rotation.
 Here, two sets of hypersurfaces can be identified as critical boundaries, as follows.
 \begin{itemize}
 \item
{\it Static (stationary) limit hypersurfaces or ergosurfaces ${\mathcal S}^{\pm}$.\/} 
These boundaries are identified by the radii $ r_{e,\pm} =  r_{e,\pm}(\theta)$
that are roots of the equation
\begin{equation}
g_{{t}{t}} = \boldsymbol{\xi}_{({t})} \cdot   \boldsymbol{\xi}_{({t})}=0
 \Longrightarrow  
 \; \; 
 r_{e,\pm}=M\pm\sqrt{M^2-a^2\cos^2\theta}
\label{eq:static-r_pm}
\; 
\end{equation}
(see Fig.~\ref{fig:Kerr-BHs}). 
 On these hypersurfaces ${\mathcal S}^{\pm}$, the norm of $ \boldsymbol{\xi}_{({t})} $ becomes null:
 thus, they are infinite-redshift hypersurfaces, by comparison of time $t$ measurements 
 near $ r_{e,\pm}$ with asymptotic infinity.

In addition, $ \boldsymbol{\xi}_{({t})} $ becomes spacelike for $r_{e,-} <r <   r_{e,+}$
 (being timelike near asymptotic infinity, and generally for
$  r > r_{e,+}$  or $r < r_{e,-} $). Then, a simple analysis shows that 
 timelike geodesics of static observers do not exist within these boundaries ($r_{e,-} <r <   r_{e,+}$)
and all paths are dragged along in the direction of the black hole's rotation. 
(Thus, the hypersurfaces ${\mathcal S}^{\pm}$ are ``{\it static or stationary limits\/}.'')
 
 Finally, these hypersurfaces are not horizons. This can be proved in two ways:
 (i) they are not null because they include a timelike direction corresponding 
  to the diagonal form of the metric~(\ref{eq:Kerr3}); 
  (ii) the geodesic analysis also shows that outgoing timelike or null geodesics do cross the outer
static limit 
$ r_{e,+}=M+\sqrt{M^2-a^2\cos^2\theta}$.
 In essence, the condition $g_{{t}{t}} =0$ does not characterize the 
horizon because it fails to account for the effect on geodesics of the black hole's
  rotational degree of freedom.
 
 \item
 {\it Outer and Inner Horizons ${\mathcal H}^{\pm}$.\/} 
Effectively, these hypersurfaces are identified by the radii $ r_{\pm} $ that give the locations of the outer ($r_{+}$)
and inner ($r_{-}$) horizons, via the roots of the equation
\begin{equation}
\D=0 \Longrightarrow 
\; \; 
r_\pm = M\pm (M^2-a^2)^{1/2}
\label{eq:horizon-r_pm}
\; 
\end{equation}
(see Fig.~\ref{fig:Kerr-BHs}). 
For the outer horizon: outgoing timelike or null geodesics do not exist crossing this surface (they are all ingoing),
when the value $r=r_{+}$ is reached; this is indeed the characterization of a stationary event horizon (see below).
\end{itemize}

The structure~(\ref{eq:horizon-r_pm}) with two horizons
 is also found for the Reissner-Nordstr\"{o}m (RN)
 case with electric charge, within the class of generalized Schwarzschild metrics~(\ref{eq:RN_metric}).
 However, for Kerr spacetimes, new features emerge due to the rotation 
 that breaks spherical symmetry and generates the additional static limit hypersurfaces. 
 Specifically, the Kerr-spacetime structure includes the regions known as ergospheres 
 (outer: $r_{+} < r< r_{\rm e,+}$ and  inner: $r_{-,e} < r< r_{\rm -}$).
See Fig.~\ref{fig:Kerr-BHs}; the outer ergosphere is most often discussed in the literature, 
 as it lies outside the event horizon, and has many surprising features associated with frame dragging, 
 including the Penrose process~\cite{GR_Carroll-2003, MTW-gravitation, GR_Wald-1984} 
 and astrophysical implications~\cite{Kerr-astrophysics, Kerr-metric_Teukolsky}.
For the analysis of interest in most practical applications, including acceleration radiation (as in this review article),
the non-extremal geometry is considered, as defined by the 
 physical condition $M>a$, 
which amounts to  $\D'_{+} \equiv \D'(r_{+}) = r_+-r_- \neq 0$, and where 
the prime stands for the radial derivative;
this excludes naked singularities  ($M< a$)~\cite{GR_Carroll-2003, GR_Wald-1984}.

Most importantly, the horizon condition~(\ref{eq:horizon-r_pm})
 can be understood by considering the black hole's rotational degree of freedom.
 A useful derived parameter is the black hole's angular velocity
 $\Omega_{H} $, which is the horizon limit
of the frame-dragging angular velocity~(\ref{eq:Kerr-angular-velocity}):
 \begin{equation}
\Omega_{H} 
= \lim_{r \rightarrow r_{+}}  \varpi =  \frac{a}{r_+^2+a^2} =
 \frac{a}{2M r_{+}}
 \, .
 \label{eq:BH-angular-velocity}
\end{equation}
This parameter is proportional to the black hole's 
angular momentum $J$ and appears in other important physical quantities. 
Moreover, $\Omega_{H}$ can be given
an operational physical meaning by considering a particle approaching the event horizon along a geodesic:
 its angular velocity $\Omega = d\phi/dt$ is restricted to an increasingly narrow range and 
is forced to become $\Omega_{H}$ as $r \rightarrow r_{+}$.
Further insight into horizon and near-horizon properties
is gained by considering the corotating Boyer-Lindquist coordinates~\cite{parkerprl},
\begin{equation}
\Tilde{t}=t  
\; , \; \; \; 
\Tilde{\phi}=\phi-\Omega_H t
\; ,
\label{eq:rotating-coords}
\end{equation}
associated with fiducial stationary observers corotating with the black hole itself.
The coordinates~(\ref{eq:rotating-coords}) 
make the metric~(\ref{eq:Kerr3}) diagonal to leading order in the near-horizon region and 
exactly diagonal at ${\mathcal H}$.
In addition, these coordinates naturally select a corotating Killing vector relevant for the spinning event horizon 
as the well-known linear combination~\cite{GR_Carroll-2003, MTW-gravitation, GR_Wald-1984} 
\begin{equation}
\boldsymbol{\xi}  
\equiv \boldsymbol{\xi}_{(\tilde{t})}
= \partial_{\tilde{t}}
=
 \boldsymbol{\xi}_{(t)} + \Omega_{H} \boldsymbol{\xi}_{(\phi)}
\; .
\label{eq:Killing-vector-corotating}
\end{equation}

The Killing vector $\boldsymbol{\xi}$ of Eq.~(\ref{eq:Killing-vector-corotating})
 is timelike outside the event horizon and becomes null at the event horizon $\mathcal{H} \equiv \mathcal{H}^{+}$,
 where
  \begin{equation}
 g_{\tilde{t}\tilde{t}}  = \boldsymbol{\xi}_{(\tilde{t})} \cdot   \boldsymbol{\xi}_{(\tilde{t})}
 = - \frac{\Delta \rho^2}{\Sigma^2} = 0
\; .
\label{eq:g_tilde-tt-null}
 \end{equation}
 This explains the criterion of Eq.~(\ref{eq:horizon-r_pm}) for the selection of Kerr horizons.
Furthermore, Eqs.~(\ref{eq:horizon-r_pm}) and (\ref{eq:g_tilde-tt-null}) select a hypersurface $ \mathcal{H}^{\pm}$, 
with a null tangent direction along $\boldsymbol{\xi} $, and two angular directions that are spacelike---this can be
deduced from the diagonal form of the metric~(\ref{eq:Kerr3}). 
 By definition, this makes $ \mathcal{H}^{\pm}$ a null hypersurface
 generated by light rays. 
 Incidentally, this finding implies that the null direction is also normal to $ \mathcal{H}^{\pm}$; in the chosen
 Boyer-Lindquist coordinates, the normal corresponds to the radial coordinate, 
and the vanishing of its norm is fixed by the following condition on the metric
 component: $g^{rr} \propto \Delta =0$.
 As for the case of generalized Schwarzschild black holes, 
 such null hypersurface is a horizon.
For the outer hypersurface at  $r=r_{+}$, this is an event
horizon, where all timelike or null geodesics are ingoing and cannot go back to infinity; 
the inner hypersurface at $r=r_{-}$ is classified as a Cauchy horizon~\cite{GR_Carroll-2003}.
 Finally, this simple set of arguments indicates that
  $\boldsymbol{\xi}$ describes a spacetime evolution in a corotating frame 
  with the same basic characteristics found in the neighborhood of a generalized Schwarzschild black hole. 
Therefore, one can predict that the near-horizon physics will exhibit an analog behavior, 
as in Sec.~\ref{sec:scalar-field-nh}, governed by scale symmetry in the form of CQM---this is verified in this appendix for 
a scalar field.

As in Sec.~\ref{sec:spacetime_CQM},
the two most relevant geometrical quantities for quantum thermodynamics
are the surface gravity and the black hole horizon area.
From Eq.~(\ref{eq:surface-gravity}), with $r =r_{+}$,
the surface gravity of a Kerr black hole takes the form 
\begin{equation}
\kappa= \frac{ \D'_{+} }{ 2 (r_{+}^2 +a^2) }
\; ,
\label{eq:surface-gravity_Kerr}
\end{equation}
which is again proportional to the Hawking temperature~(\ref{eq:Hawking-temperature}).
And the horizon area, generally defined via
\begin{equation}
A= \int \sqrt{
g_{\theta \theta} g_{\phi \phi} } \, d \theta d \phi
= 4 \pi (r_{+}^{2}+a^{2}) 
\; ,
\label{eq:area_Kerr}
\end{equation}
is also proportional to the Bekenstein-Hawking entropy~(\ref{eq:BH-entropy}).
In particular, from 
Eqs.~(\ref{eq:horizon-r_pm}) and (\ref{eq:surface-gravity_Kerr}), this implies that
the area changes are 
\begin{equation}
  \delta A  = 8 \pi   \, \frac{ \delta \tilde{M} }{ \kappa}
  \; ,
  \label{eq:BH-area-changes-geom_Kerr}
  \end{equation}
where $ \delta \tilde{M}$ is the corotating energy change of Eq.~(\ref{eq:tilde-M-change}).
Equation~(\ref{eq:BH-area-changes-geom_Kerr})
extends Eq.~(\ref{eq:BH-area-changes-geom}) and its RN generalizations to 
incorporate the black hole's angular momentum.

\subsection{Field theory and near-horizon approximation on Kerr spacetime}

For a quantum field theory on Kerr spacetime, we consider a real scalar field $\Phi$ 
as the simplest representative of the relevant behavior of acceleration radiation. This mirrors our
treatment in the main text of this article; see Sec.~\ref{sec:scalar-field_curved-ST}. 

The Klein-Gordon equation~(\ref{eq:Klein_Gordon_basic})
with mass $\mu_\Phi$ and possibly nonminimal coupling $\xi$, i.e.,
i.e.,
\begin{displaymath}
\frac{1}{ \sqrt{-g} }\partial_{\mu} \left(\sqrt{-g} \,g^{\mu \nu}\,\partial_{\nu} \Phi\right)- (\mu_\Phi^{2} + \xi R) \Phi= 0
\; ,
\end{displaymath}
 in the geometric background of a Kerr spacetime metric~(\ref{eq:Kerr1}), takes the form
\begin{equation}
\begin{aligned} 
-\frac{\Sigma^2}{\Delta} \frac{\partial^2 \Phi}{\partial t^2}
- \frac{4 M r a}{\Delta}  \frac{\partial^2 \Phi}{\partial t \, \partial \phi}
&
+ \left(\frac{1}{\s^2 \theta} - \frac{a^2}{\Delta} \right) \frac{\partial^2 \Phi}{\partial \phi^2}
\\ & 
+ \frac{\partial}{\partial r} \left( \Delta  \frac{\partial \Phi}{\partial r} \right)
+ \frac{1}{\s \theta}  \frac{\partial}{\partial \theta}
 \left( \s \theta  \frac{\partial \Phi}{\partial \theta} \right) 
- \mu_\Phi^{2}  \rho^2 \Phi = 0 \; .
\label{eq:Kerr_Klein_Gordon_spatial}
\end{aligned}
\end{equation}
 The general solution of Eq.~(\ref{eq:Kerr_Klein_Gordon_spatial}) can be written in terms of
 a set of modes $ \phi_{\tilde{\omega} l m} (t, \boldsymbol{r})$ in separable form,
\begin{equation}
\phi_{{\omega} l m} (t, \boldsymbol{r}) \equiv
    \phi_{\boldsymbol{s}} (t, r,\Omega) 
    = R_{\boldsymbol{s}} (r) S_{ lm} (\theta) e^{im\phi} e^{-i\omega t} 
\; ,
\label{eq:rotating-coords_separation_1}
\end{equation}
where 
$\boldsymbol{r}=(r,\theta, \phi)$, and
$S_{ lm} (\theta) $
 are oblate spheroidal wave functions of the first kind~\cite{frolov,spheroidal,spheroidal-2},
which satisfy the angular equation 
\begin{equation}
\frac{1}{\s \theta}  \frac{d}{d \theta} \left( \s \theta  \frac{d S}{d \theta} \right) 
+
\left[ a^2 \omega^2 \cos^2 \theta - \frac{m^2}{\sin^2 \theta} 
+  \Lambda_{{\boldsymbol{s}}}
-
a^2 \mu_\Phi^2  \cos^2 \theta \right]
 S = 0
 \label{eq:Kerr_Klein_Gordon_polar}
\; .
\end{equation}
The normalized combination $  Z_{lm}(\Omega) =(2 \pi)^{-1/2}
S_{ lm} (\theta)  e^{im\phi}  $ is usually called a spheroidal harmonic.
Equation~(\ref{eq:Kerr_Klein_Gordon_polar}) is a Sturm-Liouville problem where
the separation constant $ \Lambda_{{\boldsymbol{s}}} $ is an eigenvalue of a self-adjoint operator---thus, 
the regular solutions form a complete orthogonal set labeled by a discrete ``spheroidal'' number $l$.
In addition, $\Lambda_{{\boldsymbol{s}}}$ and the associated solutions
depend on the discrete quantum numbers $l$ and $m$ as well 
the continuous parameters $\mu_\Phi$ and $\omega$ that
appear in the dimensionless combinations $ a \mu_\Phi $ and $a \omega $.
Then, the separation of variables of Eq.~(\ref{eq:rotating-coords_separation_1}) gives the
corresponding radial function $R(r)$ subject to the equation
\begin{equation}
 \frac{d}{d r} \left( \D  \frac{d R}{d r} \right) +
\left[\frac{(r^2+a^2)^2\omega^2 - 4 Mra m \omega + a^2 m^2}{\D}
 - \Lambda_{{\boldsymbol{s}}} - a^2 \omega^2 - \mu_\Phi^2 r^2
\right] R = 0
\label{eq:Kerr_Klein_Gordon_radial}
\, .
\end{equation}

The near-horizon approximation can be implemented 
with $x \equiv r-r_+ \ll r_+$,
using the expansions
$\D(r)  \stackrel{(\mathcal H)}{\sim}  \D'_{+}  \, x \left[ 1 + O(x) \right]$,
and $\D'(r)  \stackrel{(\mathcal H)}{\sim}  \D'_{+} \left[ 1 + O(x) \right]$,
where $ \D'_{+}  = r_{+} - r_{-}$.
The derivation can proceed in either one of two ways. 
In the first approach, the radial equation~(\ref{eq:Kerr_Klein_Gordon_radial})
is the starting point for the near-horizon expansion;
then, the  leading terms as $r \rightarrow r_{+}$
are the one with radial derivatives and the ones with an inverse $\Delta $ factor,
where square completion can be used, along with the black hole angular velocity $\Omega_H$
of Eq.~(\ref{eq:BH-angular-velocity},) to get a shifted frequency 
\begin{equation}
\tilde{\omega}= \omega - \Omega_{H} m
\; .
\end{equation}
The resulting leading-order near-horizon radial equation becomes
\begin{equation}
\left[\frac{1}{x} \frac{d}{d  x} \left( x  \frac{d}{d x} \right) 
+ \left( \frac{ {\tilde{\omega}}}{f'_{+}} \right)^{2}
\frac{1}{x^2} \right] R(x)
\stackrel{(\mathcal H)}{\sim} 0 
 \; \; , \; \; \;  \text{with} \; \displaystyle f'_{+} =  \frac{ {\D'_{+}} }{ ({r_{+}^2+a^2}) } 
\; ;
\label{eq:Kerr_Klein_Gordon_conformal-R_Kerr}
\end{equation}
and, via a Liouville transformation $R(x) \propto x^{-1/2} u(x)$,
\begin{equation}
u''(x)+\frac{ \lambda_{} }{x^{2}}
\,\left[ 1 + {O}(x) \right]u (x)=0
\; \; , \; \; \;
\lambda_{} = \frac{1}{4} + \Theta^{2}\, , \; \; \; \;
 \Theta= \frac{\tilde{\omega}}{  f'_{+} } \equiv  \frac{\tilde{\omega}}{  2 \kappa  }
\;  ,
\label{eq:Klein_Gordon_conformal_Kerr}
\end{equation}  
with the Kerr surface gravity $\kappa$ of Eq.~(\ref{eq:surface-gravity_Kerr}). 
Equations~(\ref{eq:Kerr_Klein_Gordon_conformal-R_Kerr}) and (\ref{eq:Klein_Gordon_conformal_Kerr})
are the CQM expressions for a Kerr black hole, which can be viewed as the analogues of the 
Schwarzschild-like metric Eqs.~(\ref{eq:Kerr_Klein_Gordon_conformal-R})--(\ref{eq:conformal_interaction}),
with $\omega$ replaced by $\tilde{\omega}$. 

In the second approach, the alternative expression for the Kerr metric of Eq.~(\ref{eq:Kerr3}) gives 
additional insight into the conformal nature of the near-horizon equations.
Indeed, the covariant metric~(\ref{eq:Kerr3}) describes the physics in a corotating frame 
with angular velocity $\varpi$.
In this interpretation, a shifted azimuthal coordinate is implicitly given by $d \phi -  \varpi dt$.
As the near-horizon region is approached with $r \rightarrow r_{+}$,
this frame becomes the black hole's corotating frame, 
with an angular velocity $\varpi \rightarrow \Omega_{H}$, and with the coordinate transformation to the
 corotating Boyer-Lindquist coordinates~(\ref{eq:rotating-coords}), i.e.,
$\Tilde{t}=t  $
and $\Tilde{\phi}=\phi-\Omega_H t$. This transformation converts 
Eq.~(\ref{eq:rotating-coords_separation_1})
(exactly, and not just in the near-horizon region) into
\begin{equation}
\phi_{\tilde{\omega} l m} (\boldsymbol{r}, t) 
= R(r) S(\theta) e^{im\tilde{\phi}}  e^{-i\tilde{\omega} \tilde{t}}
\; , \; \; \; 
\tilde{\omega}=\omega-m\Omega_H
\; .
\label{eq:rotating-coords_separation_2}
\end{equation}
Here, the timelike behavior of the associated Killing vector 
$ \boldsymbol{\xi}_{(\tilde{t})} =  \boldsymbol{\xi} $ 
allows for positive frequency modes of the Klein-Gordon equation~(\ref{eq:Klein_Gordon_basic}) within the ergosphere
and near the horizon, via the condition
 \begin{equation}
\boldsymbol{\xi}_{(\tilde{t})} \phi_{\boldsymbol{s}}  =  - i \tilde{\omega} \phi_{\boldsymbol{s}}
 \; .
 \end{equation}
 Then, inversion of the metric gives the contravariant components needed for the Klein-Gordon equation~(\ref{eq:Klein_Gordon_basic}); moreover, making the transition to 
the near-horizon region, instead of Eq.~(\ref{eq:Kerr_Klein_Gordon_spatial}) or
Eq.~(\ref{eq:Kerr_Klein_Gordon_radial}), one can directly write
\begin{align}
&
\left[
- \frac{\Sigma^2}{\rho^2 \Delta} \frac{\partial}{\partial  \tilde{t}^2}
+
\frac{\rho^2 }{\Sigma^2 \s^2 \T } \frac{\partial}{\partial  \tilde{\phi}^2}
+
\frac{1}{\rho^2} \frac{\partial}{\partial  r} \left( \D  \frac{\partial}{\partial r} \right)
+
\frac{1}{\rho^2} \frac{\partial}{\partial  \T^2}
\right]
\Phi
\label{eq:Kerr_Klein_Gordon_conformal-with-time_0}
\\
&
\stackrel{(\mathcal H)}{\sim}
\left[
- \frac{(r^2+a^2)^2}{\rho^2 \Delta} \frac{\partial}{\partial  \tilde{t}^2}
+
\frac{1}{\rho^2} \frac{\partial}{\partial  r} \left( \D  \frac{\partial}{\partial r} \right)
\right] \Phi
\stackrel{(\mathcal H)}{\sim}
0
\; ,
\label{eq:Kerr_Klein_Gordon_conformal-with-time}
\end{align}
due to the leading behavior 
$\D(r)  \stackrel{(\mathcal H)}{\sim}  \D'_{+}  \, x$, which selects the radial-time sector of the metric.
Equation~(\ref{eq:Kerr_Klein_Gordon_conformal-with-time})
reproduces again the asymptotically exact equation~(\ref{eq:Kerr_Klein_Gordon_conformal-R_Kerr}).
It should be noted that the first three terms of Eq.~(\ref{eq:Kerr_Klein_Gordon_spatial}), 
 involving derivatives with respect to $t$ and $\phi$, actually combine to the diagonal
 form of the metric exhibited in Eq.~(\ref{eq:Kerr3}), i.e., they correspond to
 $\left[ g^{tt} \left( \partial_{t} + \varpi \partial_{\phi} \right)^{2} + (1/g_{\phi \phi} ) \partial_{\phi}^2 \right] \Phi$ (after 
 removal of an overall factor $1/\rho^2$); this provides a direct link between the two approaches (fixed vs 
 corotating frames, and their near-horizon limits).
 
The assignments leading to CQM, via
Eqs.~(\ref{eq:Kerr_Klein_Gordon_conformal-R_Kerr}) and (\ref{eq:Klein_Gordon_conformal_Kerr}),
 show that one can define an equivalent scale factor
  \begin{equation}
   f (r) \equiv \frac{ {\D} }{ ({r^2+a^2}) }
  \; ,
      \label{eq:metric-time-scale-factor}
  \end{equation}
  for comparison with the analogue Schwarzschild-like metrics.
  [For example, Eq.~(\ref{eq:surface-gravity_Kerr}) conforms to this replacement for $\kappa = f_{+}'/2$.]
   In essence, this argument shows that,
   \begin{quotation}
   \noindent
   with the factor $f(r)$ of Eq.~(\ref{eq:metric-time-scale-factor}), 
  {\it the near-horizon Kerr metric is a generalized Schwarzschild metric~(\ref{eq:RN_metric})
   in the corotating frame.\/}
  \end{quotation}
Finally, as in Sec.~\ref{sec:scalar-field-nh}, and applying Eq.~(\ref{eq:rotating-coords_separation_2}),
  a pair of linearly independent solutions is given by 
  $u(x)\propto  x^{1/2  \pm i\Theta}$, so that
  \begin{equation}
\Phi^{ \pm {\rm \scriptscriptstyle (CQM)} }_{\omega l m} \propto 
 x^{\pm i \Theta} 
 e^{im\tilde{\phi}}  S_{ lm} (\theta)  e^{-i \tilde{\omega} \tilde{t} } 
\label{eq:CQM_modes_Kerr}
\; ,
\end{equation}
 thus generalizing Eq.~(\ref{eq:CQM_modes}).
In addition, a crucially important property of this analysis is noteworthy.
The radial equation~(\ref{eq:Kerr_Klein_Gordon_conformal-R_Kerr})
 is the particular scalar case (for spin $s=0$) 
of the Teukolsky equation~\cite{frolov} valid for arbitrary field spin with the same structural form. As a result,
the conformal behavior displayed in the near-horizon approximation (see below) is universal:
 it is exhibited by all fields, with arbitrary spin, in the background of generic black holes.

The geodesic equations are also obtained by a similar analysis as in Sec.~\ref{sec:CQM},
in terms of the specific energy and 
 axial component of angular momentum, defined via Eqs.~(\ref{eq:conserved-quantities}),
 in addition to the invariant mass $\mu$ and the Carter constant $\mathcal{Q}$~\cite{MTW-gravitation}.
In this case, if the second-order geodesics are written explicitly, a near-horizon limiting procedure
yields the same for of Eqs.~(\ref{eq:tau_in_x}) and (\ref{eq:t_in_x}).
The coordinate $\phi$ also needs to be specified, and has an expansion with another logarithmic term; but
the relevant corotating coordinate has a simple linear expression
$ \tilde{\phi} \stackrel{(\mathcal H)}{\sim} \alpha x + O (x^2)$.
The coefficients $k$, $C$, and $\alpha$, which
 are expressed in terms of the conserved quantities and a limiting value of the coordinate 
$\theta$, do not play a direct role in the radiation formulas.
Instead, as in Sec.~\ref{sec:CQM}, it is the logarithmic term in the coordinate time expansion~(\ref{eq:t_in_x}),
combined with a similar term of the radial part of the mode, 
that completely determines the final Planck form of the radiation equations.

In closing, there is one final remark on the Kerr metric that involves a technicality related to the vacuum.
The choice of vacuum states is a subtle and important physical criterion in curved spacetime~\cite{birrell-davies}.
The definition of a Boulware-like vacuum needed for the HBAR thought experiment
is subtle because of the superradiant modes~\cite{Unruh-Starobinsky-rad_1,Unruh-Starobinsky-rad_2,frolov}.
This issue, along with the resolution,
are discussed in Refs.~\cite{acceler-rad-Kerr, acceler-rad-Qopt-2}.
This difficulty can be bypassed via the introduction of a boundary that excludes the regions of asymptotic infinity, e.g.,
using a mirror.  Any such properly controlled field mode would
qualify as Boulware-like and is suitable for the generation of HBAR radiation.

\end{appendix}


\end{document}